\documentclass{aastex}
\shorttitle{The ALFA ZOA Deep Survey: First Results}
\shortauthors{T. McIntyre et al.}

\begin{document}
\title{The ALFA ZOA Deep Survey: First Results}

\author{T. P. McIntyre,\altaffilmark{1} P. A. Henning,\altaffilmark{1} R. F. Minchin,\altaffilmark{2} E. Momjian,\altaffilmark{3}  Z. Butcher,\altaffilmark{4}}
\altaffiltext{1}{Department of Physics and Astronomy, University of New Mexico, 1919 Lomas Blvd NE, Albuquerque, NM 87131.}
\email{tpm@unm.edu}
\altaffiltext{2}{Arecibo Observatory, HC03 Box 53995, Arecibo, PR 00612.}
\altaffiltext{3}{NRAO, Domenici Science Operations Center, PO Box O, 1003 Lopezville Rd, Socorro, NM 87801.}
\altaffiltext{4}{Department of Astronomy, University of Massachusetts Amherst, 710 N Pleasant St., Amherst, MA 01003-9305.}

\begin{abstract}
The Arecibo L-Band Feed Array Zone of Avoidance (ALFA ZOA) Deep Survey is the deepest and most sensitive blind H\,{\sc{i}} survey undertaken in the ZOA. ALFA ZOA Deep will cover about 300 square degrees of sky behind the Galactic plane in both the inner ($30^{\circ} \le l \le 75^\circ; b \le |2^\circ|$) and outer ($175^\circ \le l \le 207^\circ; -2^\circ \le b \le +1^\circ$) Galaxy, using the Arecibo Radio Telescope. First results from the survey have found 61 galaxies within a 15 square degree area centered on $l = 192^\circ$ and $b = -2^\circ$. The survey reached its expected sensitivity of rms = 1 mJy at 9 km s$^{-1}$ channel resolution, and is shown to be complete above integrated flux, $F_{HI}$ = 0.5 Jy km s$^{-1}$. The positional accuracy of the survey is $28^{\prime\prime}$ and detections are found out to a recessional velocity of nearly 19,000 km s$^{-1}$. The survey confirms the extent of the Orion and Abell 539 clusters behind the plane of the Milky Way and discovers expansive voids, at 10,000 km s$^{-1}$ and 18,000 km s$^{-1}$. 26 detections (43\%) have a counterpart in the literature, but only two of these have known redshift. Counterparts are 20\% less common beyond $v_{hel}$ = 10,000 km s$^{-1}$ and 33\% less common at extinctions higher than $A_{B}$ = 3.5 mag. ALFA ZOA Deep survey is able to probe large scale structure beyond redshifts that even the most modern wide-angle surveys have been able to detect in the Zone of Avoidance at any wavelength.
\end{abstract}


\section{Introduction}

It has been suggested (Loeb \& Narayan 2008) that undiscovered mass behind the Milky Way may explain the discrepancy between the cosmic microwave background dipole and the mass density dipole (Erdogdu et al. 2006), which drives the peculiar velocity of the Local Group. Neutral hydrogen (H\,{\sc{i}}) surveys of the Zone of Avoidance (ZOA) can successfully map large scale structure in the Universe behind the Milky Way (ALFA ZOA; Henning et al. 2010, HIZOA; Henning et al. 2005). While shallow H\,{\sc{i}} surveys have uncovered a large, spiral galaxy at $\sim$ 3 Mpc (Kraan-Korteweg et al. 1994), and more sensitive surveys have discovered nearby dwarf galaxies (McIntyre et al. 2011, Massey et al. 2003, Begum et al. 2005), H\,{\sc{i}} ZOA surveys have shown that there are no hidden Local Group galaxies with neutral hydrogen mass, $M_{HI}$, greater than $M_{HI} \sim 10^6$ M$_{\odot}$ in the southern sky (Henning et al. 2000) and $M_{HI} \sim 10^7$ M$_{\odot}$ in the northern sky (Henning et al. 1998). A nearby massive spiral ($10^{12} M_\odot$) behind the Milky Way can now largely be ruled out as a way to recover the mass dipole vector, though H\,{\sc{i}} surveys are not sensitive to giant elliptical galaxies (Roberts \& Haynes 1994). A galaxy cluster ($10^{15} M_\odot$) at 20 Mpc could affect the mass dipole vector, and results from HIZOA in the southern sky are forthcoming. Most of the northern sky is only starting to be surveyed deeply enough to detect a cluster at 20 Mpc (EBHIS; Kerp et al. 2011).

More recent results have shown that our knowledge of the bulk flow of peculiar velocities within 60 h$^{-1}$ Mpc (h$^{-1}$ = H$_{0}$/100, H$_{0}$ is the Hubble constant in km s$^{-1}$ Mpc$^{-1}$) is complete enough to be constrained by cosmic variance and so no new galaxy discoveries within this distance will result in a significant correction to the dipole vector (Watkins et al. 2009). Asymmetry in large scale structure beyond 100 h$^{-1}$ Mpc remains as a likely explanation (Lavaux et al. 2010), though even the most modern all sky redshift surveys retain a Zone of Avoidance at these distances (Huchra et al. 2012), forcing bulk motion studies to ``fill in'' the ZOA by cloning galaxies above and below the Plane (Lavaux \& Hudson 2011).

The Arecibo L-Band Feed Array Zone of Avoidance (ALFA ZOA) Deep Survey is a blind survey for galaxies in the ZOA out to distances of 200 h$^{-1}$ Mpc. The full survey will cover nearly 300 square degrees behind both the inner and outer Galaxy regions visible with the Arecibo Telescope, with a sensitivity of 1 mJy per beam (at 9 km s$^{-1}$ velocity resolution). The sensitivity and depth of ALFA ZOA will probe large scale structure out to 200 h$^{-1}$ Mpc in an area of sky that intersects with known overdensities: the Orion Cluster, Abell 539, Abell 634, and  CIZA J0603.8+2939 as well as C1, C5, C10, C21, and C31 (Erdogdu et al. 2006). It will also probe the Orion, Canis Major, and Lepus Voids.
The survey will detect H\,{\sc{i}} to greater depth than recent, large H\,{\sc{i}} surveys (ALFALFA; Haynes et al. 2011, HIPASS; Zwaan et al. 2005), allowing ALFA ZOA to probe environments at farther distances. The low mass end of the H\,{\sc{i}} mass function (HIMF) has been shown by some studies to have a significantly steeper slope when measured at high sensitivity at distances outside the local Universe (Arecibo Galaxy Environment Survey ``AGES''; Davies et al. 2011, Arecibo Ultra Deep Survey ``AUDS''; Freudling et al. 2011). If confirmed, this could have a major impact on the amount of hydrogen gas available for ongoing star formation and galaxy evolution. ALFA ZOA Deep is similar in sensitivity to AGES and surveys $10^3$ times more area than AUDS, putting it in a unique position to check the latest HIMF results as an unbiased, blind survey.

This paper describes the observation, data reduction, and source detection and parameterization techniques for the ALFA ZOA Deep Survey. It presents a catalog of the first results of the survey, coming from a 15 square degree region, and uses these results to analyze survey performance. Section \ref{survey} describes the observation technique of the survey. Section \ref{datared} discusses the data reduction process. Section \ref{search} describes the survey search method and source parameterization. Section \ref{catalog} presents the first results catalog. Section \ref{performance} discusses survey accuracy, sensitivity, and effectiveness. Section \ref{LSS} discusses large scale structure uncovered by the survey. Section \ref{conclusion} is the conclusion.

\section{\label{survey} Survey Design}

\subsection{Receiver}

ALFA ZOA utilizes the ALFA receiver on the 305m Arecibo Radio Telescope in Puerto Rico. The ALFA receiver consists of seven independent beams with two orthogonal linear polarizations each, allowing the survey to cover the same area seven times faster than a single beam receiver. The center beam is surrounded by six outer beams in a hexagonal pattern. The receiver covers frequencies 1 - 2 GHz and the FWHM at 1.4 GHz is approximately 3.4$^{\prime}$ per beam. The gain of the seven ALFA beams ranges between 8.5 and 11 K Jy$^{-1}$, and the system temperatures range between 26 and 30 K.

Observations are recorded in one-second integrations using the Mock Spectrometer, which performs an ``on the fly'' Fast Fourier Transform each time the voltage is sampled. The digital spectrometer covers 300 MHz from 1225 MHz to 1525 MHz using two overlapping 172 MHz sub-bands centered on 1300 MHz and 1450 MHz. The sub-bands overlap between 1364 MHz and 1386 MHz so that there is no loss of sensitivity due to roll off at the edges of the bandpass. Each sub-band is divided into 8192 channels, producing a spectral resolution of 21 kHz. This equates to a velocity resolution of $\sim$ 4.5 km s$^{-1}$ for neutral hydrogen emission. 

\subsection{Observations}
The ALFA ZOA Deep survey covers nearly 300 square degrees through both the inner ($30^{\circ} \le l \le 75^\circ; b \le |2^\circ|$) and outer ($175^\circ \le l \le 207^\circ; -2^\circ \le b \le +1^\circ$) Galaxy. ALFA ZOA takes data simultaneously with a survey for pulsars in the Milky Way (PALFA; e.g. Cordes et al. 2006) and a Galactic radio recombination line survey (SIGGMA; Liu et al. 2013), as well as several SETI groups that receive data from the ALFA receiver (Astropulse; Von Korff et al. 2013, SETI@home; Anderson et al. 2002, Serendip V.v; Cobb et al. 2000). In particular, the first fast radio burst detected with an instrument other than the 13-beam receiver of the Parkes Radio Telescope (Spitler et al. 2014) was discovered by PALFA using the same data taken by ALFA ZOA for this paper.

\begin{figure}[h]
\centering
\includegraphics[scale=0.4]{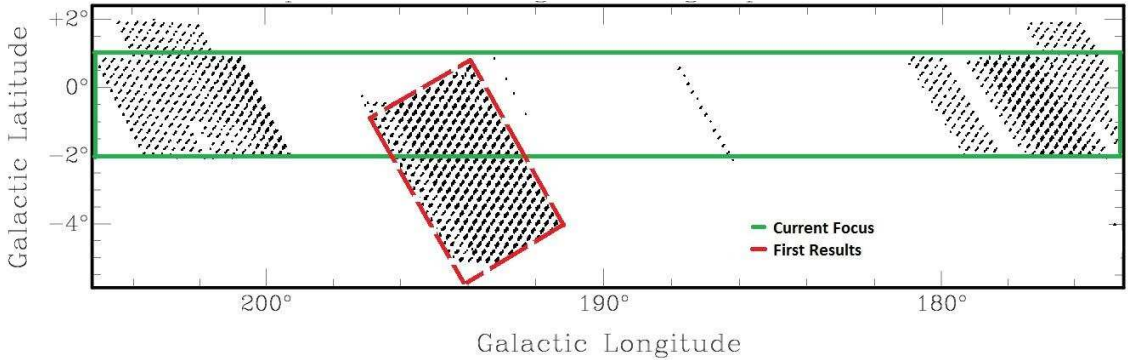}
\caption{\label{deepcoverage} \footnotesize Sky coverage of the survey in the Outer Galaxy in Galactic coordinates. Each black dot is a central beam pointing. The area inside the dashed, red box is the focus of the First Results Catalog. The area inside the solid, green box is the focus for the full survey.}
\end{figure}

Observations in the outer Galaxy are controlled by ALFA ZOA, and the inner Galaxy is controlled by PALFA. The setup of observations is slightly different between the inner and outer regions and this is discussed in the next section. The first results presented in this paper come from a completely surveyed, 15 square degree area in the outer Galaxy centered on l = 192$^\circ$ and b = -2$^\circ$. The area was chosen to intersect large scale structure predicted from known structure above and below the Galactic Plane. The dimensions are 330$^{\prime}$ $\times$ 164$^{\prime}$, ranging across right ascensions 05:55:30 to 06:18:20 and declinations +14:30:00 to +17:12:00. Observations were conducted from December 2010 to March 2012. Figure \ref{deepcoverage} shows a map of the survey area in the outer Galaxy including pointings that have been observed so far.

\subsection{Tiling Pattern}
Commensal observations are an incredibly efficient use of telescope time (e.g. five other projects listed above take data simultaneously with ALFA ZOA) but require compromises between commensal partners. While drift scans have been shown to produce superior baselines for spectral line surveys (Briggs et al. 1997), the ALFA ZOA Deep Survey is constrained to tracking single pointings due to commensal obligations. One method to bandpass correct a single pointing is self-subtraction of a bandpass by its median filter, though this was shown to be inadequate for use as a data reduction procedure in a blind H\,{\sc{i}} galaxy survey (McIntyre 2013a). As such, the survey uses the position switching method for bandpass correction, a non-standard observing mode for extragalactic ALFA surveys, requiring ALFA ZOA Deep to observe position switched pairs and develop unique software for data reduction. Position switching uses a total power on-off technique to remove the bandpass from a source (ON) by subtracting the bandpass from a position off-source (OFF). ALFA ZOA uses the geometry of the ALFA pattern to generate a list of pointings that most efficiently covers the survey area and is designed so that several different pointings can be tracked over the same path of alt-az coordinates in order to produce flat baselines.

\begin{figure}[h]
\centering
\includegraphics[scale=.75]{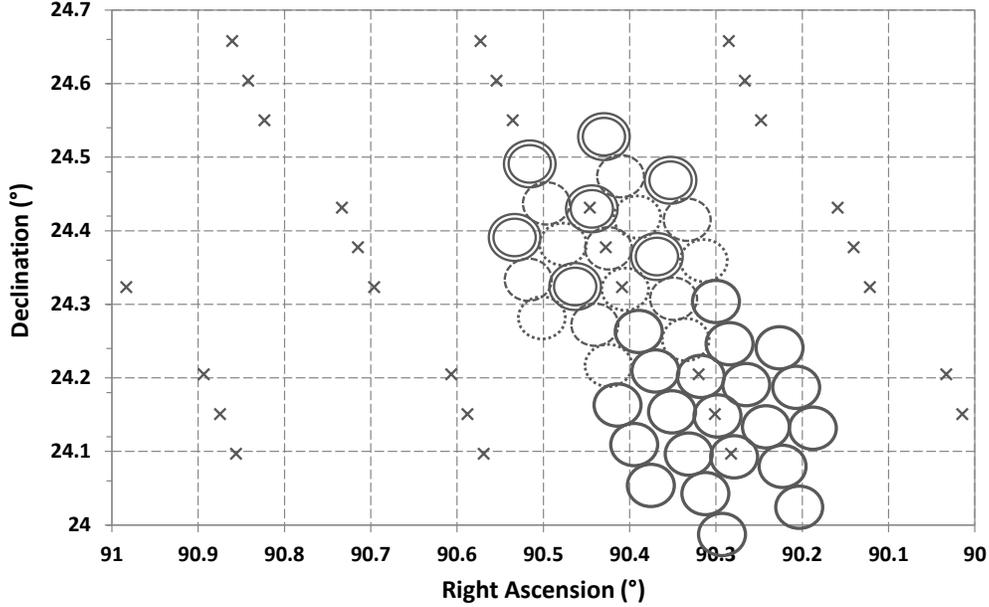}
\caption{\label{tiling} \footnotesize Tiling pattern used by the ALFA ZOA Deep Survey shown for a small region of sky. Clusters of three pointings fully cover their area and fit into surrounding clusters. The seven beams of the ALFA pattern are included for two of the clusters to demonstrate how the geometry of the ALFA beams is used to cover the survey area. One cluster shows the beams coded separately for the three different pointings.}
\end{figure}

The survey tiling pattern is made from clusters of three beam patterns, the centers of which are separated by one FWHM along an axis tilted 19$^\circ$ from north-south, as shown in Figure \ref{tiling}. Clusters fully cover their area and fit into surrounding clusters in order to fill in the survey area without Nyquist sampling. The axis of the ALFA receiver is rotated counter-clockwise, 19$^\circ$ from north-south, so that adjacent clusters to the east and west are centered on the same declination roughly 60 seconds of right ascension apart. This allows the survey to integrate one pointing for 270 seconds, for instance, and have time to slew to another pointing 300 seconds of right ascension away so that it can then track over the same alt-az path (i.e. create an ON-OFF pair). Every two observations must be observed in pairs in this way in order to apply bandpass correction, and the full survey area is filled out by leapfrogging through this tiling pattern.

The observing setup is slightly different between the inner and outer Galaxy regions because of commensal obligations. In the inner Galaxy, each pointing is integrated for 270 seconds on sky and then 6 seconds on the ALFA high flux noise diode for calibration. A single ALFA beam pattern covers a 60 square arcminute area, and about 10 pointings can be observed in an hour. At this rate, the survey covers approximately 10 square degrees per 60 hours of telescope time, though the area is not covered at the Nyquist sampling rate.

Outer Galaxy observations are taken for 180 second integrations and 6 seconds on the high flux noise diode. Each pointing has two Nyquist counterparts, each 1/2 FWHM away on either side, perpendicular to the axis of the three pointing cluster. The effective integration time for each spatial position is the same as in the inner Galaxy. However, the survey coverage rate is slower by a factor of two because three 180 second integrations are taken in the outer Galaxy for every one 270 second integration taken in the inner Galaxy. Surveying an area that is Nyquist sampled allows the integrated flux of a detection to be fully recovered.

\section{\label{datared} Data Reduction}

\subsection{Process}

The ALFA ZOA Deep Survey developed unique software for data reduction in the IDL programming language because of the total power on-off observing technique. Each one-second integration of a telescope pointing needs to be identified with its off-source counterpart, within a small angular separation, $\delta\theta$. In an effort to define an upper limit for $\delta\theta$, several observations were taken during June 2010 to test the effect of ON-OFF alt-az positional difference on the quality of spectral baselines. The results from this test show that the upper limit is $\delta\theta_{lim}$ = 1.7$^{\prime}$ (McIntyre 2013b), or 1/2 FWHM. If a one-second integration does not have an off-source integration within $\delta\theta_{lim}$, then it is rejected. For an ON that has at least one OFF within $\delta\theta_{lim}$, then the OFF with the smallest $\delta\theta$ for each ON integration is used to create a reduced bandpass, (ON-OFF)/OFF. This is divided by the median of the OFF bandpass for normalization. All the reduced spectra from one pointing (maximum 270 in the inner Galaxy, 180 in the outer) are median-combined to reject spurious RFI and recover the total integration time of the pointing. Most ON-OFF pairs have integrations rejected due to separation larger than $\delta\theta_{lim}$, though it is very rare to have more than 20 integrations rejected. The noise diode from the OFF is used to calibrate from receiver units into temperature, and the gain, G(beam, ZA, polarization), converts temperature into janskys. The frequency is resampled with a linear interpolation from topocentric to barycentric and converted relativistically into heliocentric velocity using the optical velocity convention. The two polarizations are averaged and the spectrum is Hanning smoothed, which mitigates ringing effects and lowers noise by a factor of $\sqrt{2}$, but worsens spectral resolution by a factor of two.

The final, reduced spectrum for a 270 second integration has a noise level of 1 mJy per channel at a velocity resolution of 9 km s$^{-1}$. The velocity range of the high frequency sub-band is cut to heliocentric velocities, $v_{hel}$ = -2,000 km s$^{-1}$ to 12,000 km s$^{-1}$ and a third order polynomial baseline is auto-fitted and subtracted over this spectrum. The low frequency sub-band is cut to a velocity range of $v_{hel}$ = 10,000 km s$^{-1}$ to 21,000 km s$^{-1}$ and also fit by a third order polynomial. The bandpass at velocities beyond 21,000 km s$^{-1}$ contains a significant amount of RFI and has not been incorporated into the survey at this point. The reduced spectra are then made into data cubes of right ascension, declination, and velocity by the program Gridzilla  (Barnes et al. 2001).

Gridzilla is part of the AIPS++ software package, and is used to create a data cube by spatially gridding spectra. For ALFA ZOA Deep, Gridzilla is set to clip spectra between -50 mJy and 500 mJy, and make a cube containing 1$^{\prime} \times 1^{\prime}$ pixels. Each input spectrum is assigned to all pixels within a 4$^{\prime}$ diameter of the spectrum's coordinates using a Top-Hat kernel. In pixels where multiple spectra contribute they are median-combined, weighted by the angular distance of the pixel from each spectrum's original coordinates using a beam power pattern with FWHM = 3.4$^{\prime}$. The end result is a 3-dimensional data structure with axes of right ascension, declination, and heliocentric velocity. Position-velocity planes can be iteratively viewed to search for galaxy detections.

\section{\label{search} Search Method and Source Parameterization}

The data cube is searched for detections by visually inspecting over the usable velocity range using the visualization tool, Karma KVIS (Gooch 1996).  This is done by examining image planes in position-velocity slices and using a greyscale to represent flux intensity. The data cube is searched by three authors (two authors and a UNM graduate student in the case of this First Results paper), who look for profile shapes consistent with known galaxies, e.g. Gaussian, double-horn, etc. Lists of galaxy candidates are prepared independently by each searcher. These lists are compared with the astronomical coordinate comparison tool, Starlink TOPCAT (Tool for OPerations on Catalogues and Tables) version 3.9 (Taylor 2005), in order to match sources with RA, Dec, and velocity coordinates within tolerances of 5$^{\prime}$, 5$^{\prime}$, and 300 km s$^{-1}$, respectively. The matched lists are adjudicated by a separate author who re-examines the position of each candidate source and either accepts it into the catalog or rejects it as a false detection. These adjudicated sources are adopted into a working catalog and each source is parameterized using MBSPECT in the software package, MIRIAD (Sault et al. 1995).

\begin{figure}[h]
\centering
\includegraphics[scale=.6]{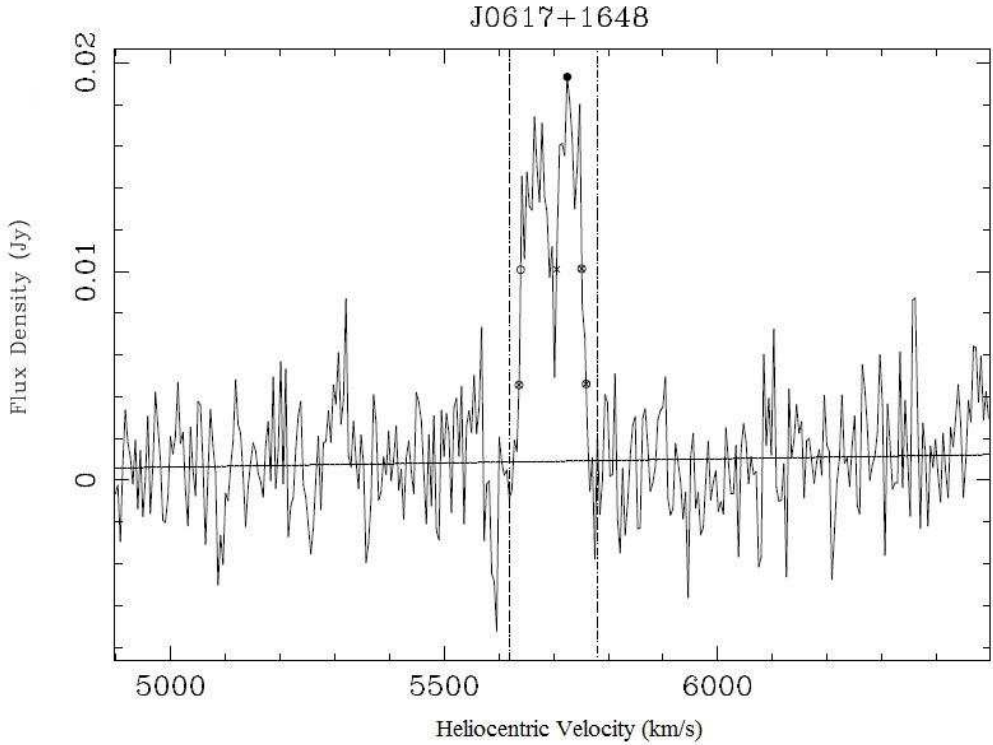}
\caption{\label{J0617+1648.ps} \footnotesize H\,{\sc{i}} profile of ALFA ZOA J0617+1648, created from MIRIAD output. The vertical dotted lines are user defined and create a profile window that source parameters are extracted from. The open circles on the outer edges of the detection represent the $W_{20}$ and $W_{50}$ velocity widths, and the closed circle at the top is the location of the peak flux.}
\end{figure}

For unresolved sources, a spectral profile is created from the weighted emission inside a 5$^{\prime}$ x 5$^{\prime}$ box surrounding the position of the source. The emission is weighted by spatially fitting a Gaussian to a moment 0 map integrated over a user-defined range of velocities, and a profile is created as shown in Figure \ref{J0617+1648.ps}. In an iterative process, the profile shape is visually inspected and a new velocity range is chosen for the profile (shown as the vertical dotted lines in Figure \ref{J0617+1648.ps}). The profile window is also used to create a mask for autofitting a polynomial to the baseline within 2000 km s$^{-1}$ of the profile, usually to first order but up to fourth order in rare cases. The integrated flux, $F_{HI}$, is computed by subtracting the polynomial fit and integrating the flux inside the profile window. The open circles on the outer edges of the profile in Figure \ref{J0617+1648.ps} represent the width-maximizer velocity widths, $W_{50}$ and $W_{20}$, at 50\% and 20\% peak flux, respectively. The closed circle at the top is the location of the peak flux. The x near the center of the profile indicates a width-minimizer parameterization, which is not used for ALFA ZOA. The central heliocentric velocity, $v_{hel}$, is taken as the midpoint of the $W_{50}$ value.

All sources go through the MIRIAD parameterization process as unresolved, but then are re-examined to determine the existence of sources with extended emission. A first list of potentially extended sources is constructed from the size of the Gaussian fit for each source as output by MBSPECT. All sources with major axes greater than 3.4$^{\prime}$ make a candidate list of extended sources. Moment 0 maps are then created for each candidate using the MIRIAD task, MOMENT. These maps are inspected visually using Karma KPVSLICE (Gooch 1996) and a spatial profile for each source is created by integrating flux across a user-defined major axis. These spatial profiles are saved as an array and fed into the IDL program GAUSSFIT, which computes a non-linear least-squares Gaussian fit. From this fit, the FWHM of the source is determined.

\begin{figure}[h]
\centering
\includegraphics[scale=.7]{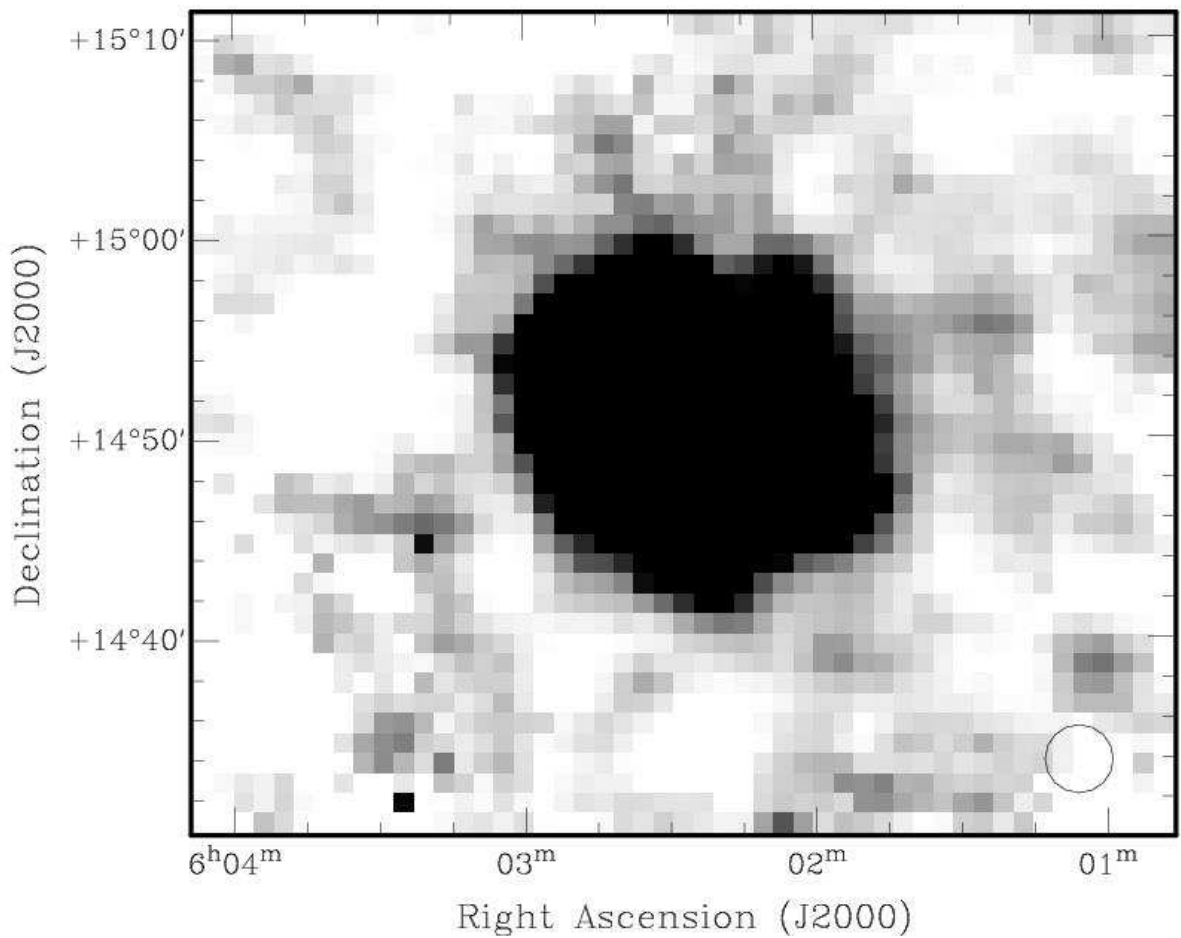}
\caption{\label{deepmoment} \footnotesize Moment map of ALFA ZOA J0602+1452. Beam size (FWHM = 3.4$^{\prime}$) is shown in bottom right corner. A clear extension beyond $5^{\prime} \times 5^{\prime}$ can be seen on this resolved source.}
\end{figure}

Every source with FWHM $> 3.4^{\prime}$ makes it into a shorter candidate list and all of these are re-examined again using KPVSLICE in order to determine an individual aperture size to use for each galaxy. KPVSLICE shows the velocity spectrum from the data cube that corresponds with each spatial pixel in the moment map. The user counts how many single spatial pixels contain a clear detection in the velocity spectrum, and decides on a corresponding aperture box height and width. These new box sizes are then used to re-determine source parameterization using MIRIAD, where spectral profiles are created by summing flux over all channels and pixels equally inside the aperture and within the profile window. The position of the source is determined from the center pixel of the aperture, as opposed to the center of a Gaussian fit, but all other source parameterization techniques are the same as for unresolved sources. The rms of resolved sources is higher than for unresolved sources because of the greater number of pixels summed over. A moment map for resolved source, ALFA ZOA J0602+1452, is shown in Figure \ref{deepmoment}. The box size of the aperture used is 23$^{\prime} \times 23^{\prime}$ (RA x Dec).

\section{\label{catalog} The ALFA ZOA Deep Survey: First Results Catalog}

The ALFA ZOA Deep Survey: First Results catalog contains 61 galaxies. Profiles are shown in Figure \ref{profiles}. Table 1 presents the following information on the catalog: 

Column (1). ALFA ZOA source name. Sources with extended emission are labeled with the 'e' superscript.

Column (2). Right Ascension of the fitted position in hours, minutes, seconds, J2000 epoch.

Column (3). Declination of the fitted position in degrees, arcminutes, arcseconds, J2000 epoch.

Columns (4) and (5). $l$ and $b$, Galactic longitude and latitude in degrees, respectively, of the fitted position in degrees.

Column (6). $F_{HI}$, integrated flux in Jy km s$^{-1}$. 

Column (7). $v_{hel}$, heliocentric velocity in km s$^{-1}$.

Columns (8) and (9). $W_{50}$ and $W_{20}$, velocity width in km s$^{-1}$ of the profile at 50\% and 20\% of the peak flux level, respectively.

Column (10). $D_{LG}$, distance to the galaxy in Mpc in the Local Group reference frame (Courteau \& van den Bergh 1999), using Hubble's Law for cosmological redshift distance and taking H$_{0}$ = 70 km s$^{-1}$ Mpc$^{-1}$.

Column (11). Log $M_{HI}$, logarithm of the total H\,{\sc{i}} mass in $M_{\odot}$ calculated from,
\begin{equation}
\label{MHI}
M_{HI} = 2.36 \times 10^5 \; D_{LG}^2 \; F_{HI},
\end{equation}
where $D_{LG}$ is the distance to the galaxy and $F_{HI}$ is the integrated flux as described above.

\begin{figure}
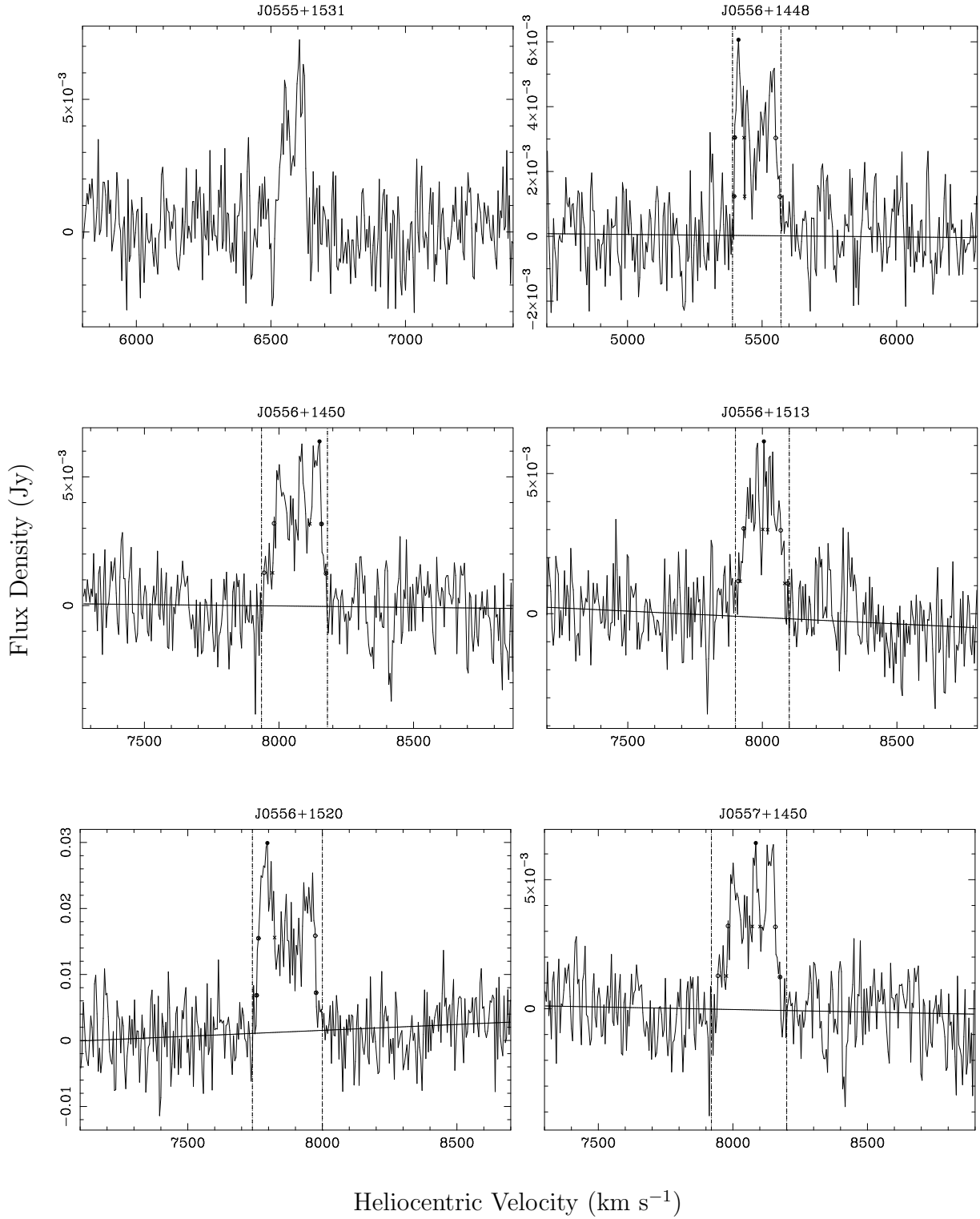

\centering{\includegraphics[scale=0.29, angle=270]{fig5_1.ps}
\includegraphics[scale=0.29, angle=270]{fig5_2.ps}
\\[1cm]
\includegraphics[scale=0.29, angle=270]{fig5_3.ps}
\includegraphics[scale=0.29, angle=270]{fig5_4.ps}
\\[1cm]
\includegraphics[scale=0.29, angle=270]{fig5_5.ps}
\includegraphics[scale=0.29, angle=270]{fig5_6.ps}}
\\[1cm]
\leavevmode\smash{\makebox[0pt]{\hspace{0em}
  \rotatebox[origin=l]{90}{\hspace{22em}
    Flux Density (Jy)}%
}}\hspace{0pt plus 1filll}\null
\leavevmode\smash{\makebox[0pt]{\hspace{0em}
  {\hspace{0em}
    Heliocentric Velocity (km s$^{-1}$)}%
}}\hspace{0pt plus 1filll}\null
\caption{\label{profiles} H\,{\sc{i}} profiles of ALFA ZOA detections. The description of the profiles is the same as in Figure \ref{J0617+1648.ps}.}
\end{figure}
\clearpage

\begin{figure}
\addtocounter{figure}{-1}
\centering{\includegraphics[scale=0.29, angle=270]{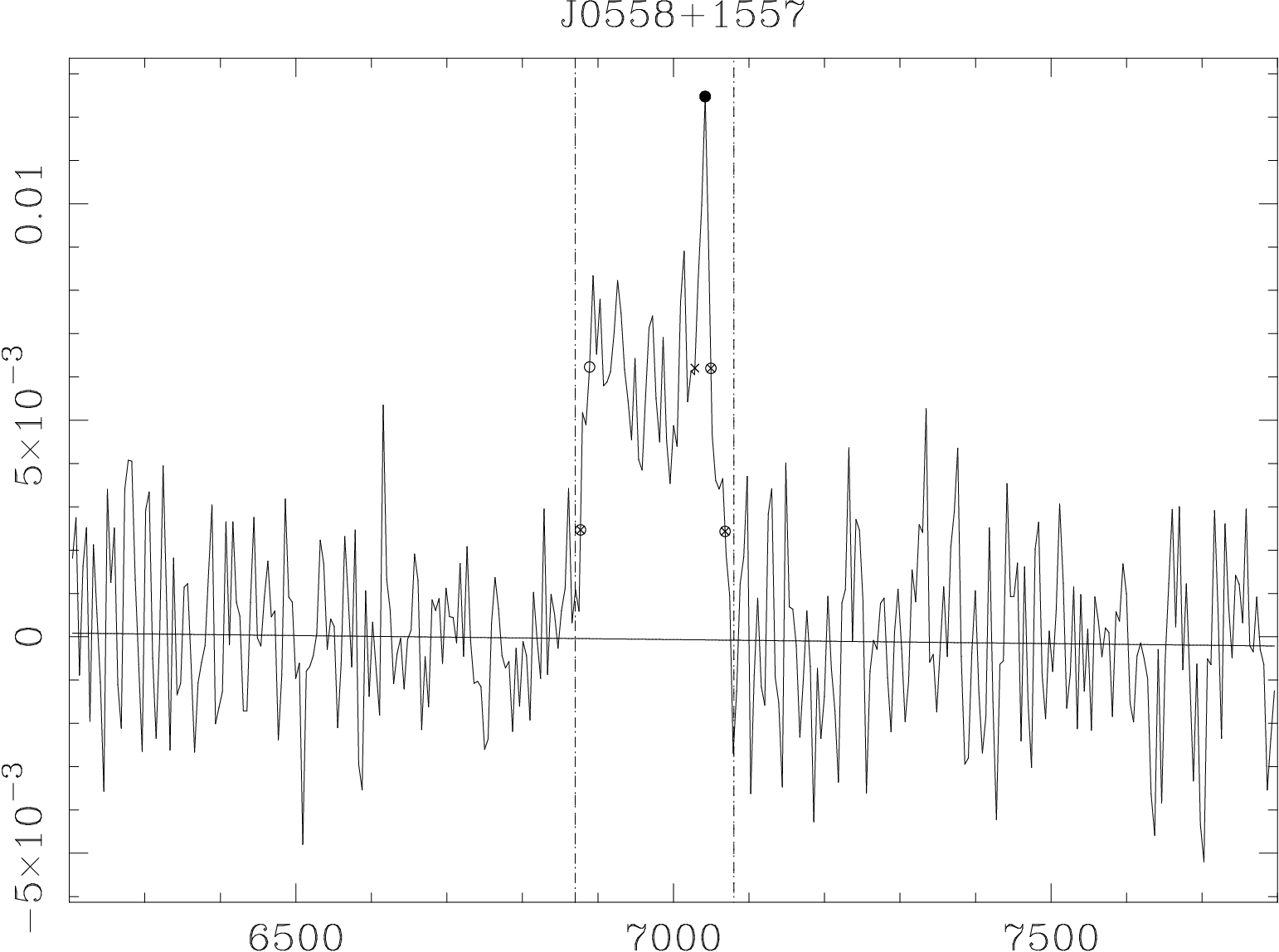}
\includegraphics[scale=0.29, angle=270]{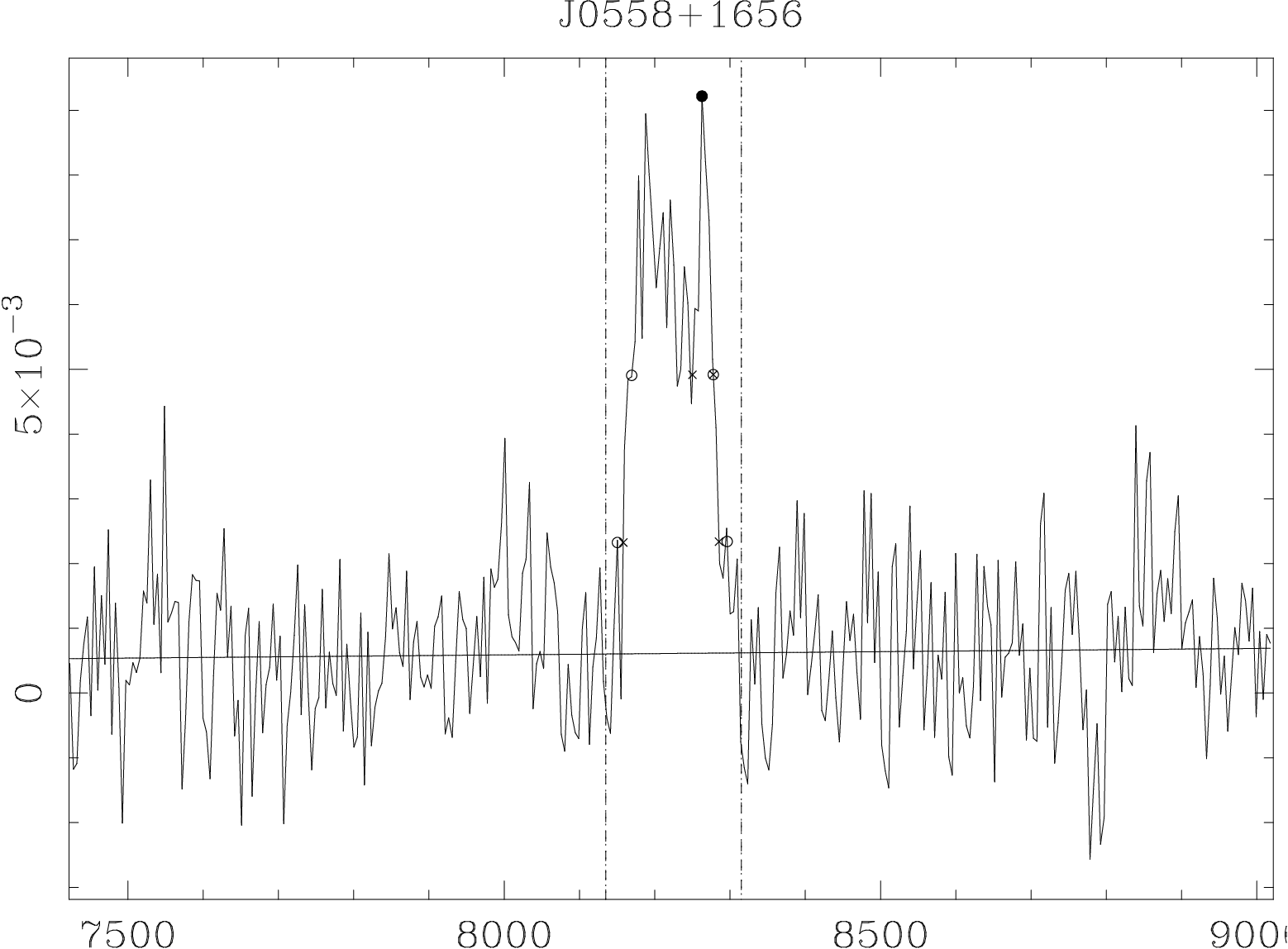}
\\[1cm]
\includegraphics[scale=0.29, angle=270]{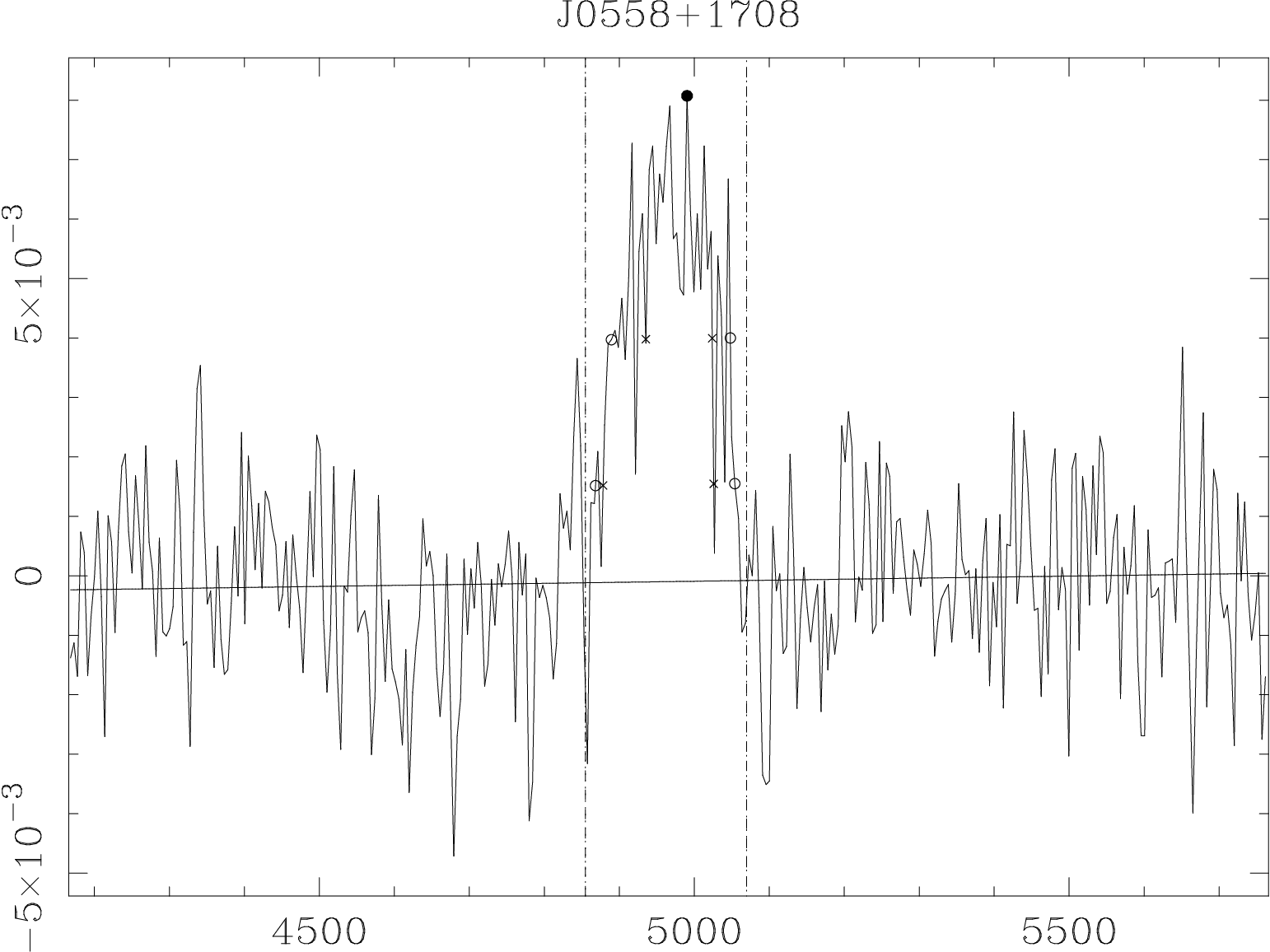}
\includegraphics[scale=0.29, angle=270]{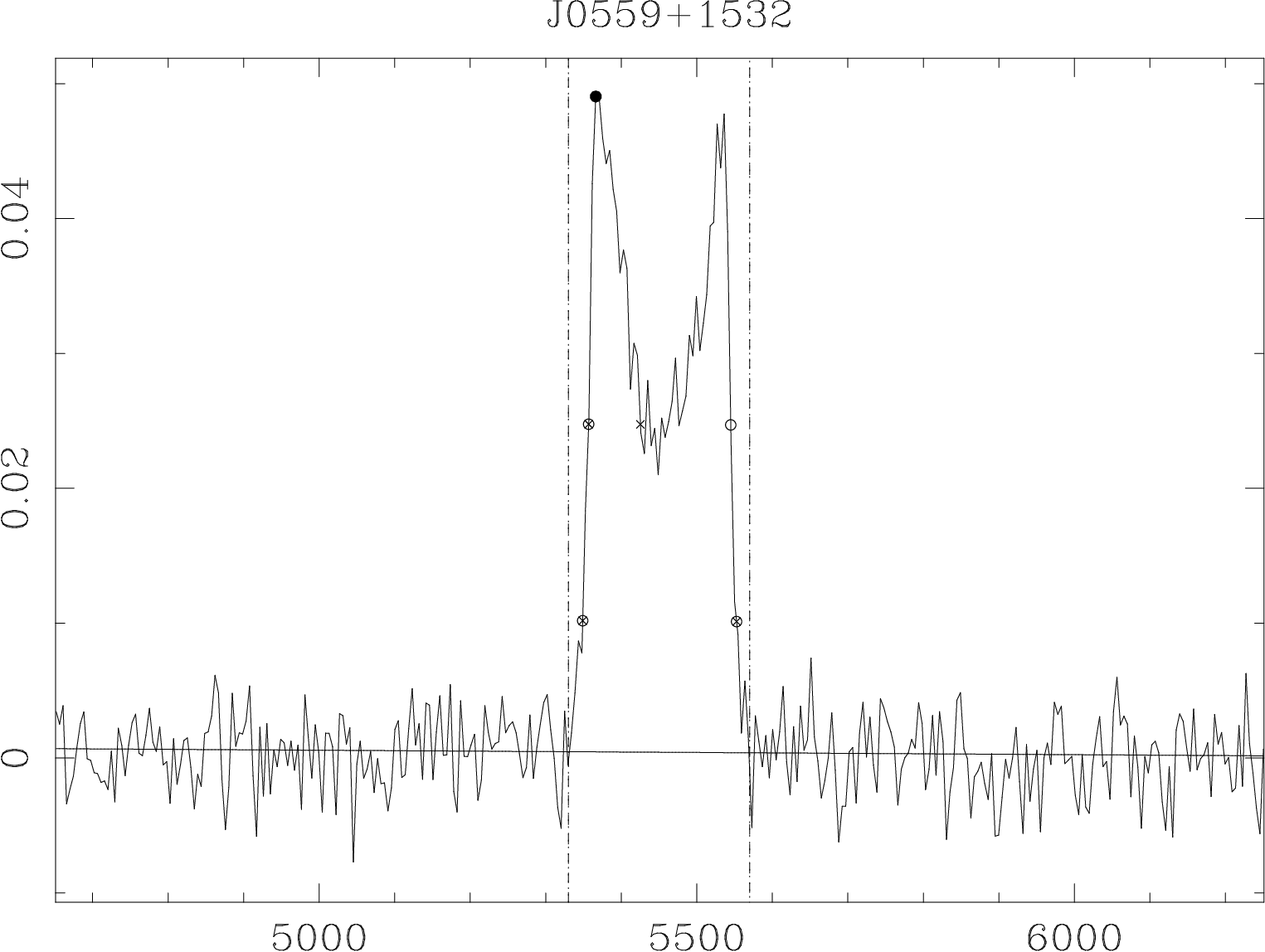}
\\[1cm]
\includegraphics[scale=0.29, angle=270]{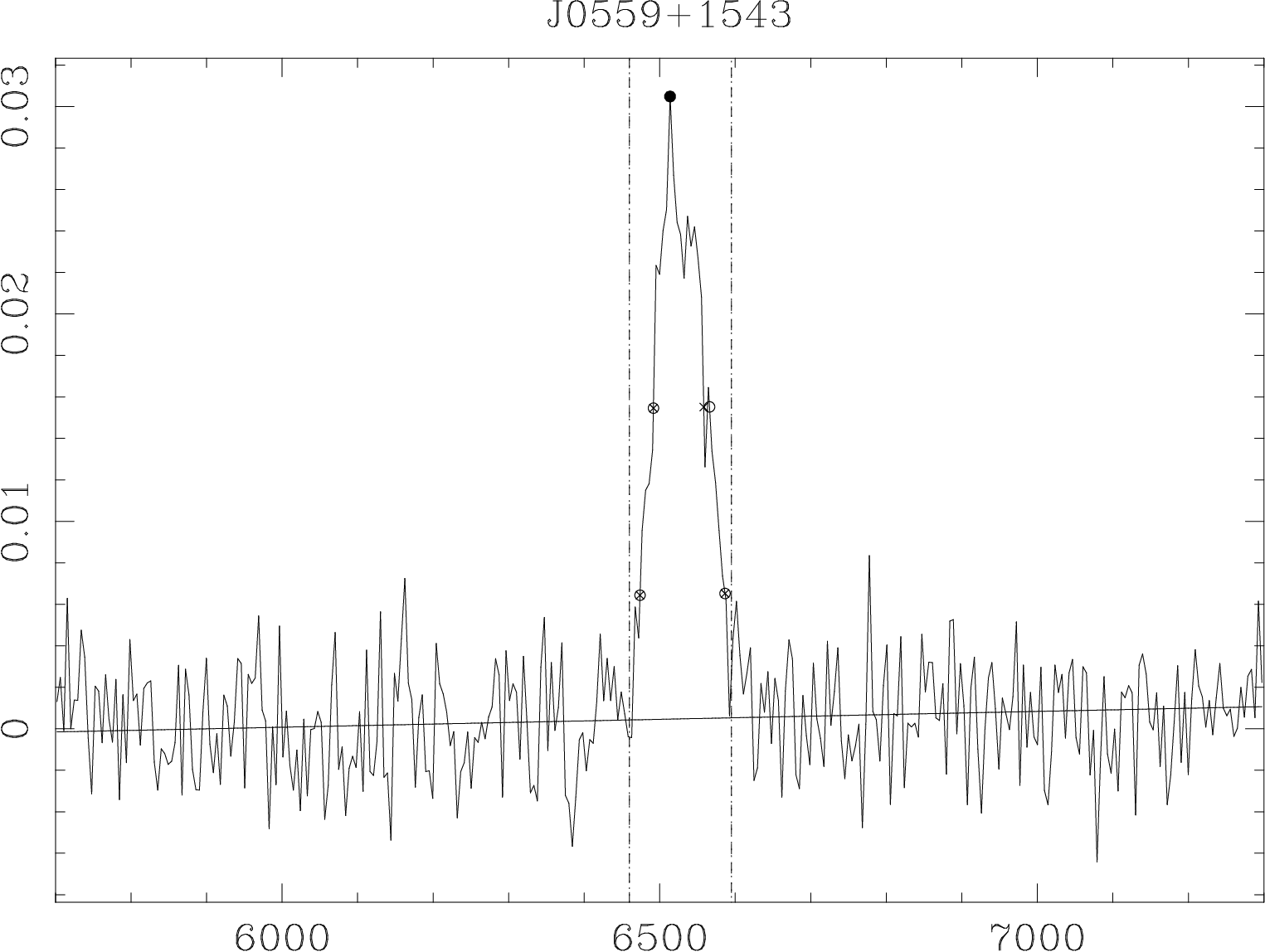}
\includegraphics[scale=0.29, angle=270]{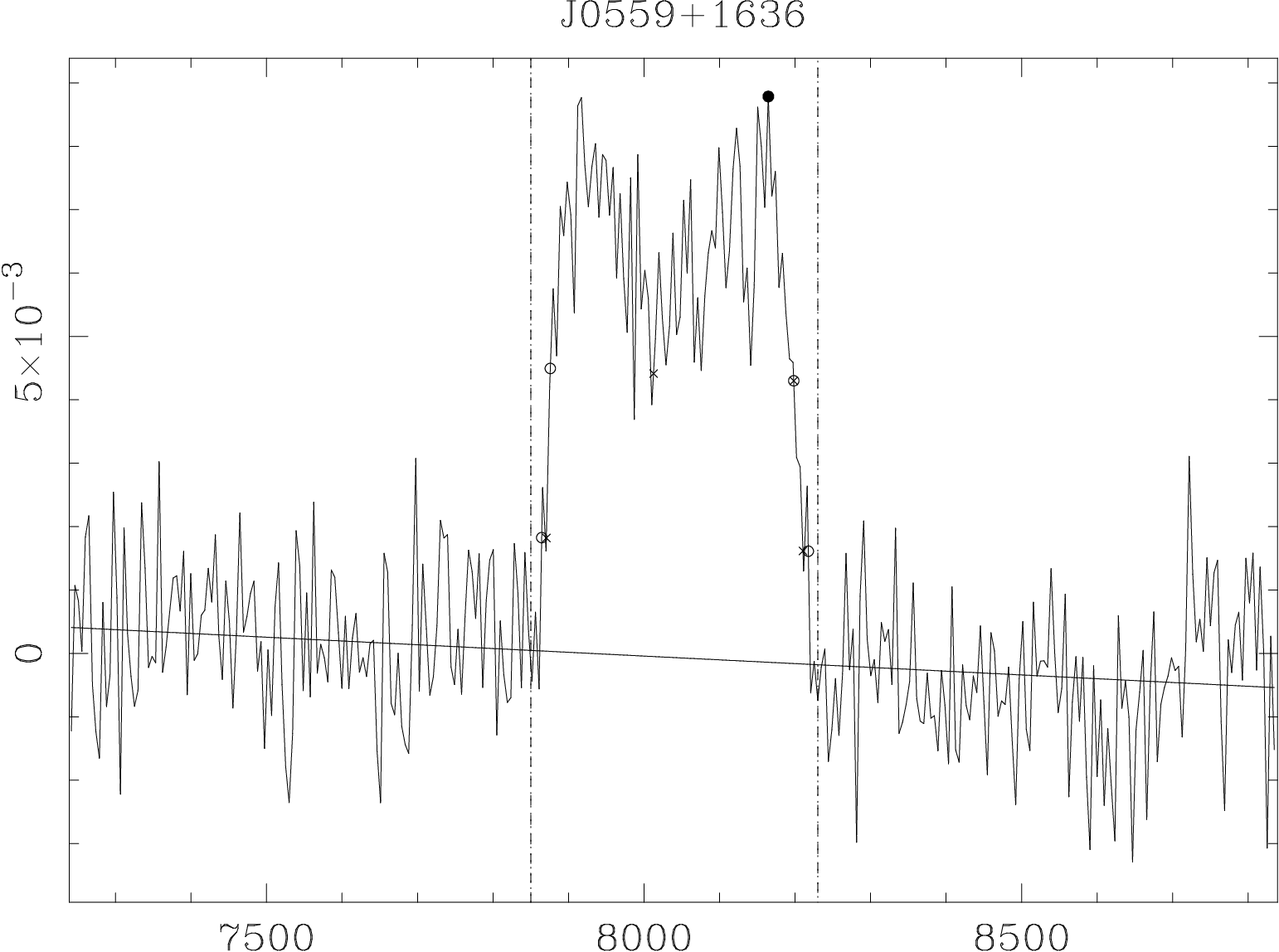}}
\\[1cm]
\leavevmode\smash{\makebox[0pt]{\hspace{0em}
  \rotatebox[origin=l]{90}{\hspace{22em}
    Flux Density (Jy)}%
}}\hspace{0pt plus 1filll}\null
\leavevmode\smash{\makebox[0pt]{\hspace{0em}
  {\hspace{0em}
    Heliocentric Velocity (km s$^{-1}$)}%
}}\hspace{0pt plus 1filll}\null
\caption{H\,{\sc{i}} profiles of ALFA ZOA detections. The description of the profiles is the same as in Figure \ref{J0617+1648.ps}. (Continued)}
\end{figure}
\clearpage

\begin{figure}
\addtocounter{figure}{-1}
\centering{\includegraphics[scale=0.29, angle=270]{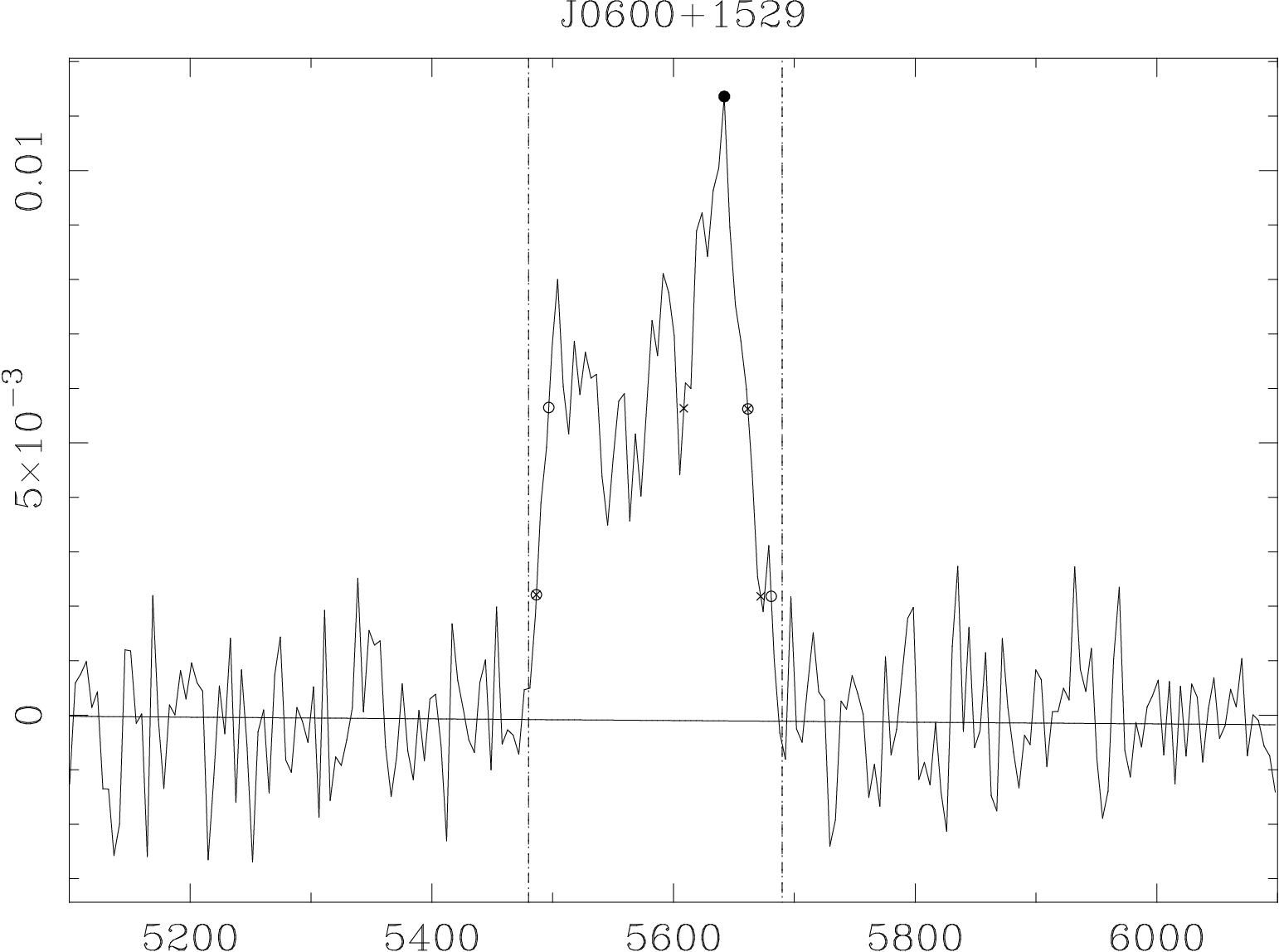}
\includegraphics[scale=0.29, angle=270]{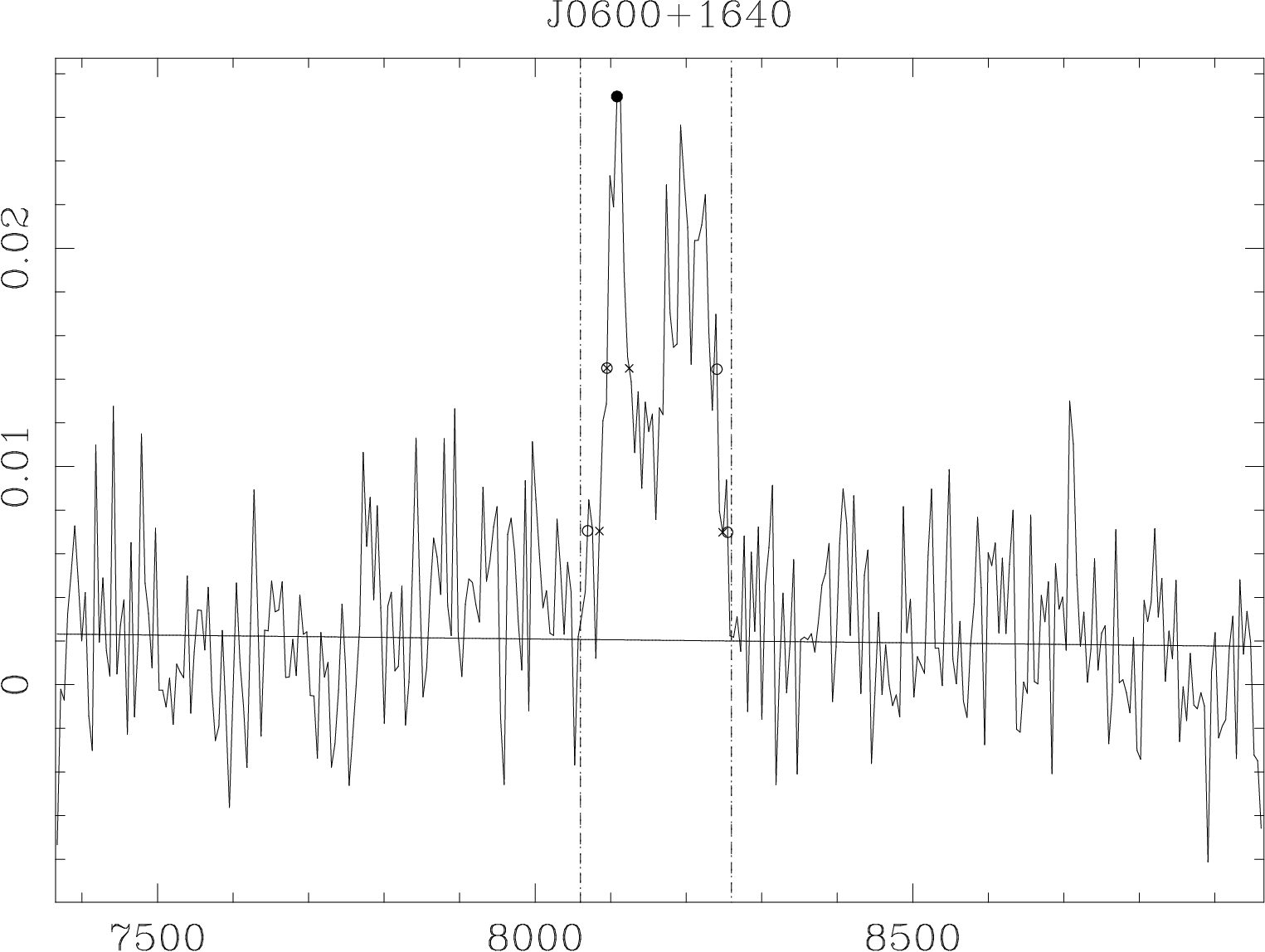}
\\[1cm]
\includegraphics[scale=0.29, angle=270]{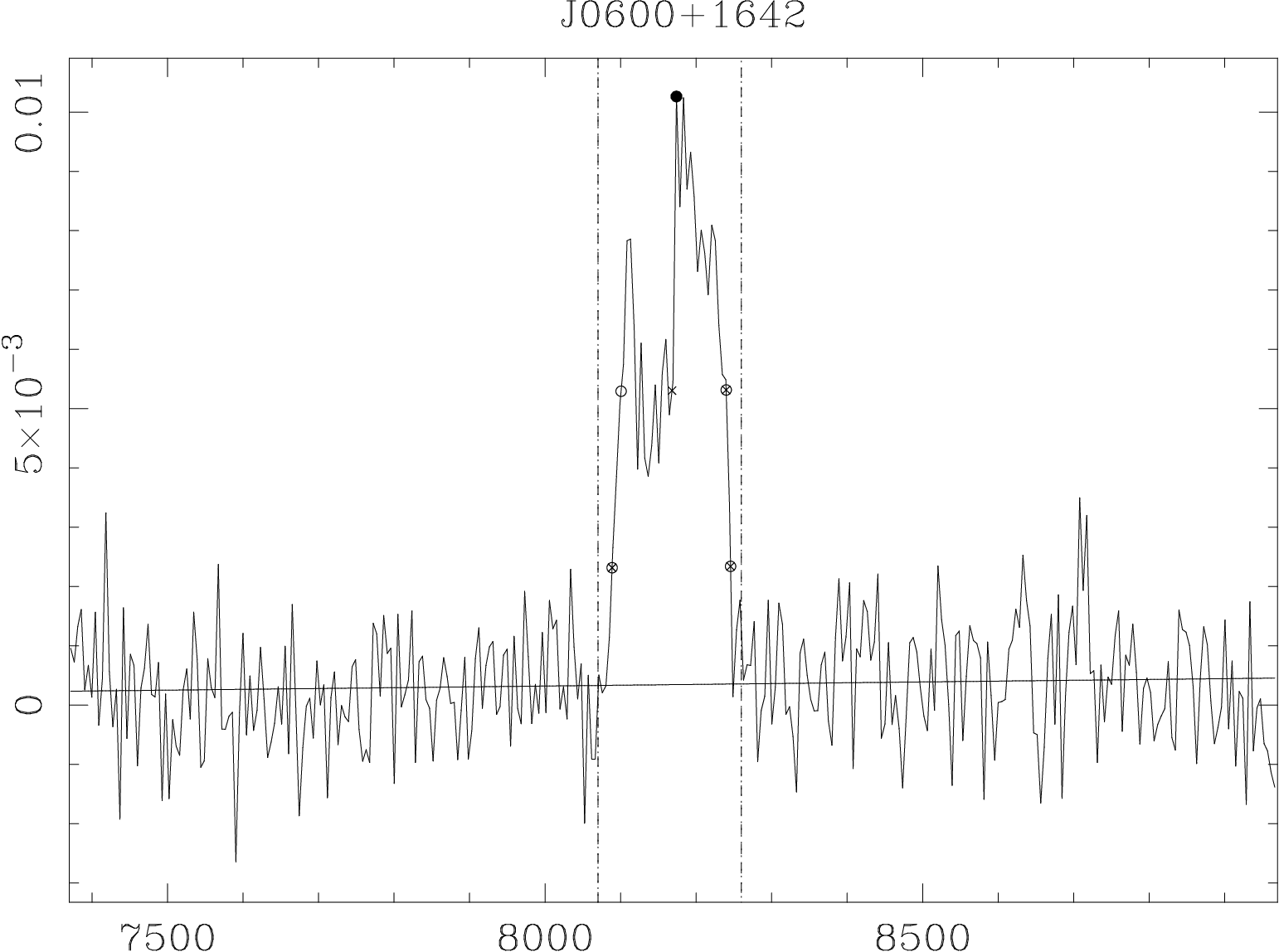}
\includegraphics[scale=0.29, angle=270]{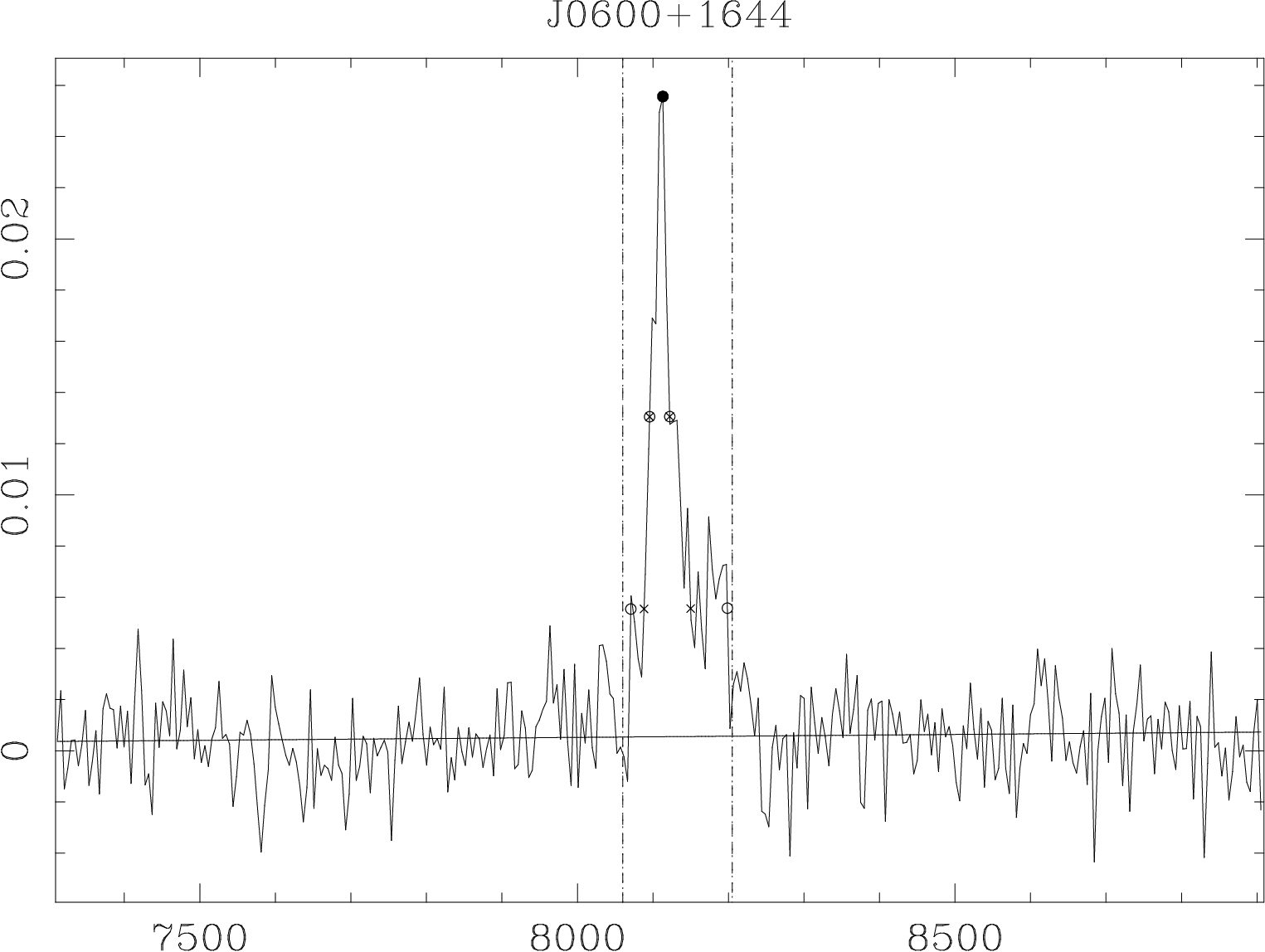}
\\[1cm]
\includegraphics[scale=0.29, angle=270]{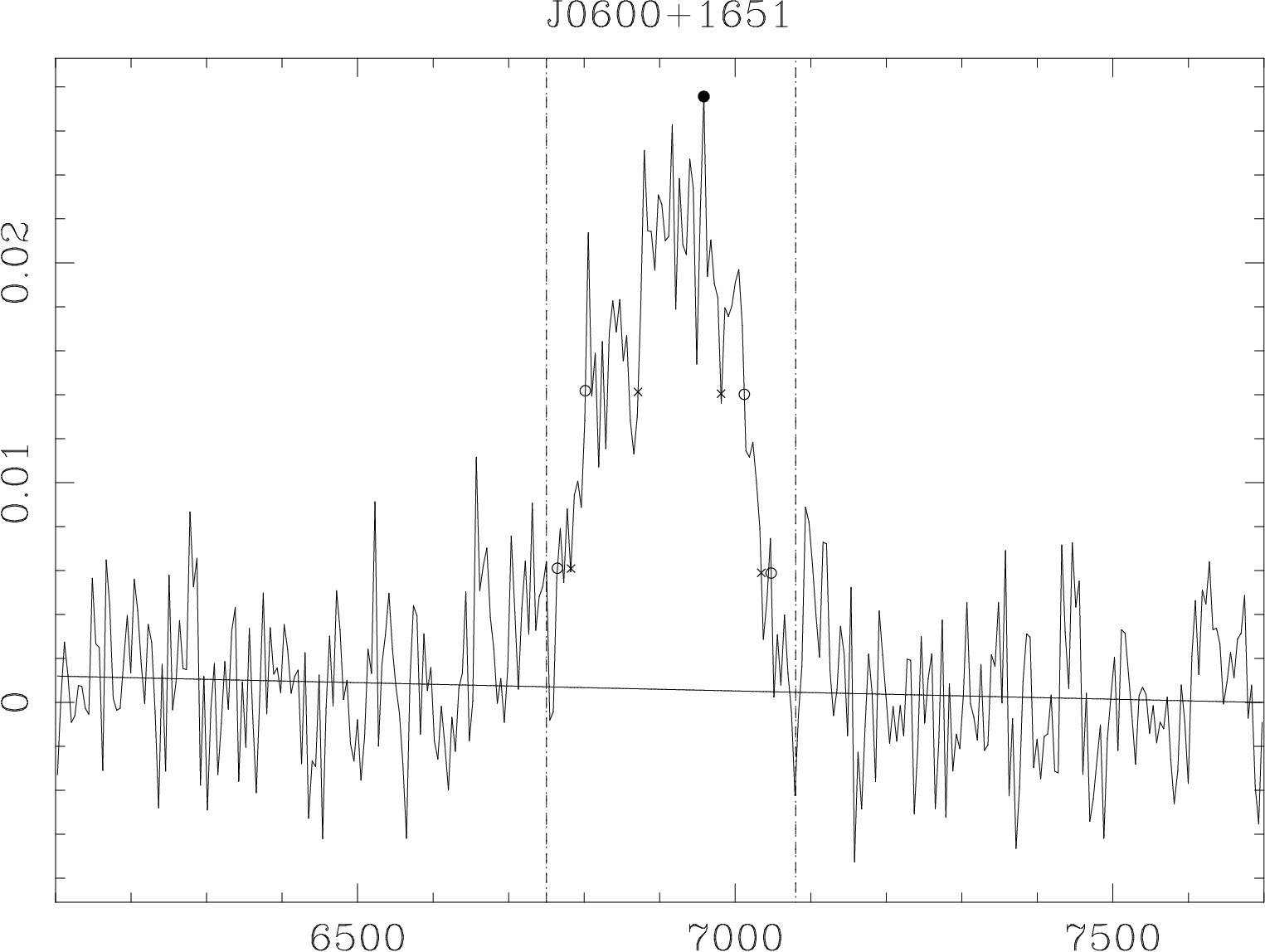}
\includegraphics[scale=0.29, angle=270]{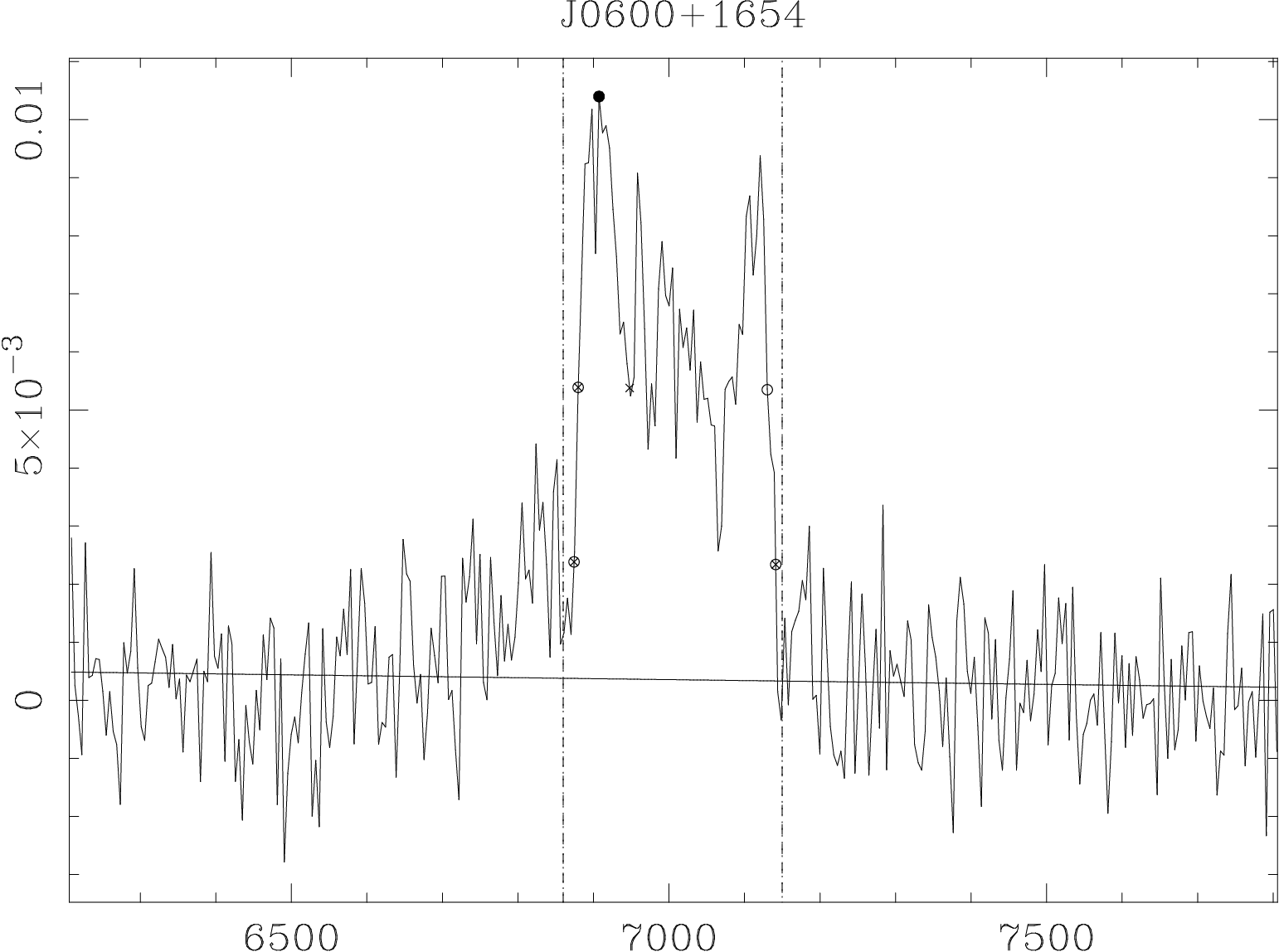}}
\\[1cm]
\leavevmode\smash{\makebox[0pt]{\hspace{0em}
  \rotatebox[origin=l]{90}{\hspace{22em}
    Flux Density (Jy)}%
}}\hspace{0pt plus 1filll}\null
\leavevmode\smash{\makebox[0pt]{\hspace{0em}
  {\hspace{0em}
    Heliocentric Velocity (km s$^{-1}$)}%
}}\hspace{0pt plus 1filll}\null
\caption{H\,{\sc{i}} profiles of ALFA ZOA detections. The description of the profiles is the same as in Figure \ref{J0617+1648.ps}. (Continued)}
\end{figure}
\clearpage

\begin{figure}
\addtocounter{figure}{-1}
\centering{\includegraphics[scale=0.29, angle=270]{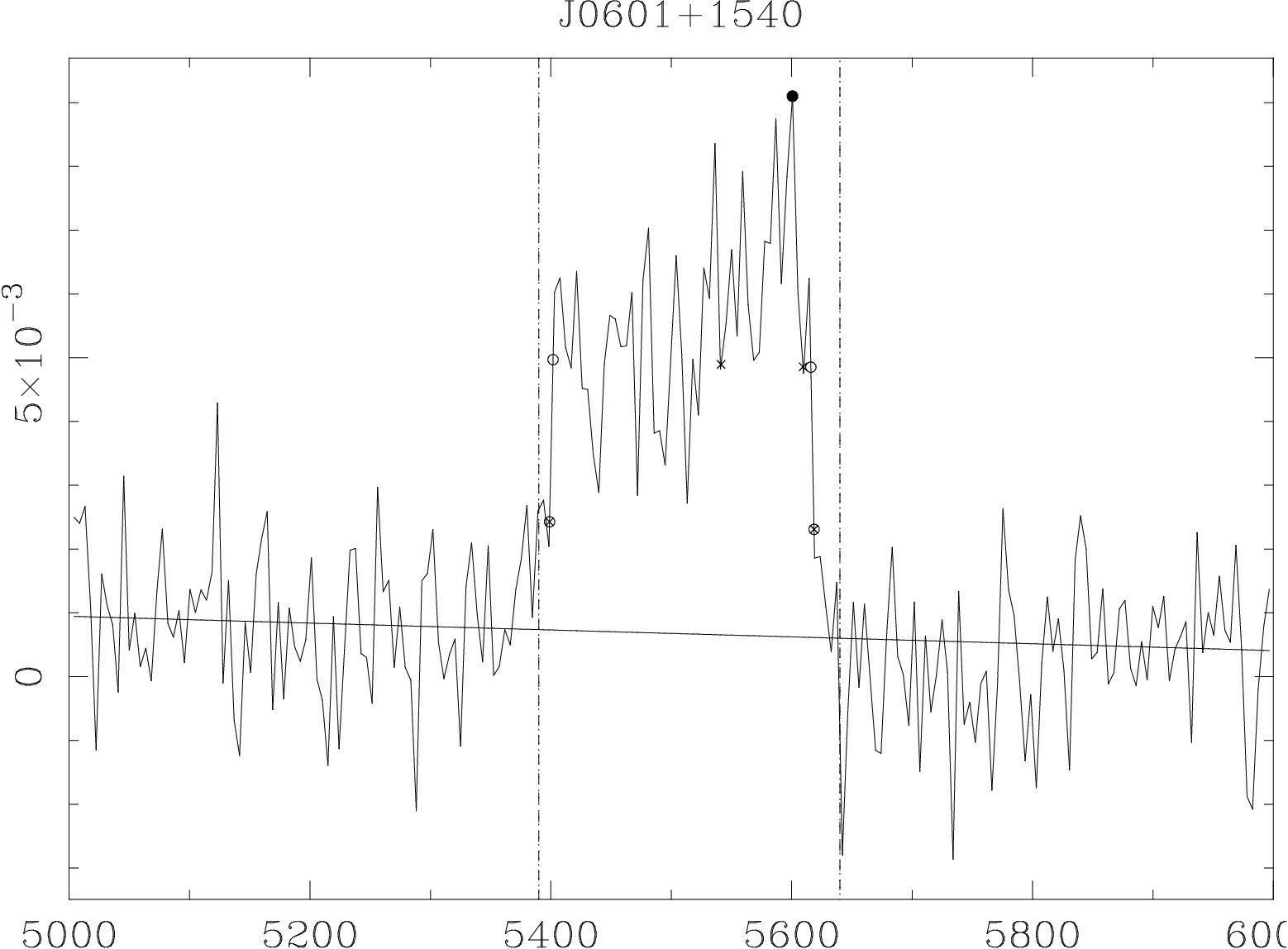}
\includegraphics[scale=0.29, angle=270]{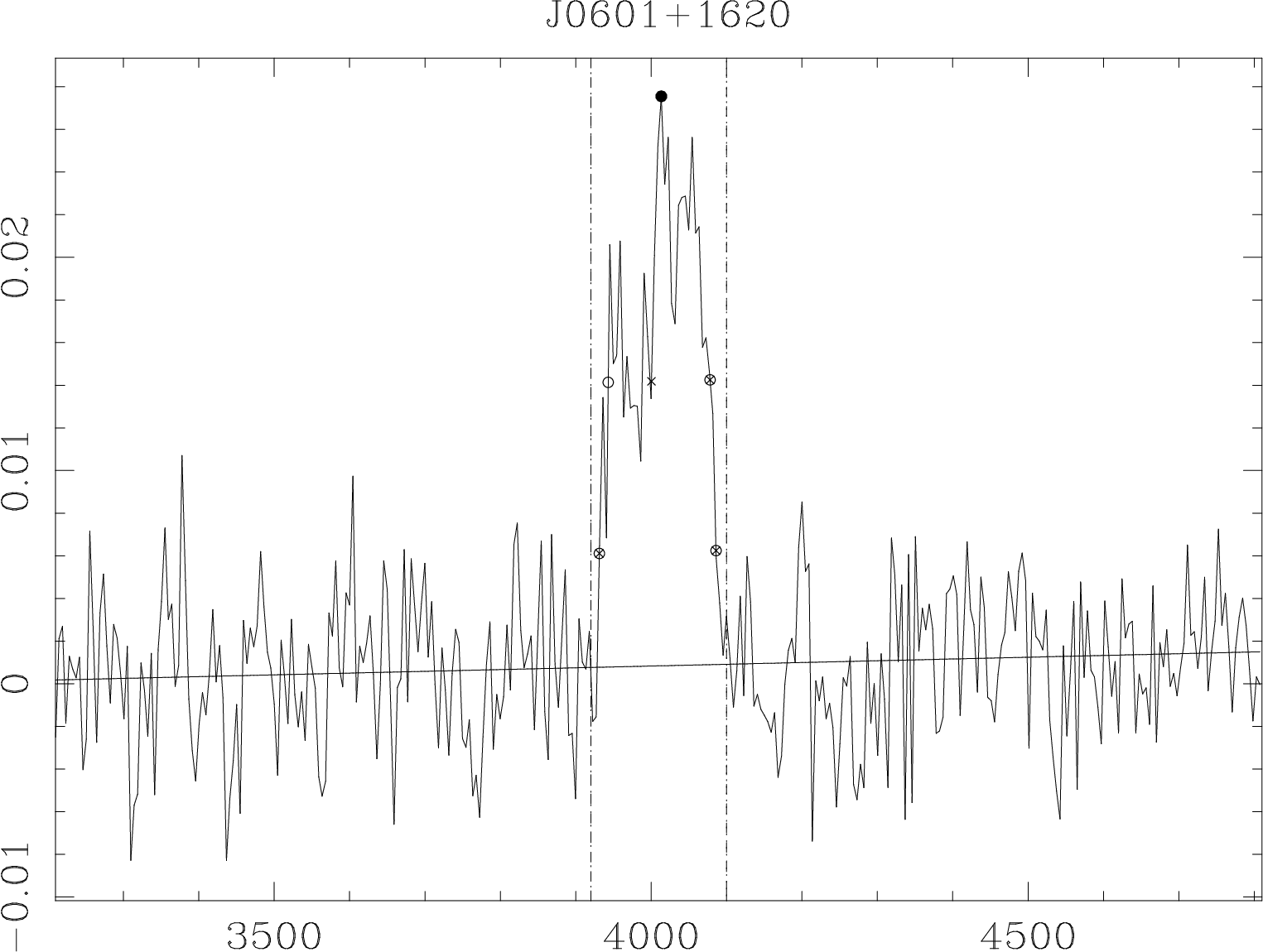}
\\[1cm]
\includegraphics[scale=0.29, angle=270]{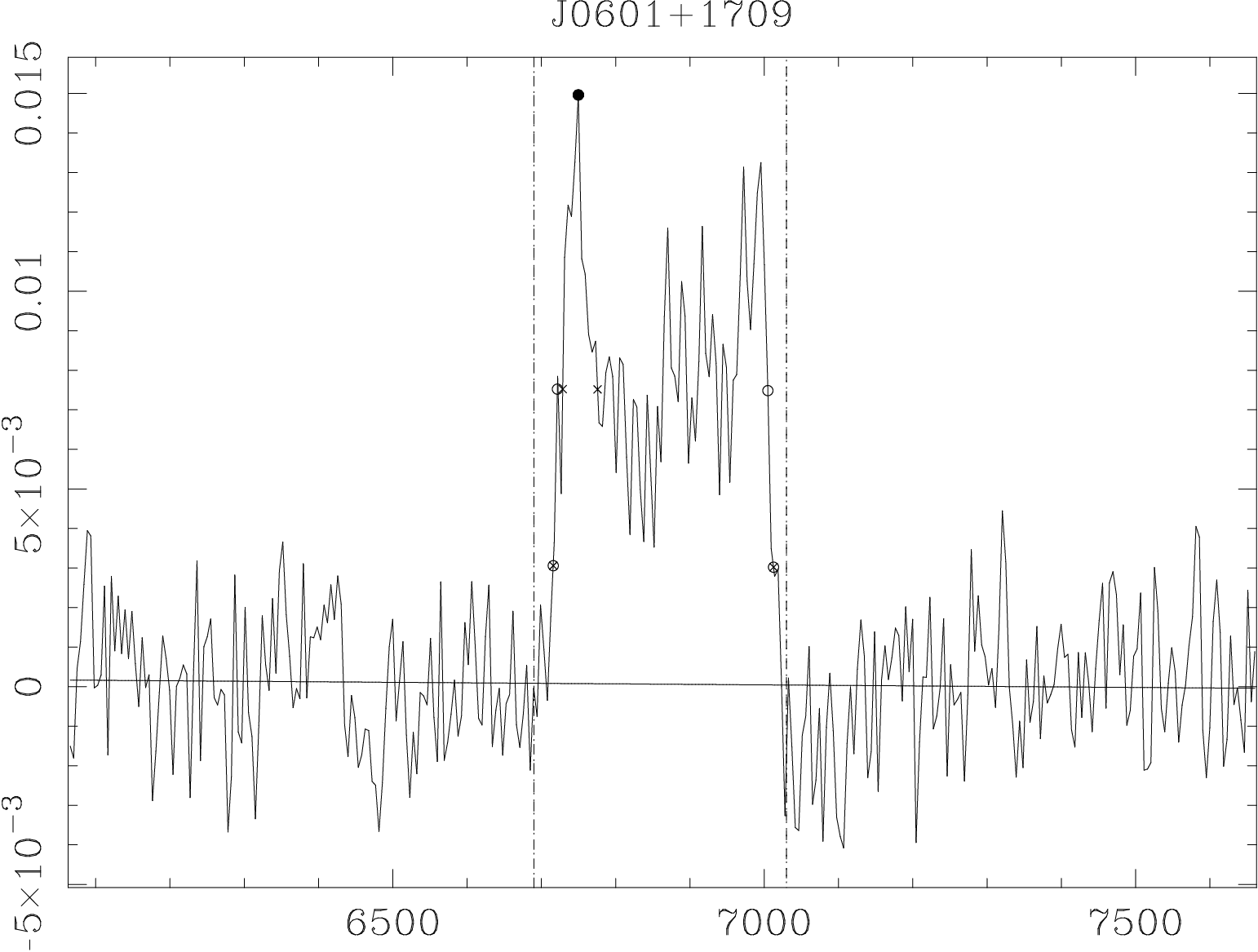}
\includegraphics[scale=0.29, angle=270]{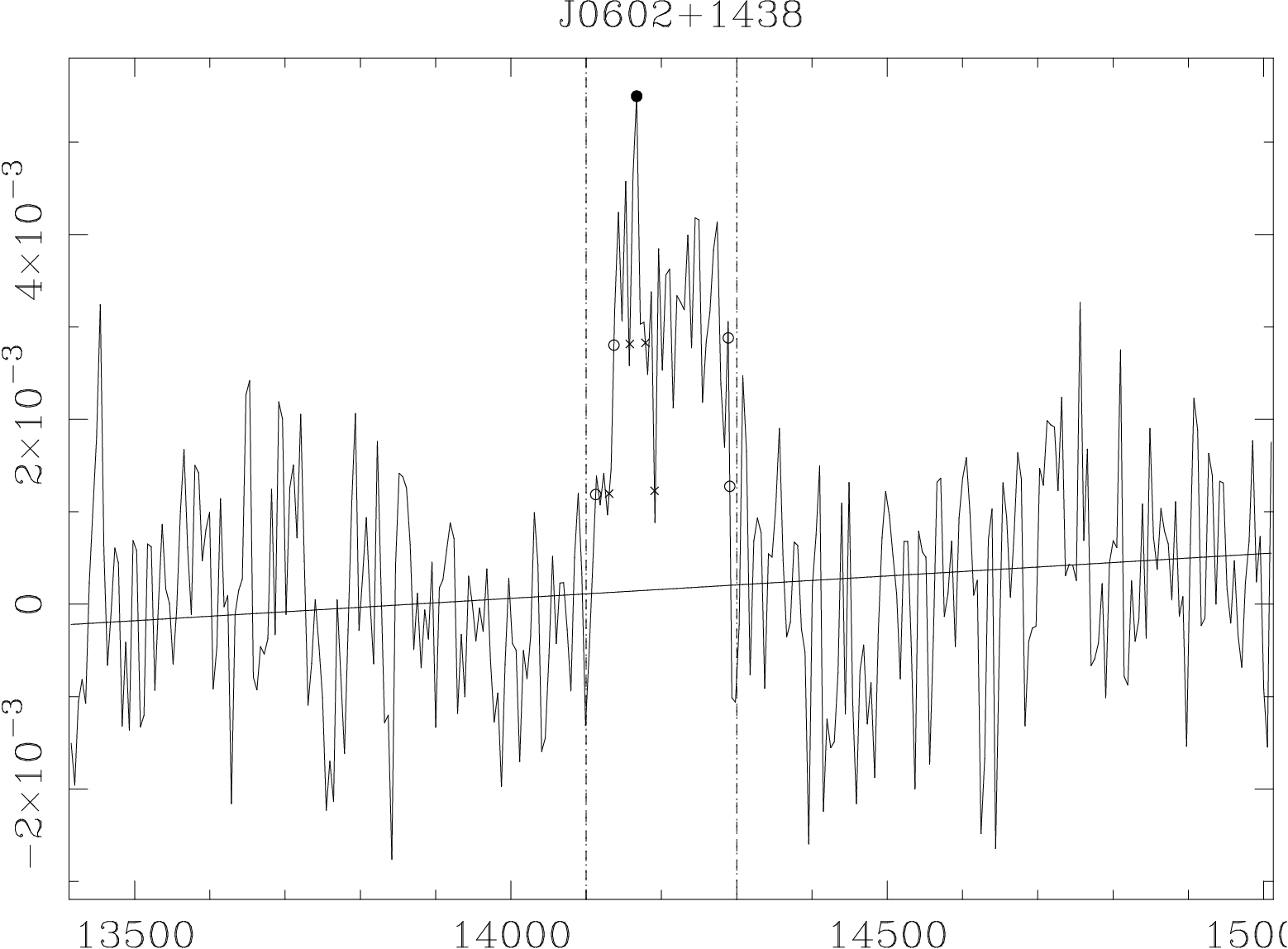}
\\[1cm]
\includegraphics[scale=0.29, angle=270]{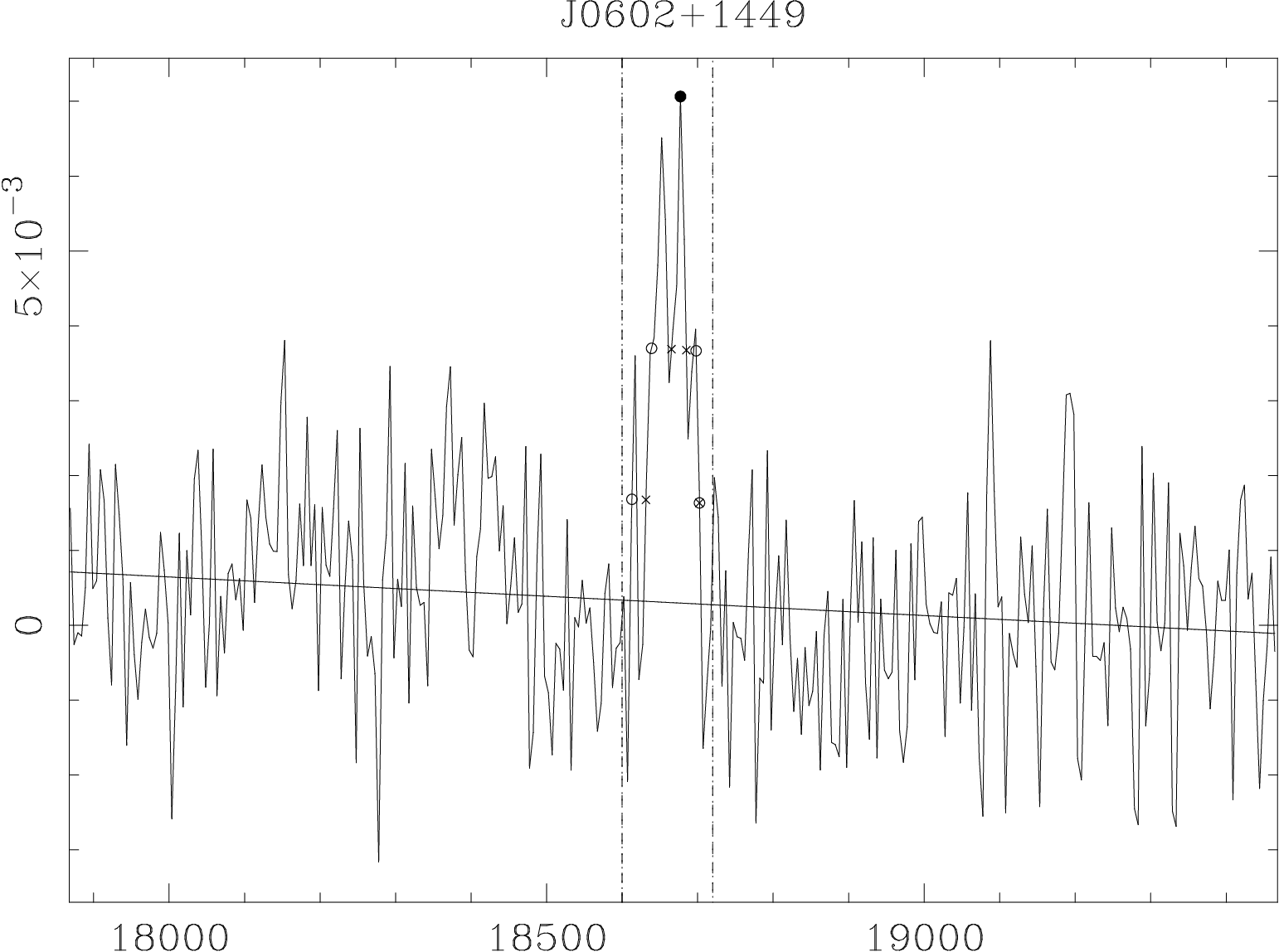}
\includegraphics[scale=0.29, angle=270]{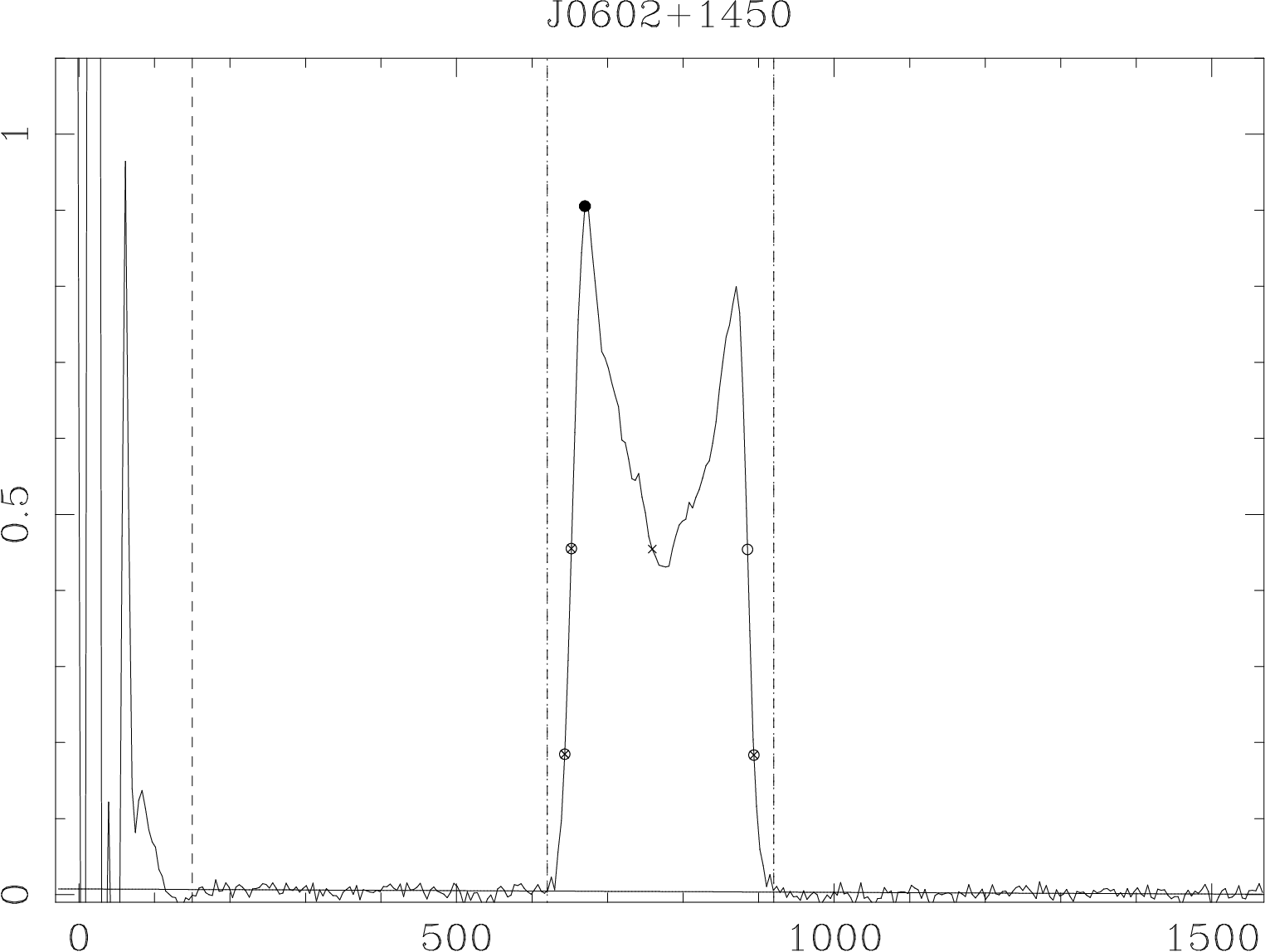}}
\\[1cm]
\leavevmode\smash{\makebox[0pt]{\hspace{0em}
  \rotatebox[origin=l]{90}{\hspace{22em}
    Flux Density (Jy)}%
}}\hspace{0pt plus 1filll}\null
\leavevmode\smash{\makebox[0pt]{\hspace{0em}
  {\hspace{0em}
    Heliocentric Velocity (km s$^{-1}$)}%
}}\hspace{0pt plus 1filll}\null
\caption{H\,{\sc{i}} profiles of ALFA ZOA detections. The description of the profiles is the same as in Figure \ref{J0617+1648.ps}. (Continued)}
\end{figure}
\clearpage

\begin{figure}
\addtocounter{figure}{-1}
\centering{\includegraphics[scale=0.29, angle=270]{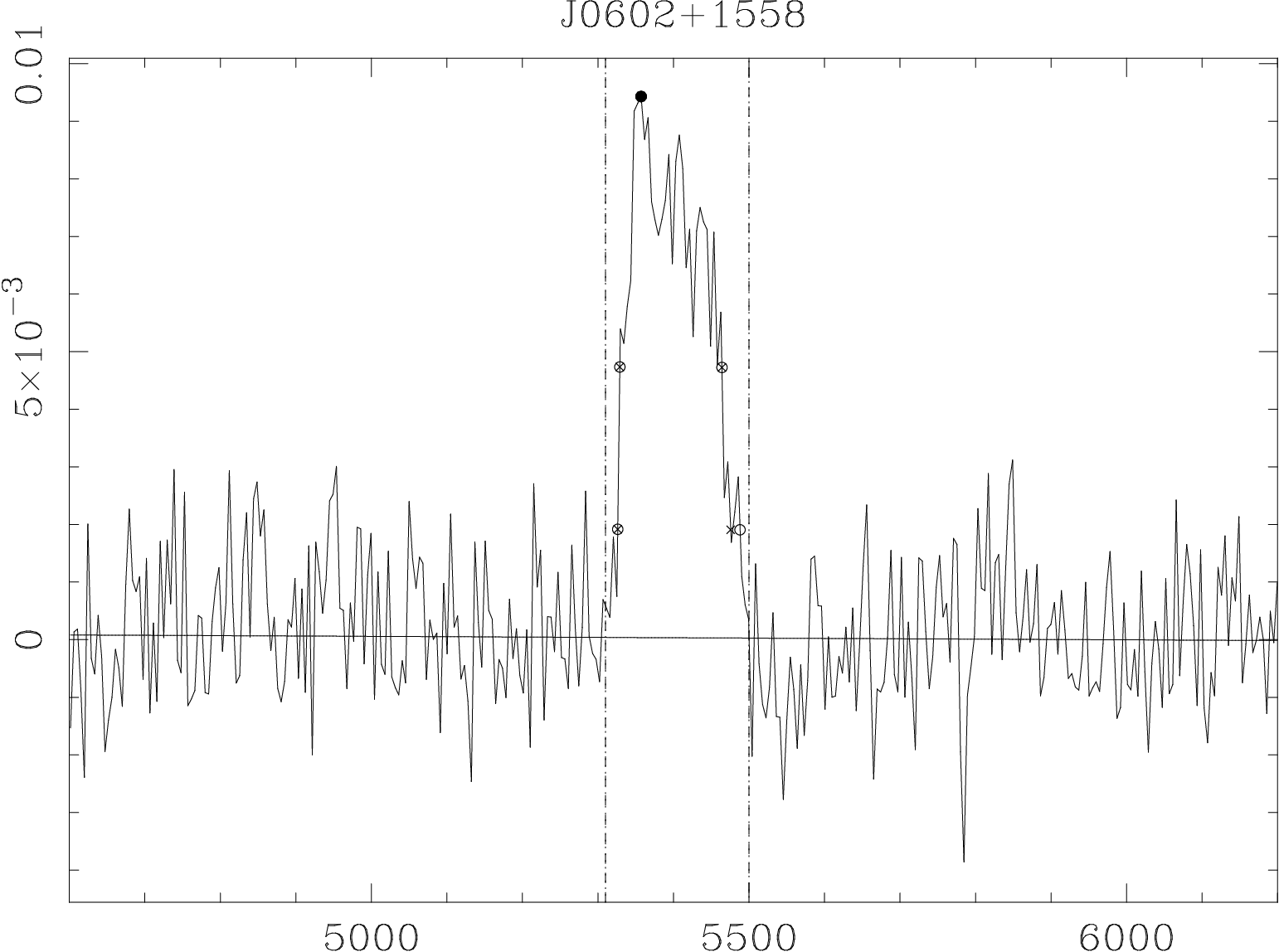}
\includegraphics[scale=0.29, angle=270]{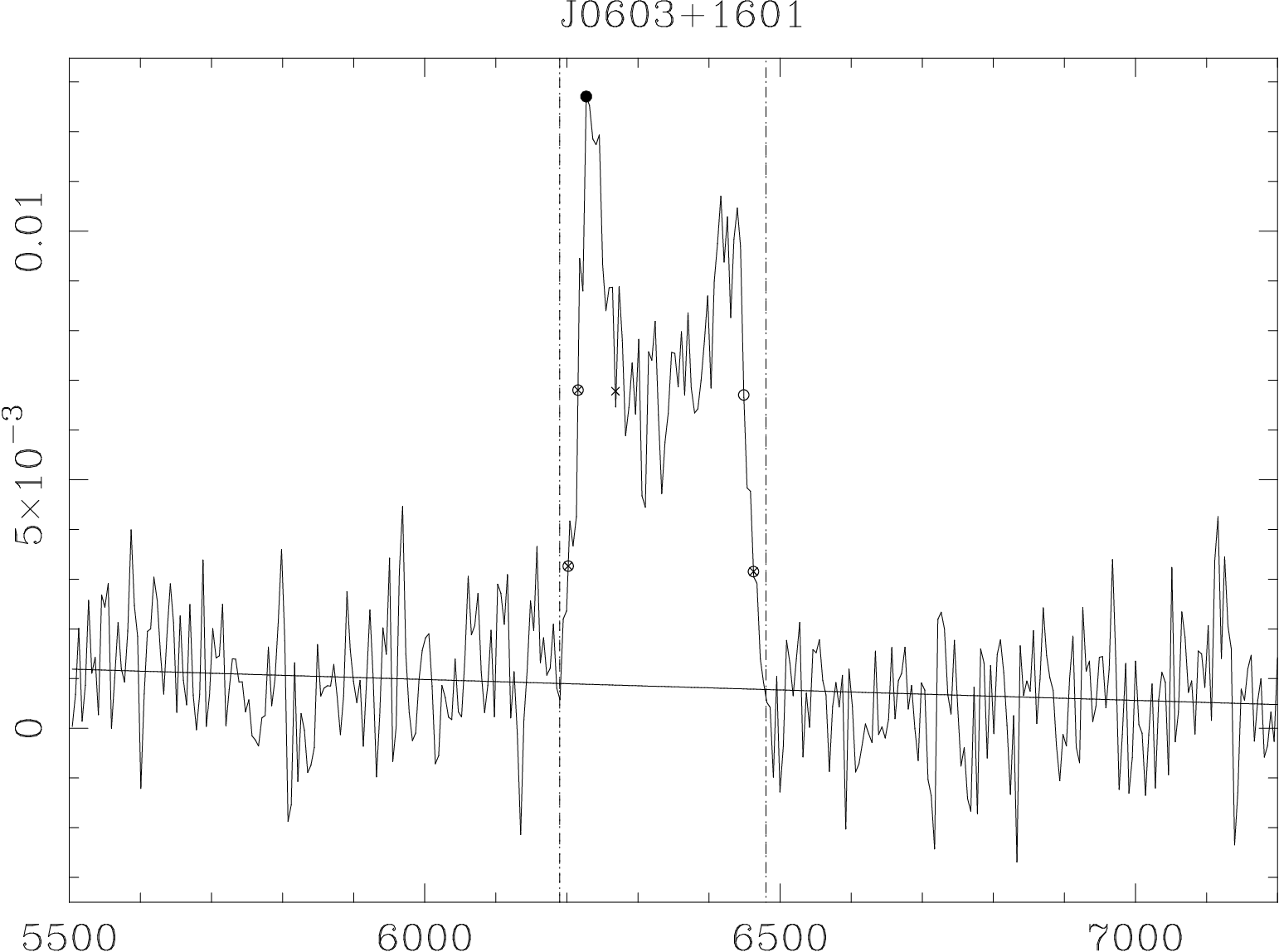}
\\[1cm]
\includegraphics[scale=0.29, angle=270]{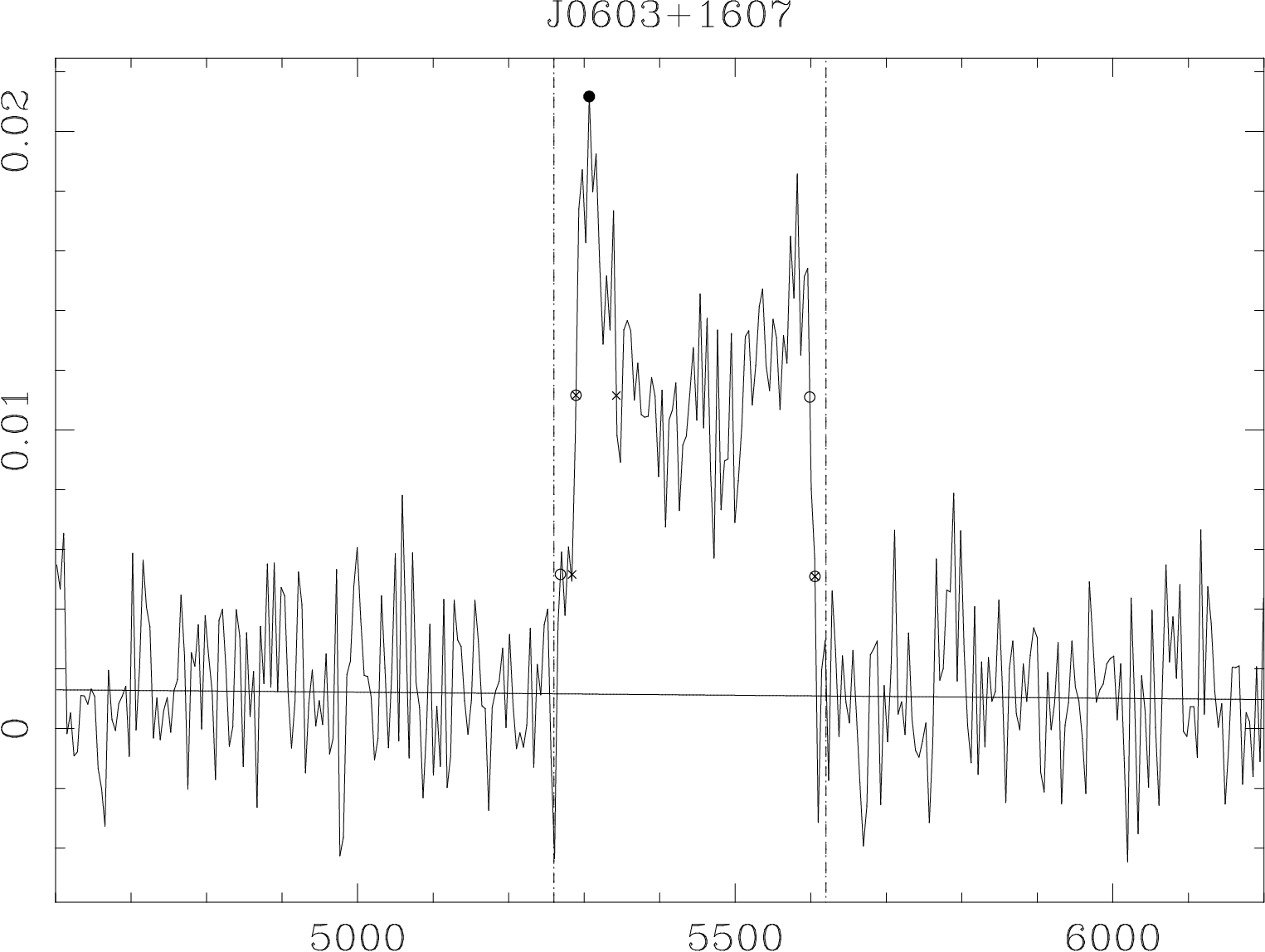}
\includegraphics[scale=0.29, angle=270]{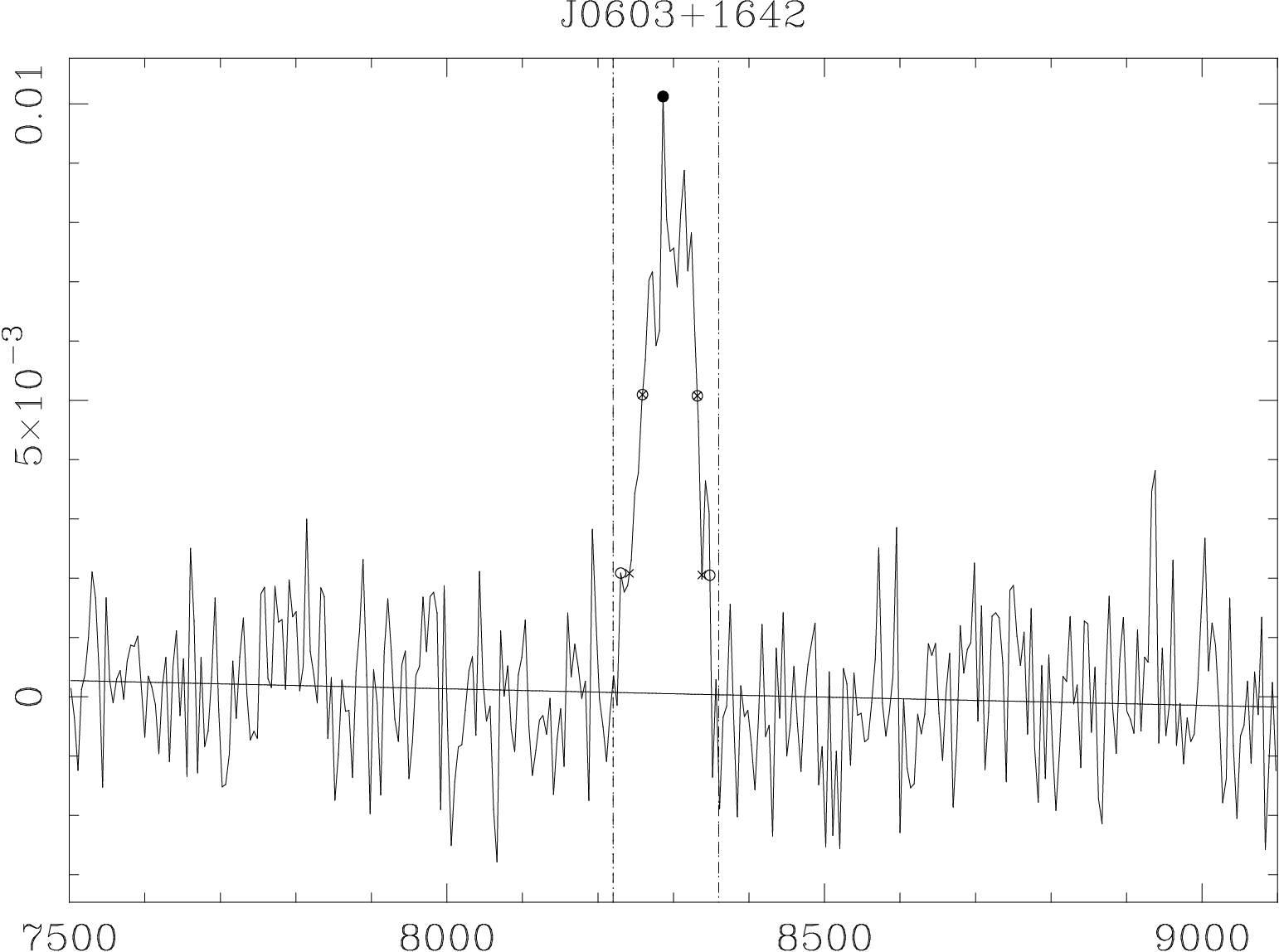}
\\[1cm]
\includegraphics[scale=0.29, angle=270]{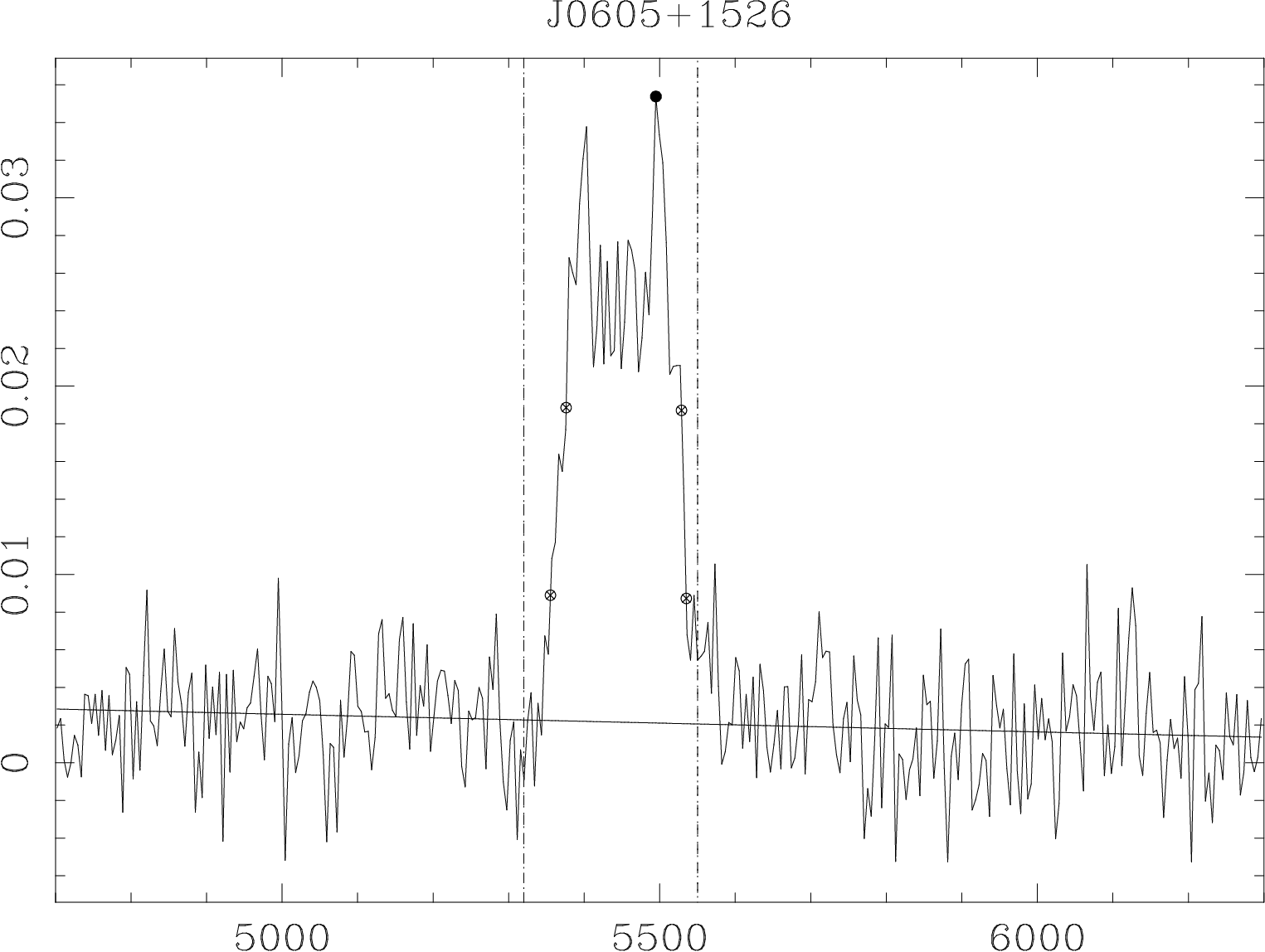}
\includegraphics[scale=0.29, angle=270]{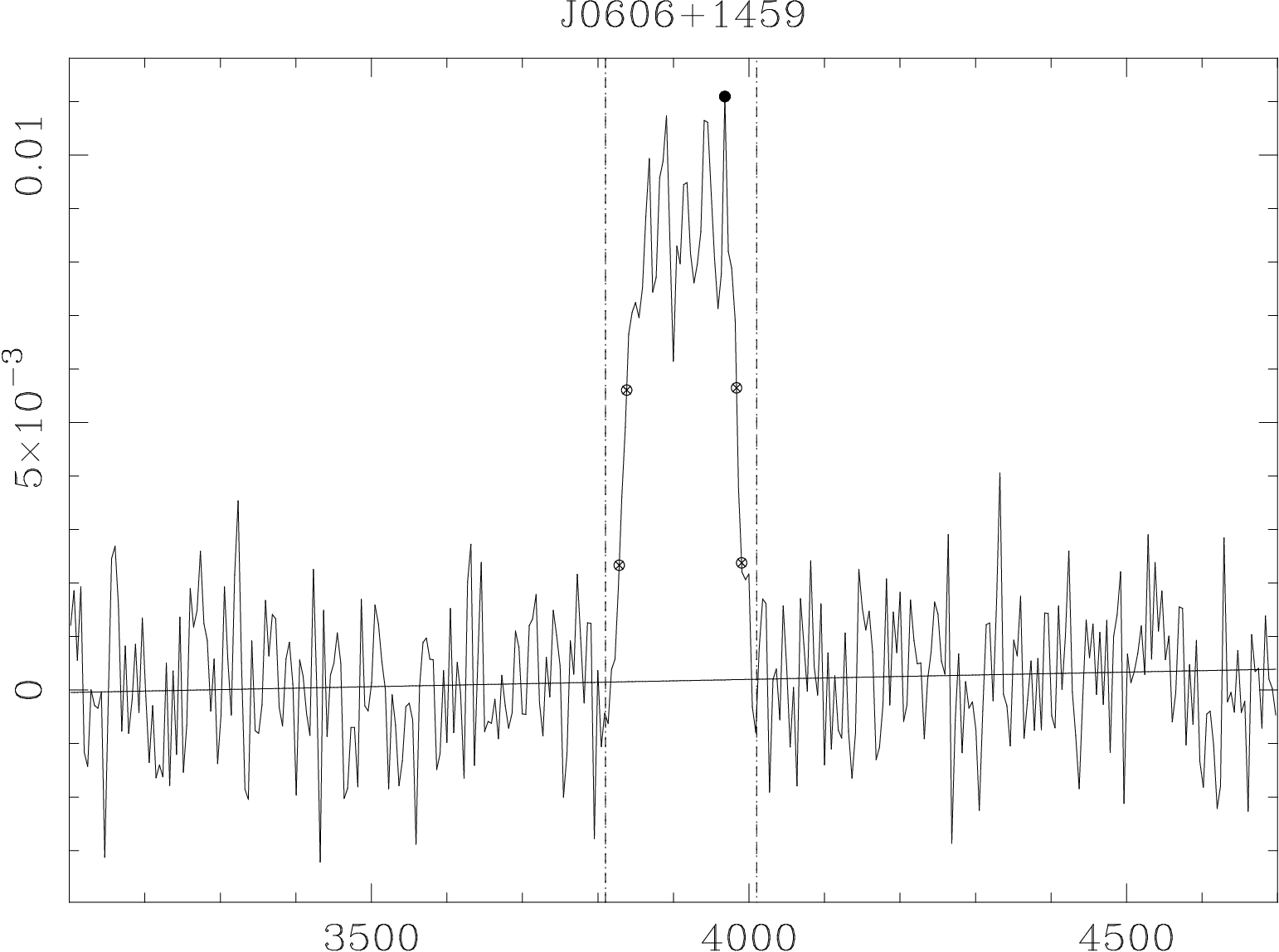}}
\\[1cm]
\leavevmode\smash{\makebox[0pt]{\hspace{0em}
  \rotatebox[origin=l]{90}{\hspace{22em}
    Flux Density (Jy)}%
}}\hspace{0pt plus 1filll}\null
\leavevmode\smash{\makebox[0pt]{\hspace{0em}
  {\hspace{0em}
    Heliocentric Velocity (km s$^{-1}$)}%
}}\hspace{0pt plus 1filll}\null
\caption{H\,{\sc{i}} profiles of ALFA ZOA detections. The description of the profiles is the same as in Figure \ref{J0617+1648.ps}. (Continued)}
\end{figure}
\clearpage

\begin{figure}
\addtocounter{figure}{-1}
\centering{\includegraphics[scale=0.29, angle=270]{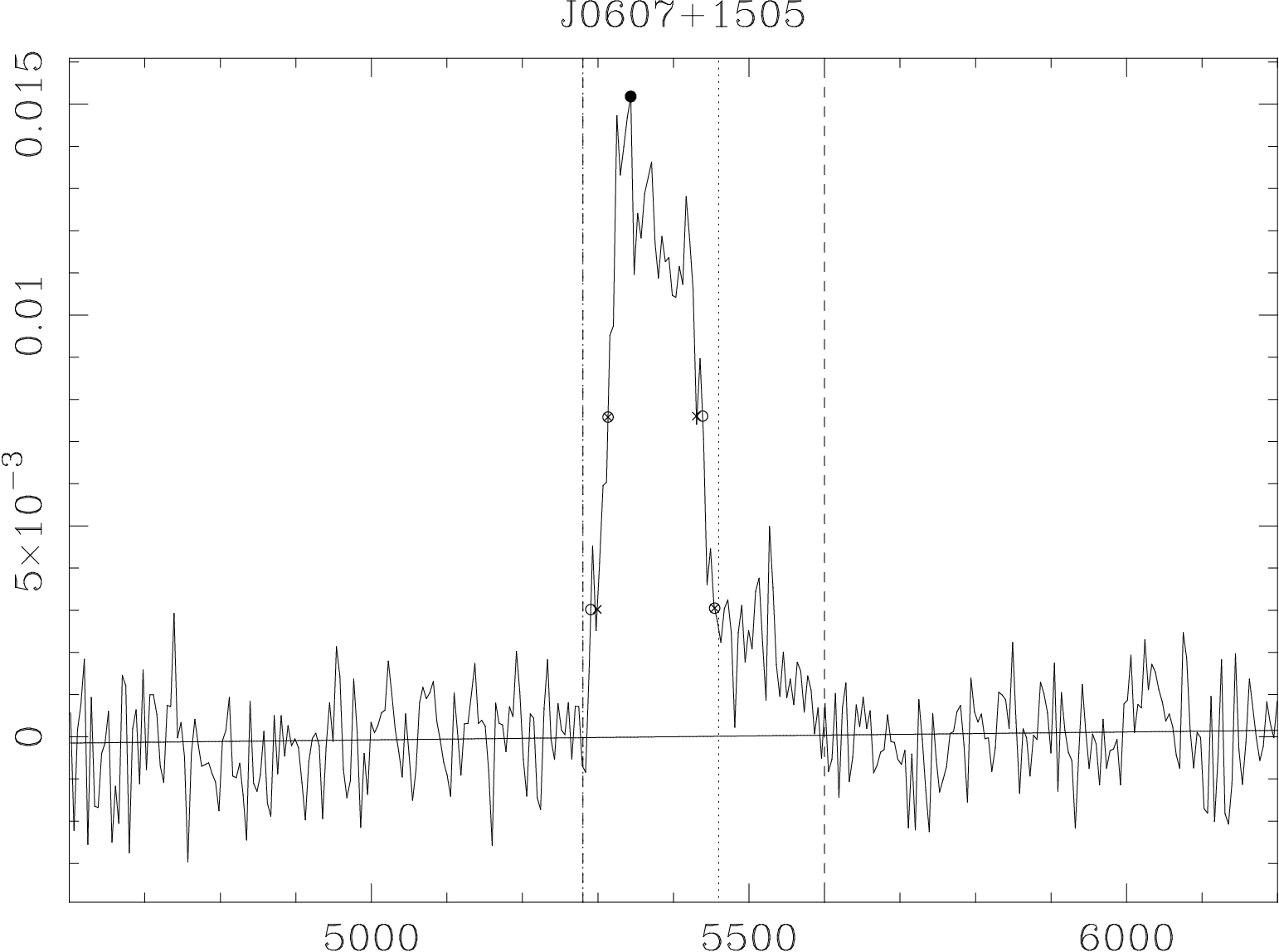}
\includegraphics[scale=0.29, angle=270]{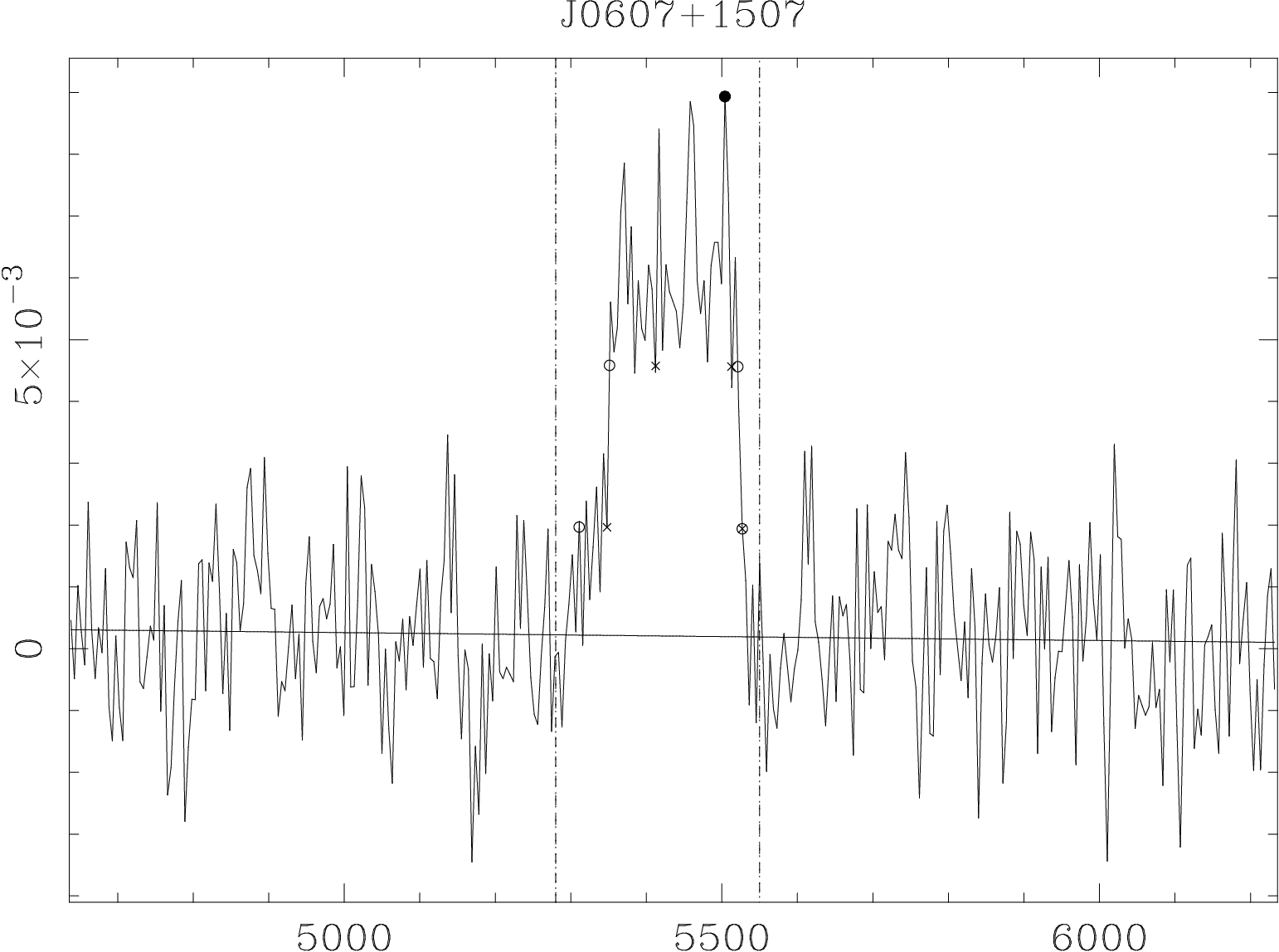}
\\[1cm]
\includegraphics[scale=0.29, angle=270]{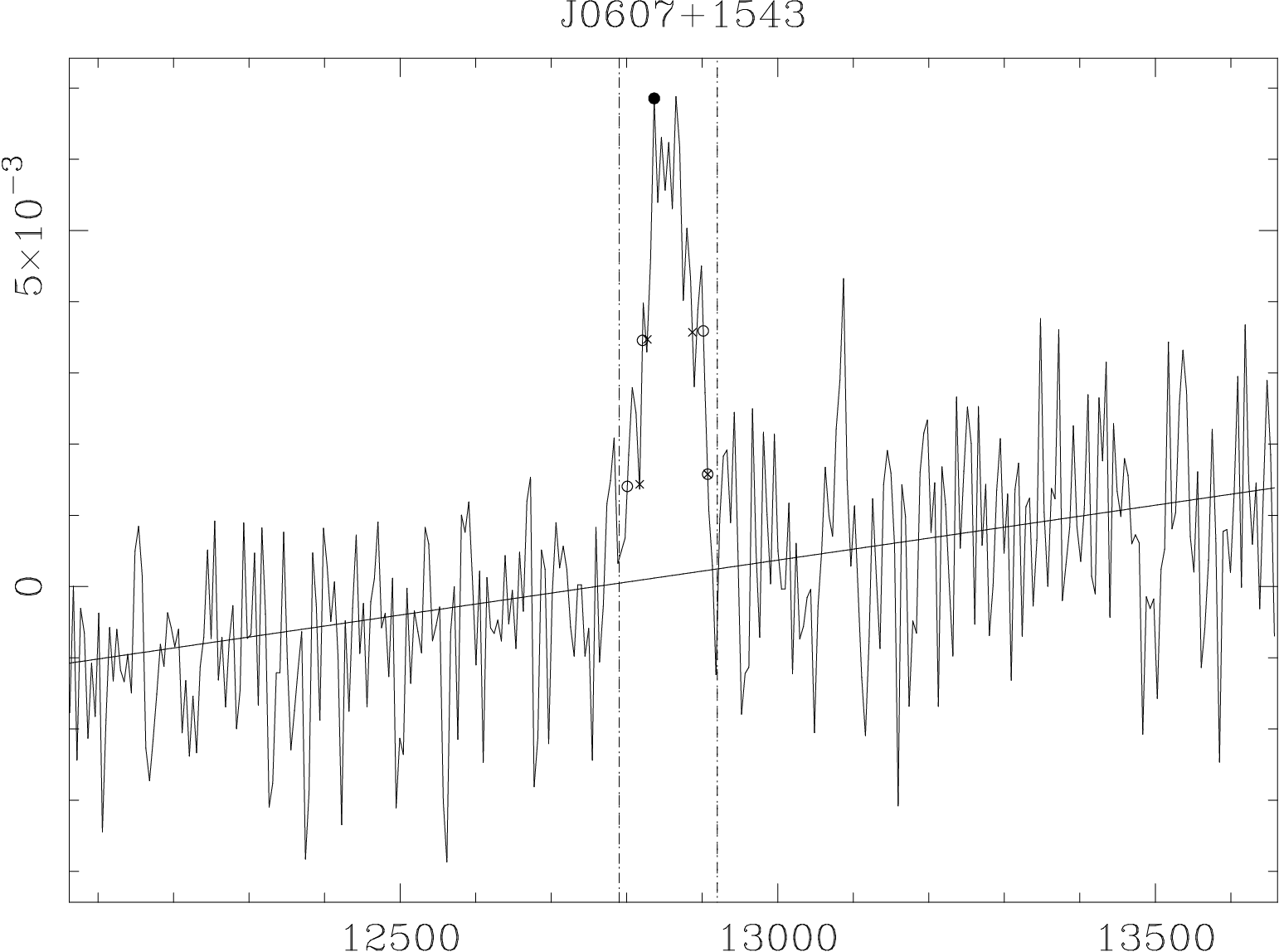}
\includegraphics[scale=0.29, angle=270]{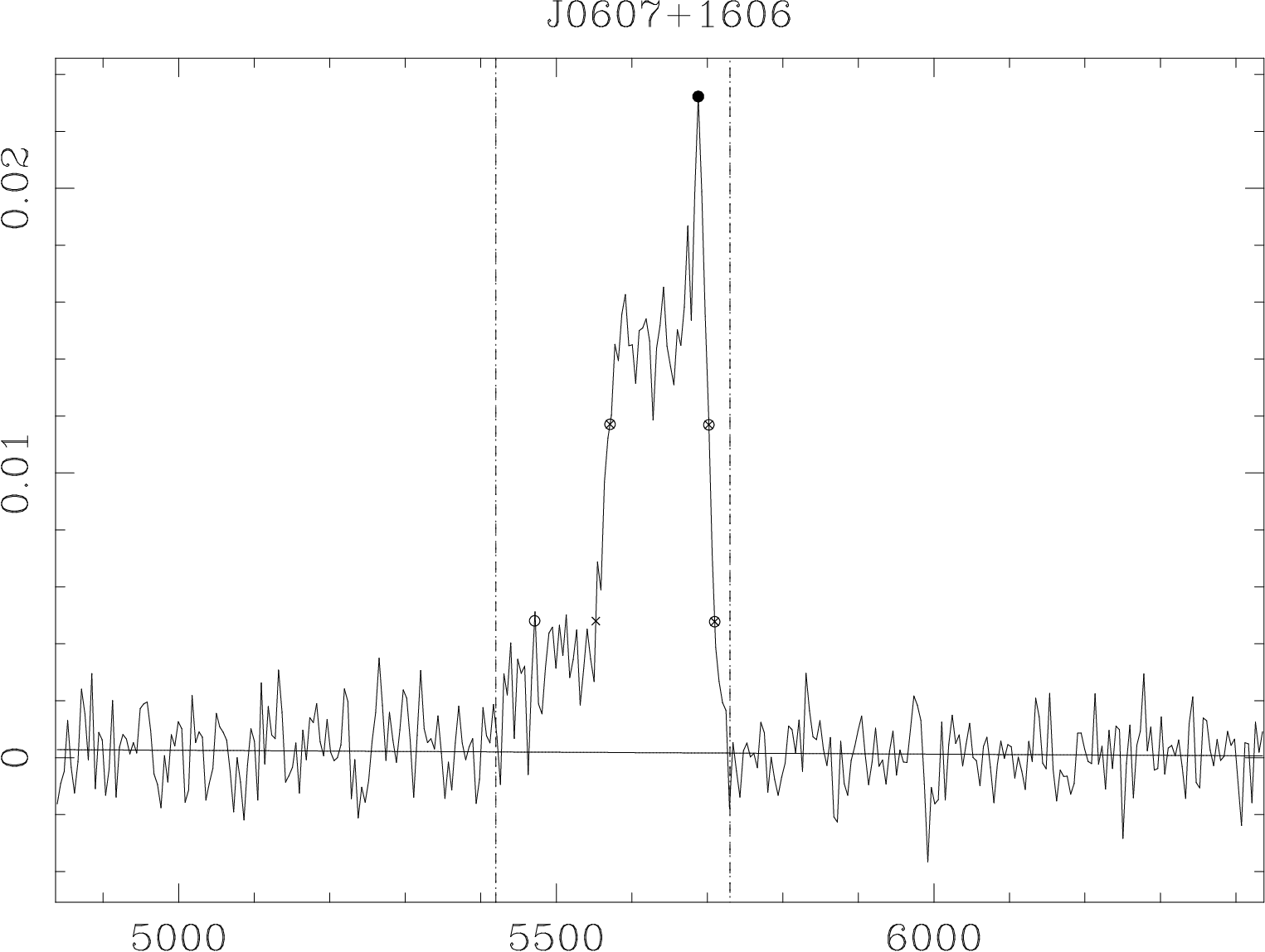}
\\[1cm]
\includegraphics[scale=0.29, angle=270]{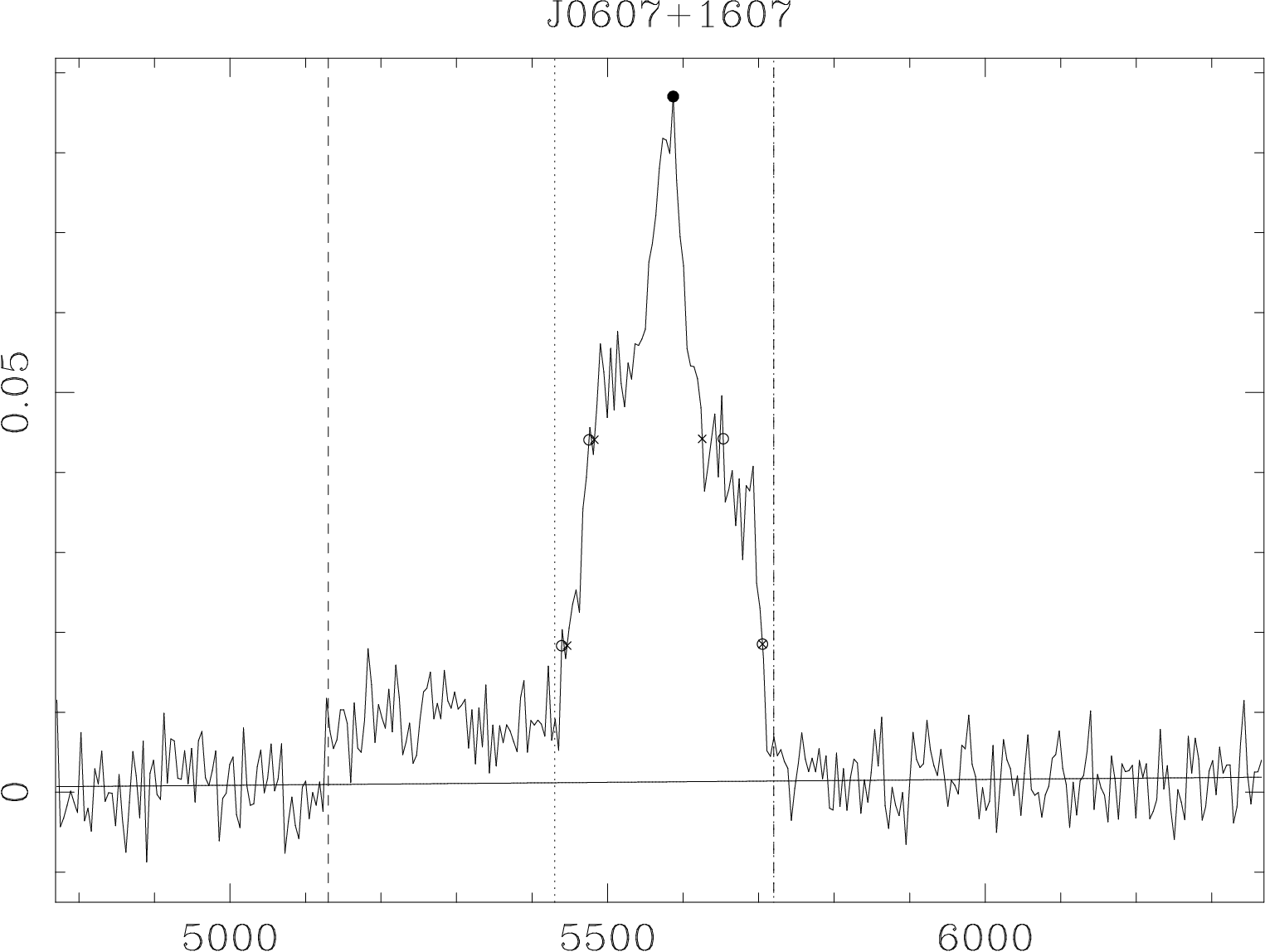}
\includegraphics[scale=0.29, angle=270]{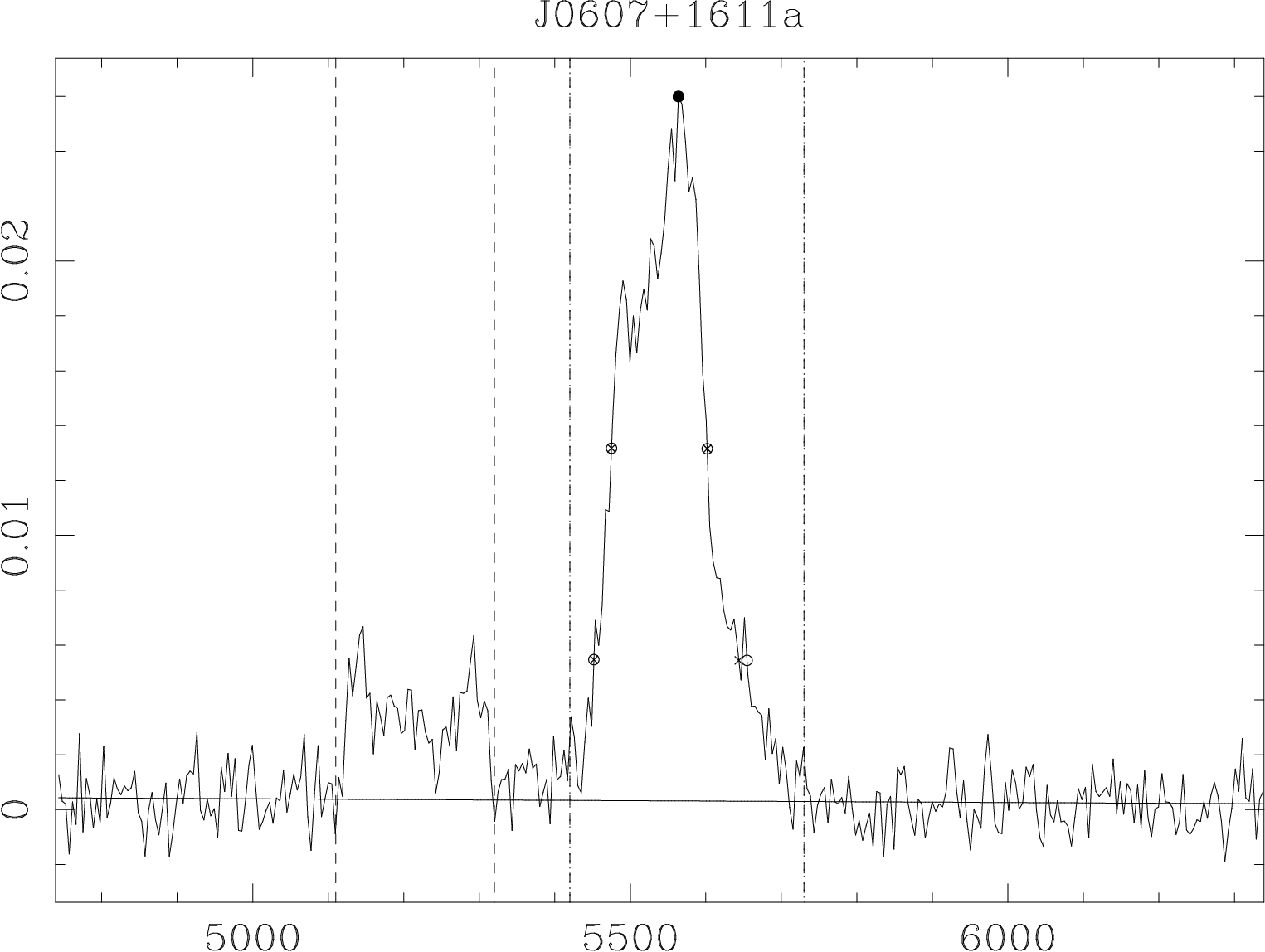}}
\\[1cm]
\leavevmode\smash{\makebox[0pt]{\hspace{0em}
  \rotatebox[origin=l]{90}{\hspace{22em}
    Flux Density (Jy)}%
}}\hspace{0pt plus 1filll}\null
\leavevmode\smash{\makebox[0pt]{\hspace{0em}
  {\hspace{0em}
    Heliocentric Velocity (km s$^{-1}$)}%
}}\hspace{0pt plus 1filll}\null
\caption{H\,{\sc{i}} profiles of ALFA ZOA detections. The description of the profiles is the same as in Figure \ref{J0617+1648.ps}. (Continued)}
\end{figure}
\clearpage

\begin{figure}
\addtocounter{figure}{-1}
\centering{\includegraphics[scale=0.29, angle=270]{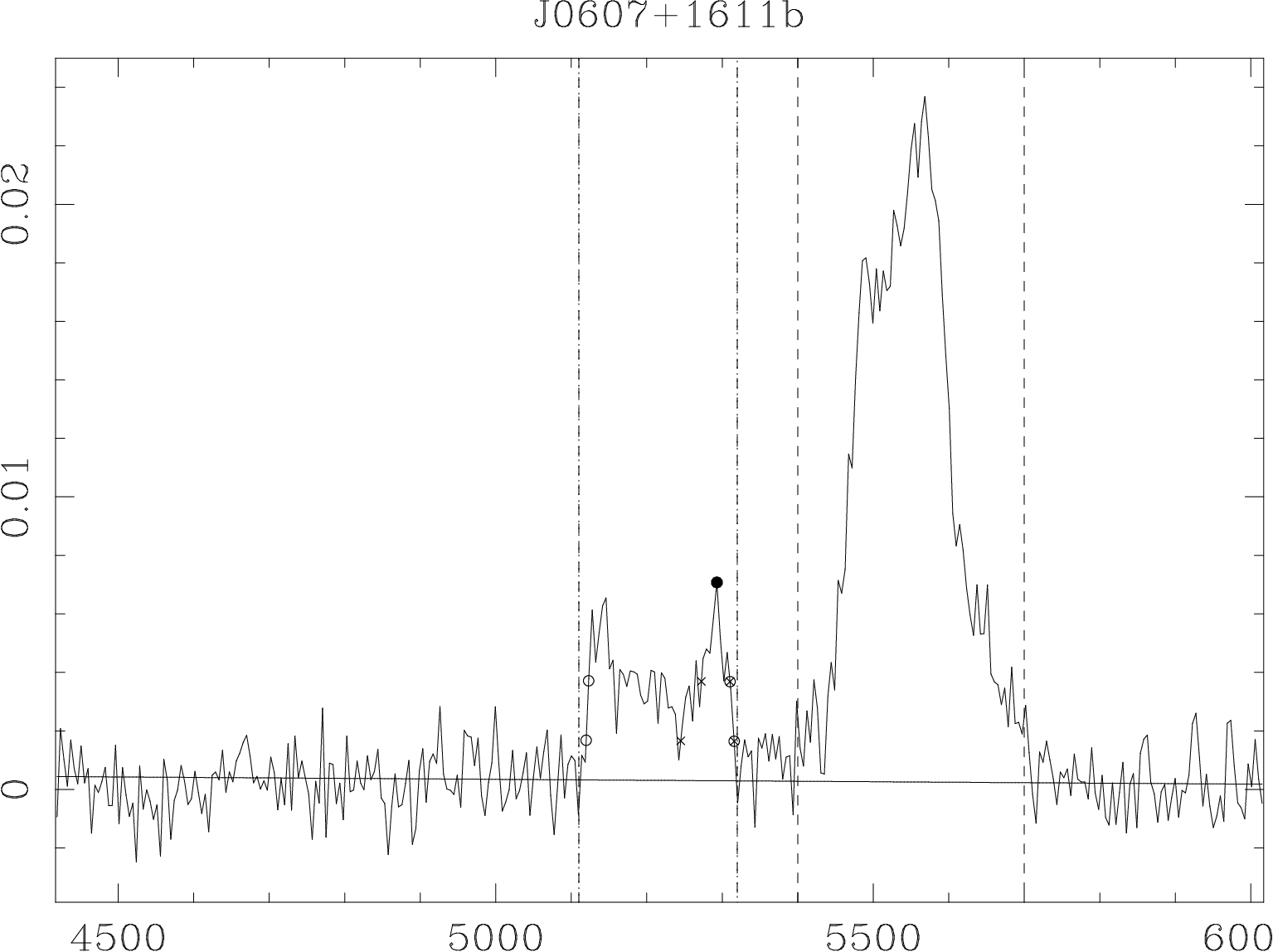}
\includegraphics[scale=0.29, angle=270]{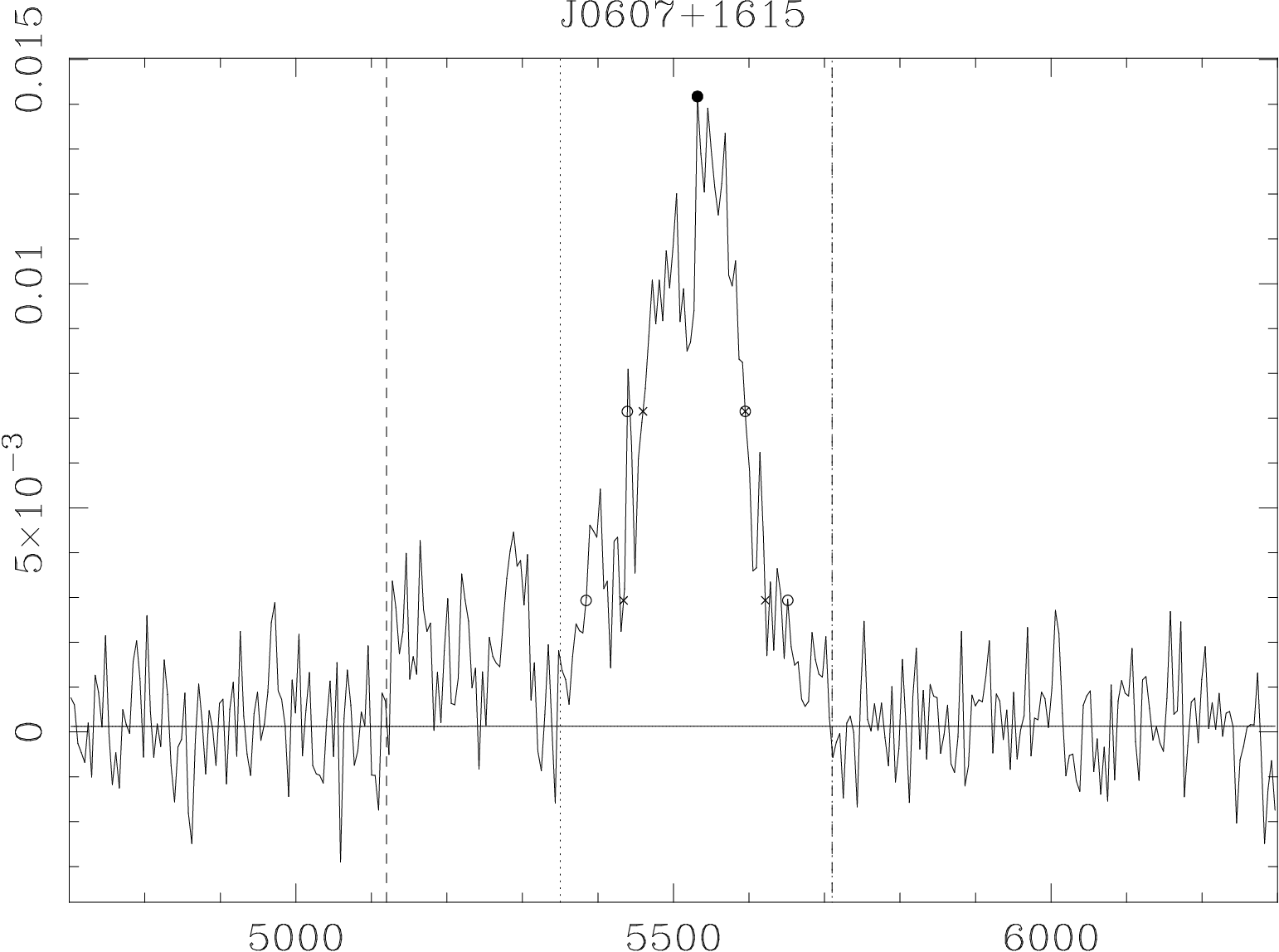}
\\[1cm]
\includegraphics[scale=0.29, angle=270]{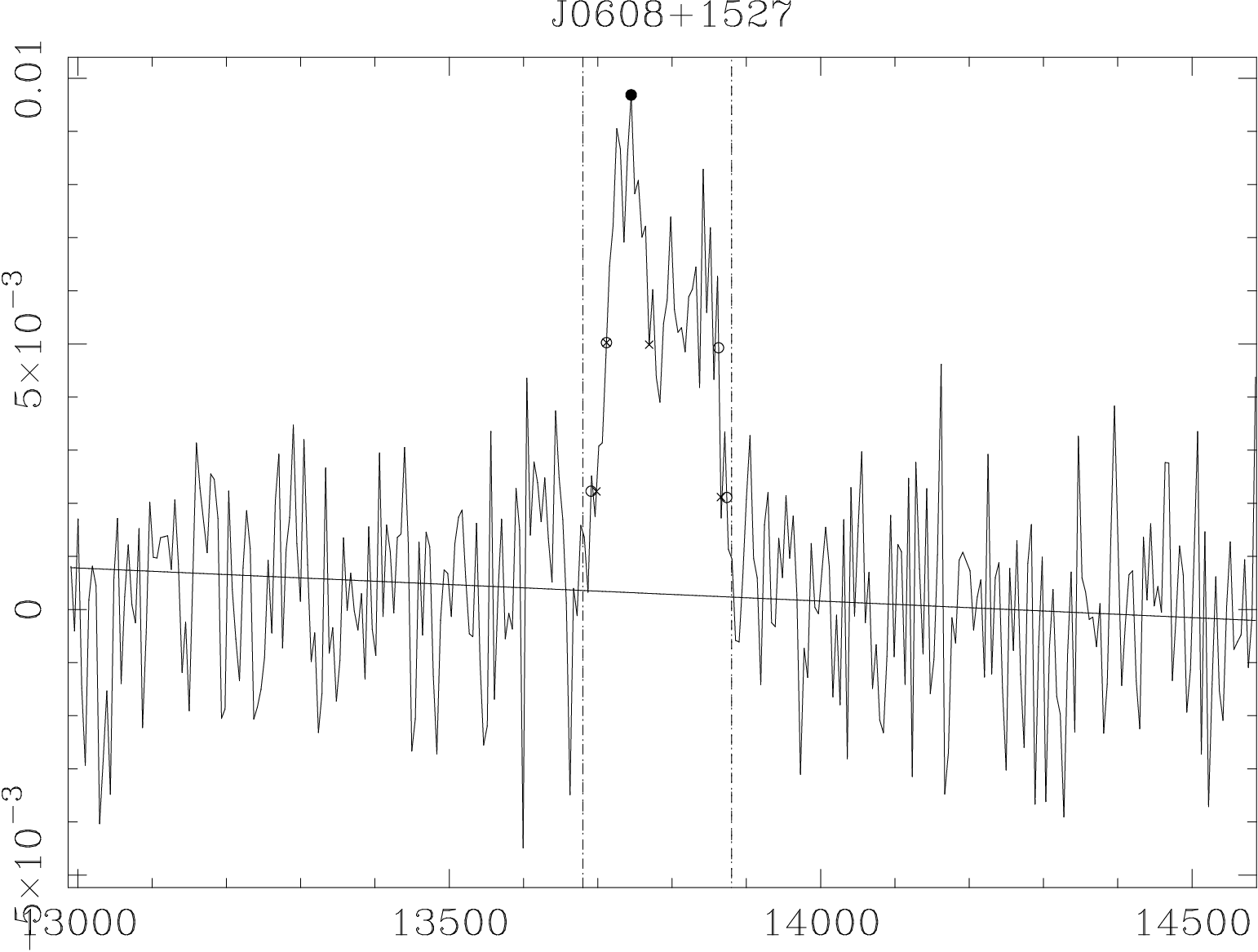}
\includegraphics[scale=0.29, angle=270]{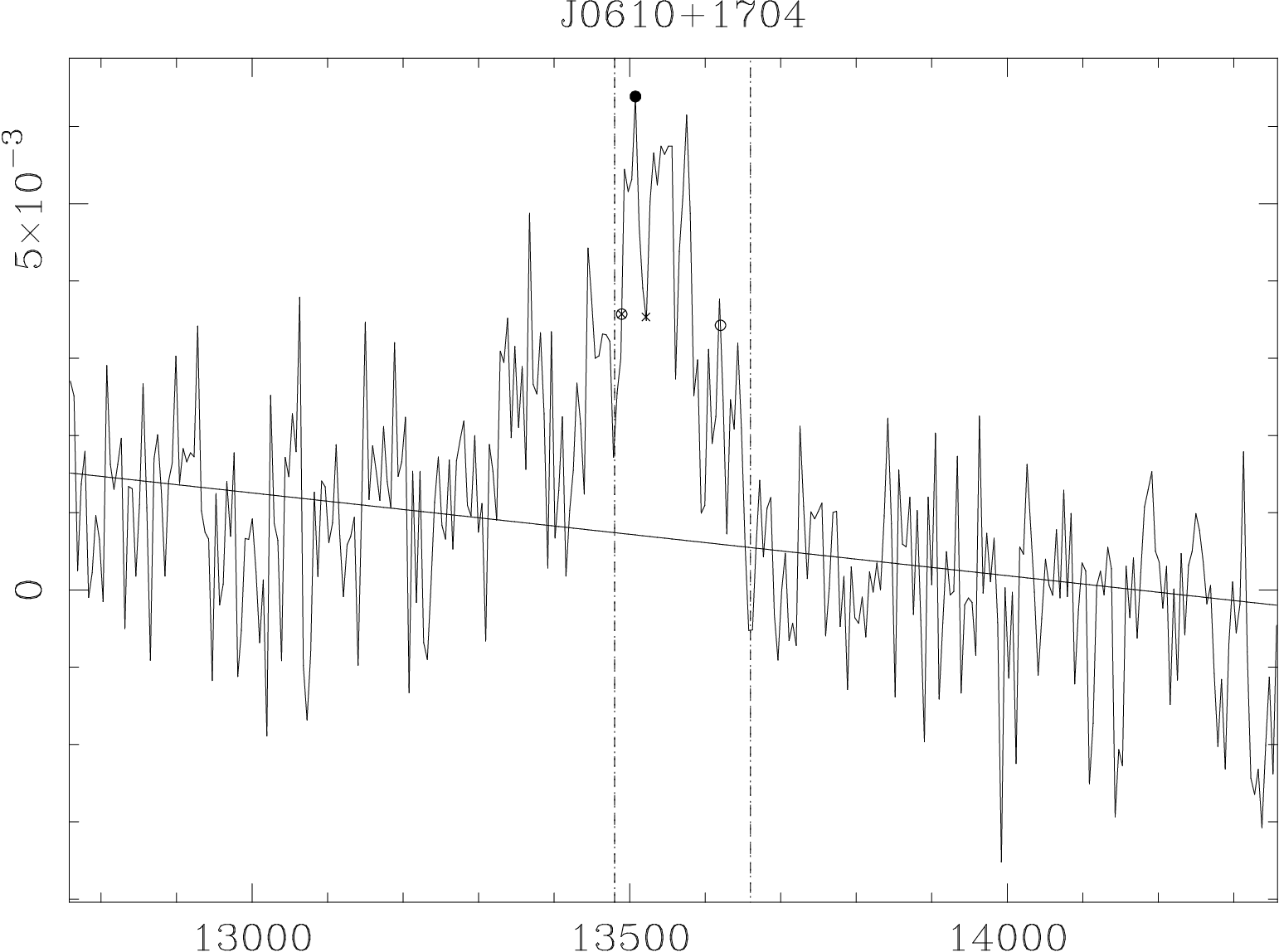}
\\[1cm]
\includegraphics[scale=0.29, angle=270]{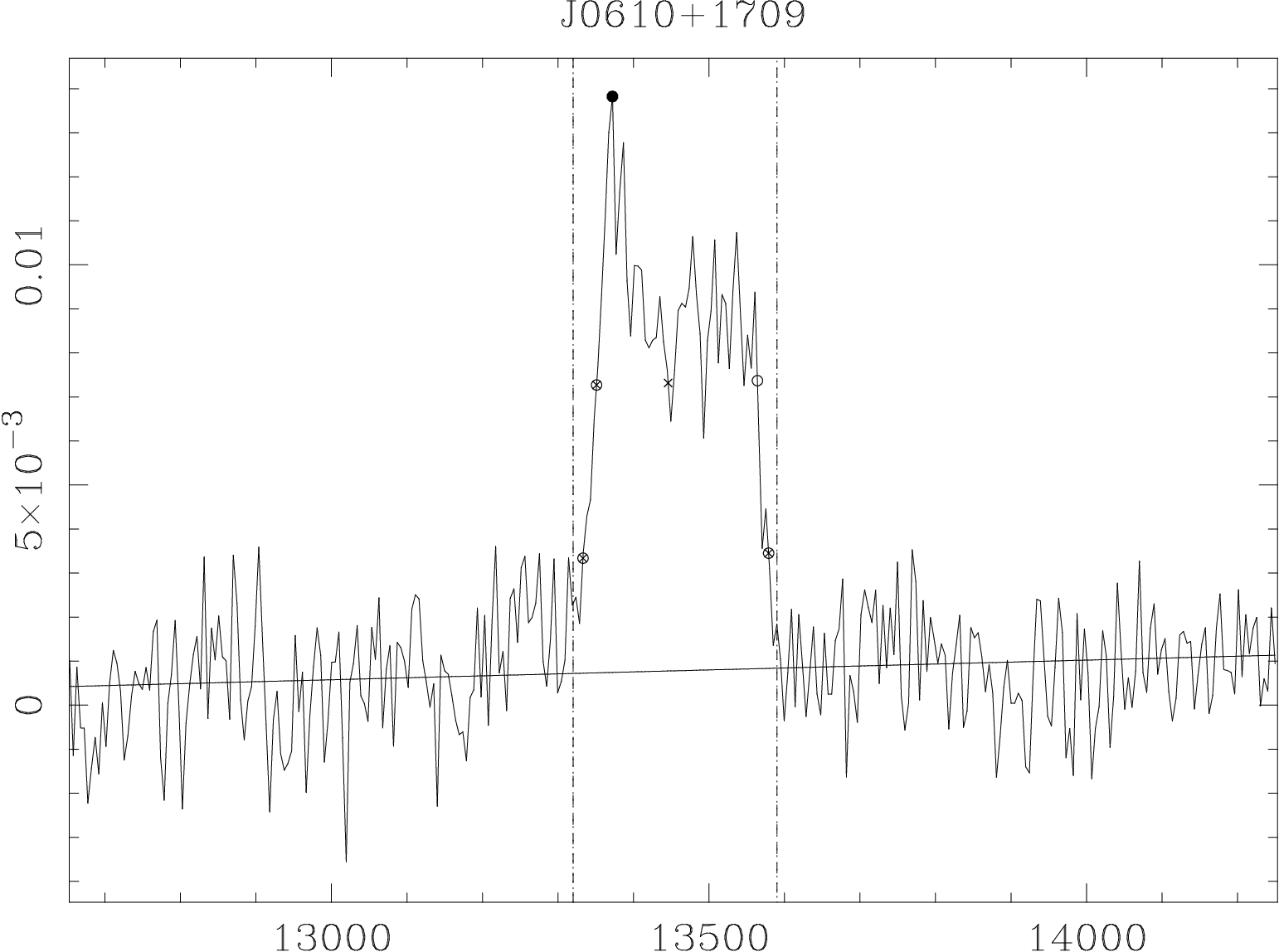}
\includegraphics[scale=0.29, angle=270]{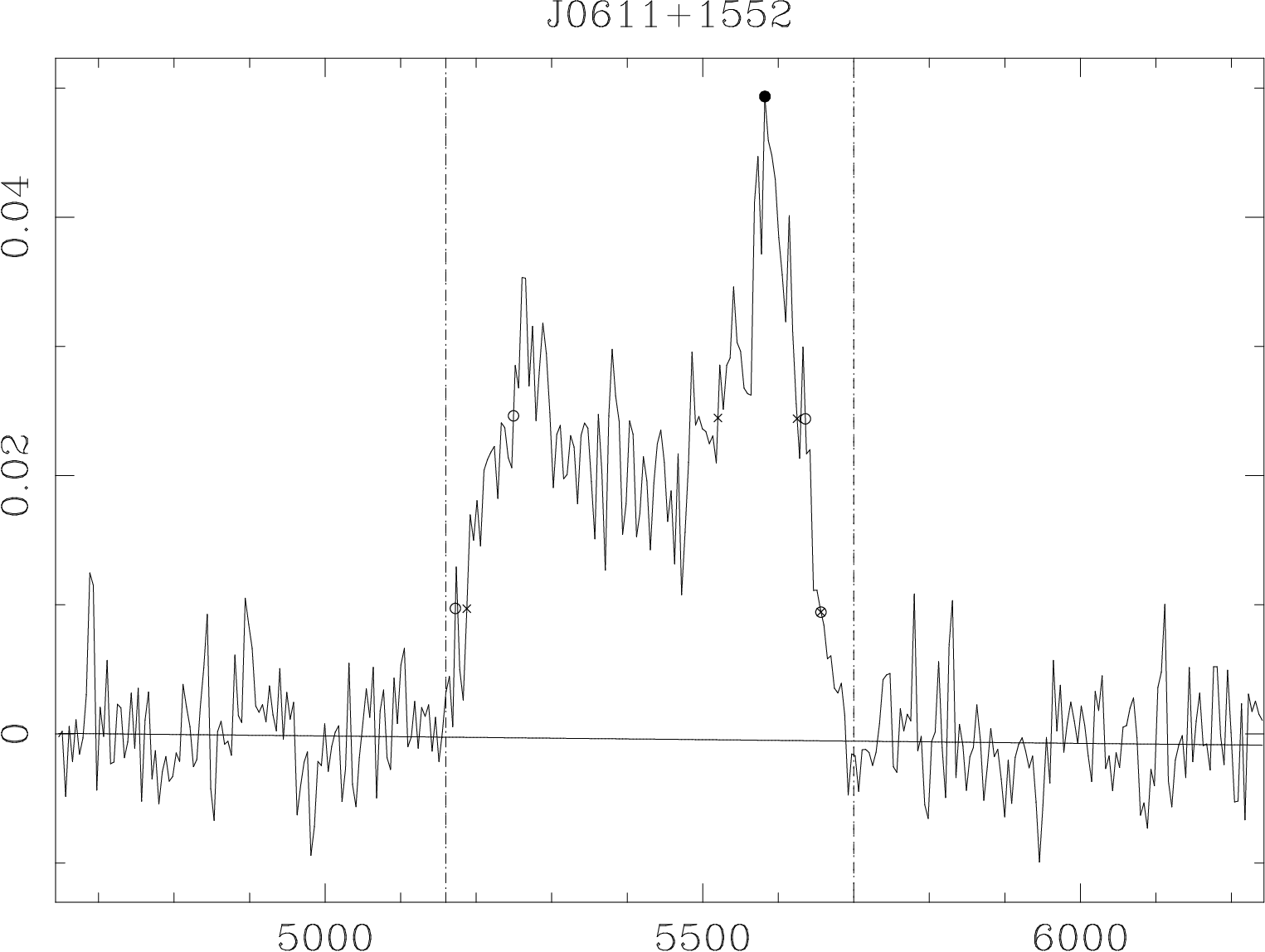}}
\\[1cm]
\leavevmode\smash{\makebox[0pt]{\hspace{0em}
  \rotatebox[origin=l]{90}{\hspace{22em}
    Flux Density (Jy)}%
}}\hspace{0pt plus 1filll}\null
\leavevmode\smash{\makebox[0pt]{\hspace{0em}
  {\hspace{0em}
    Heliocentric Velocity (km s$^{-1}$)}%
}}\hspace{0pt plus 1filll}\null
\caption{H\,{\sc{i}} profiles of ALFA ZOA detections. The description of the profiles is the same as in Figure \ref{J0617+1648.ps}. (Continued)}
\end{figure}
\clearpage

\begin{figure}
\addtocounter{figure}{-1}
\centering{\includegraphics[scale=0.29, angle=270]{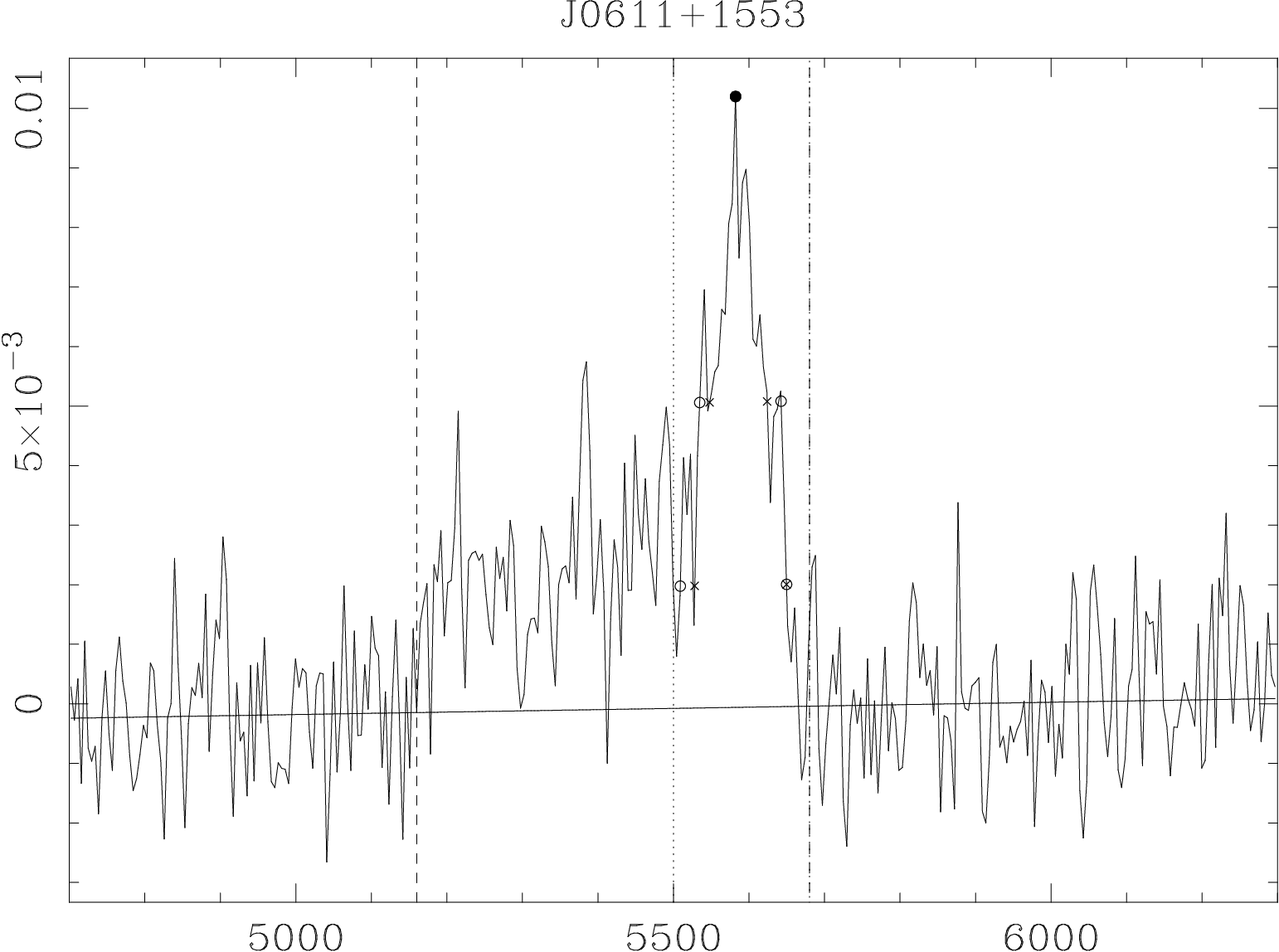}
\includegraphics[scale=0.29, angle=270]{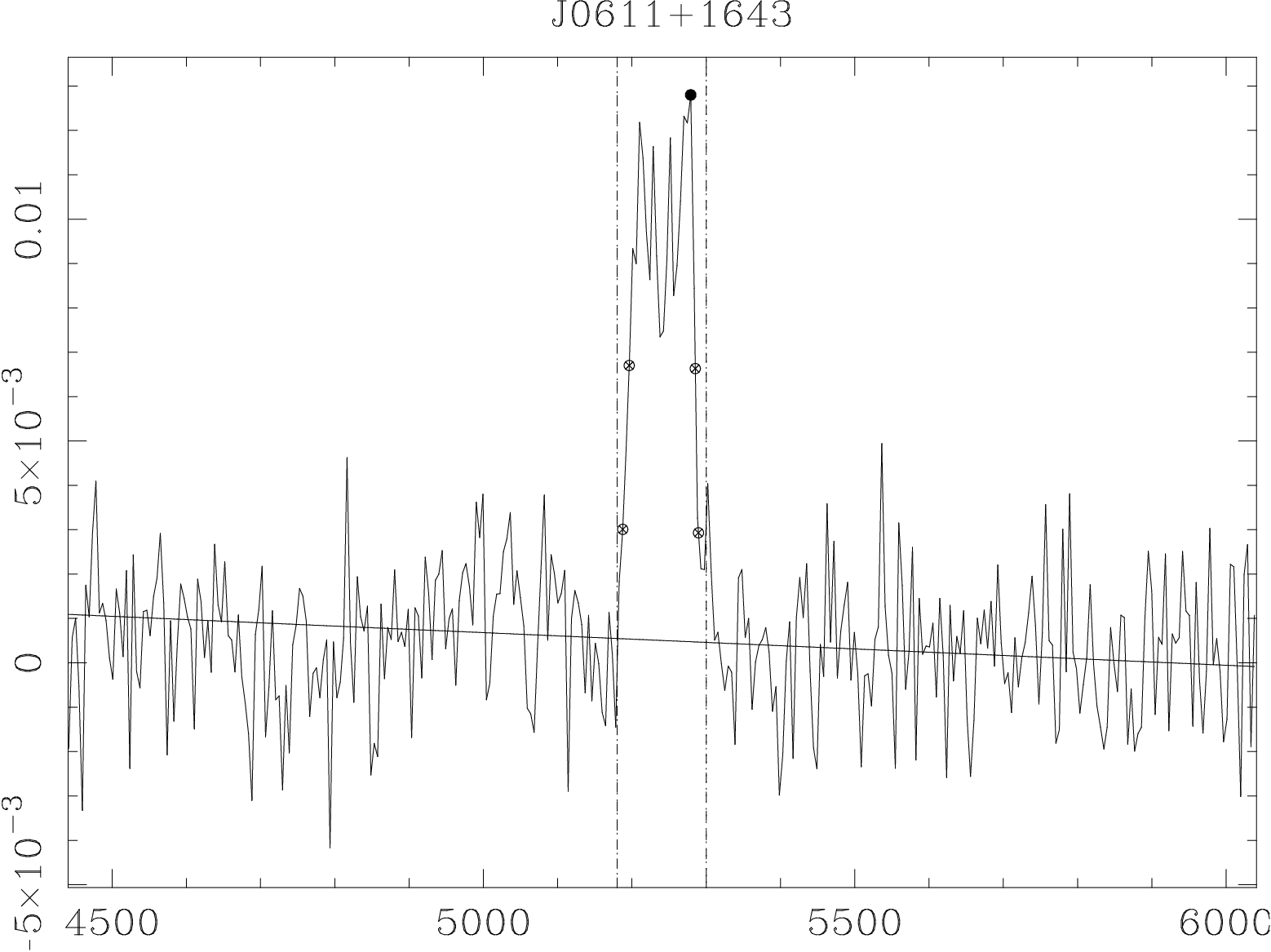}
\\[1cm]
\includegraphics[scale=0.29, angle=270]{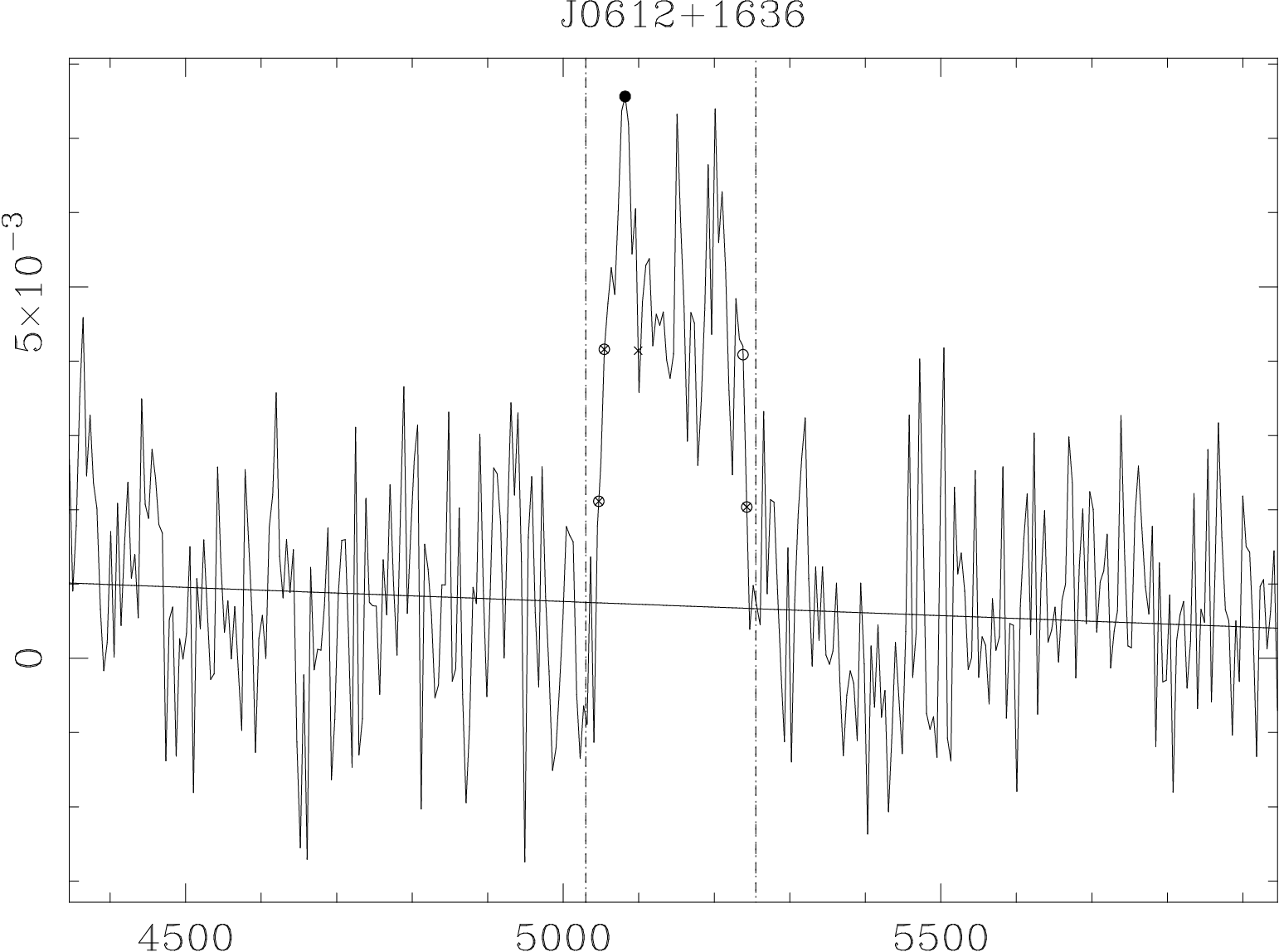}
\includegraphics[scale=0.29, angle=270]{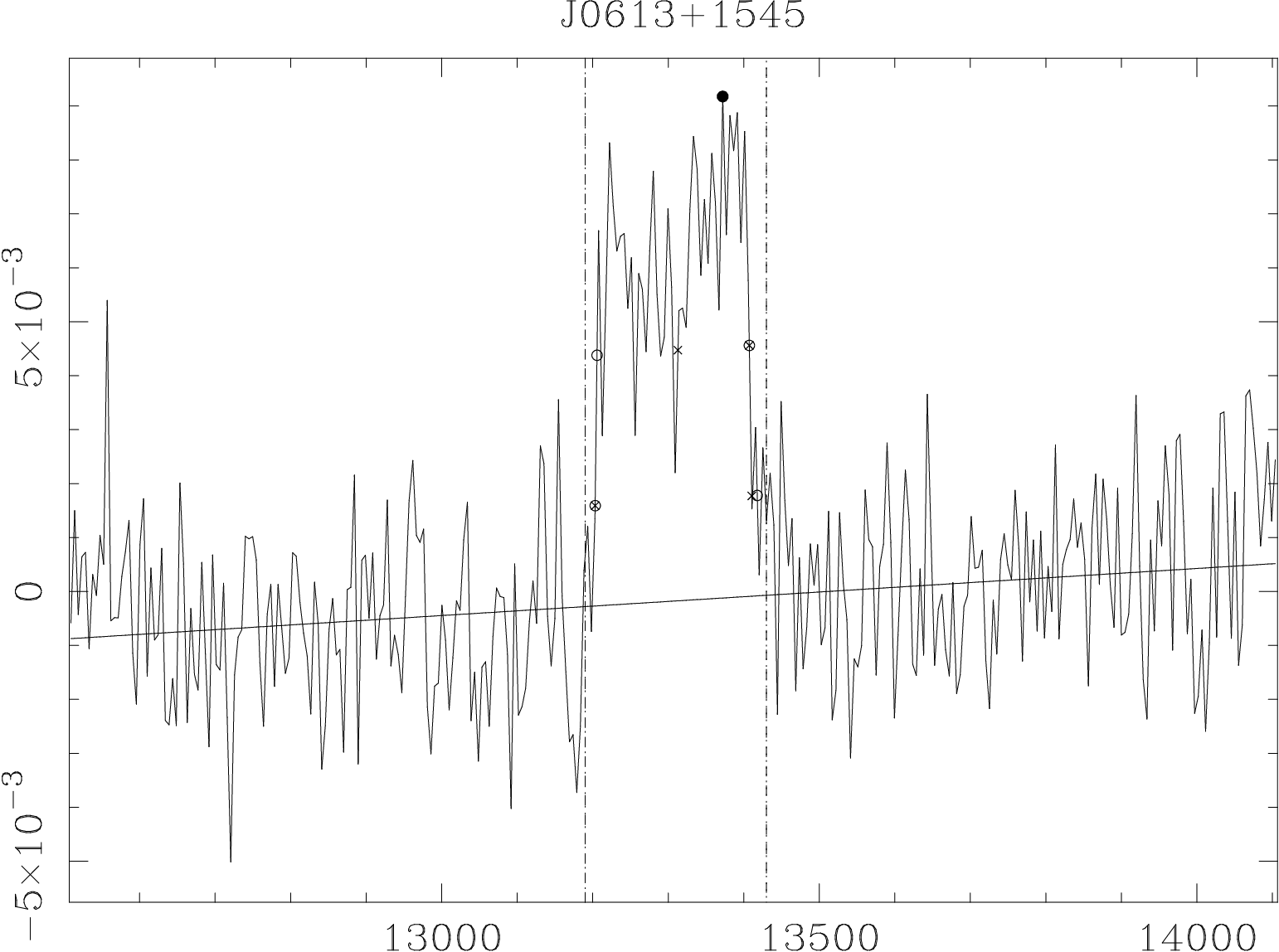}
\\[1cm]
\includegraphics[scale=0.29, angle=270]{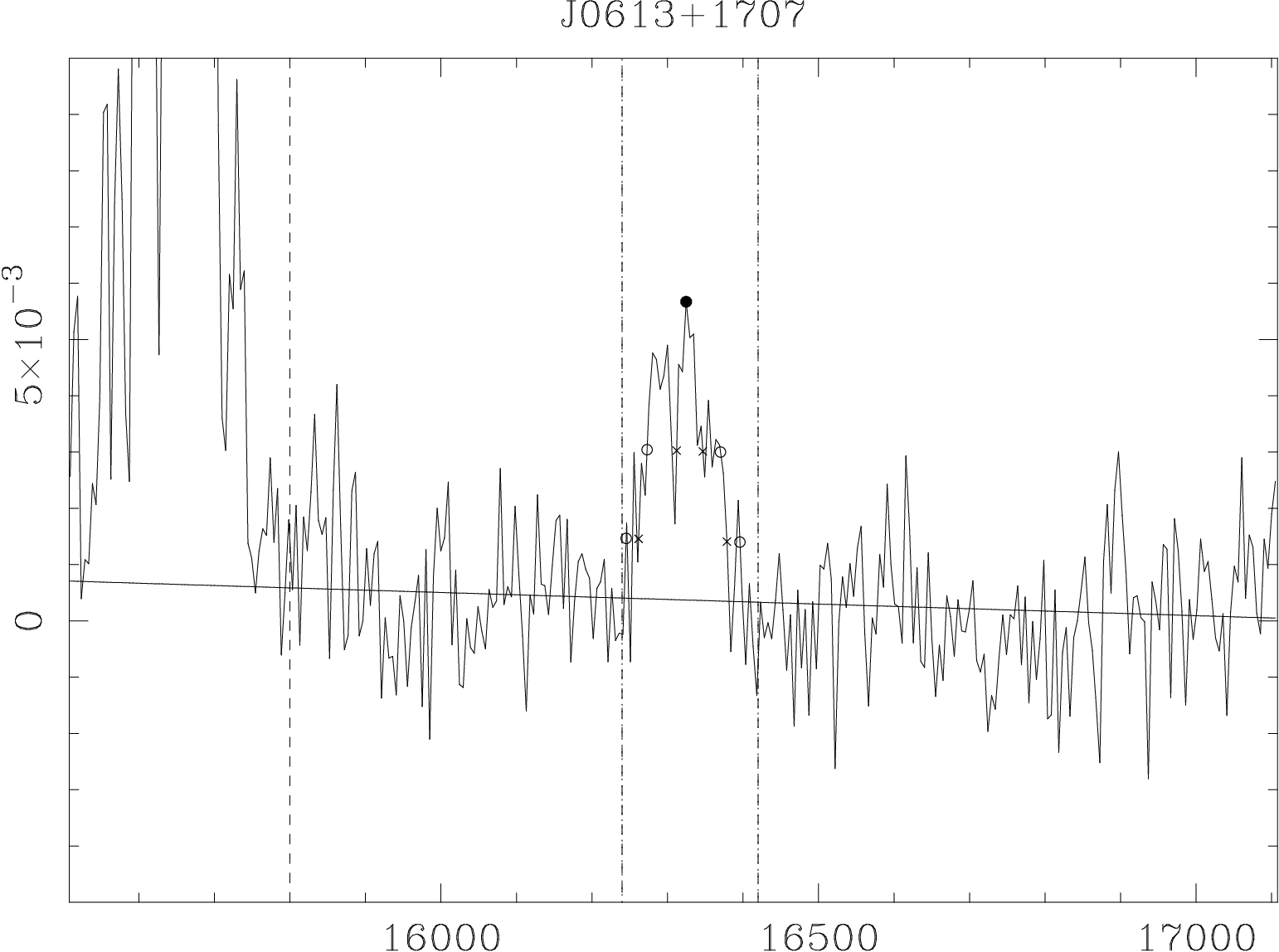}
\includegraphics[scale=0.29, angle=270]{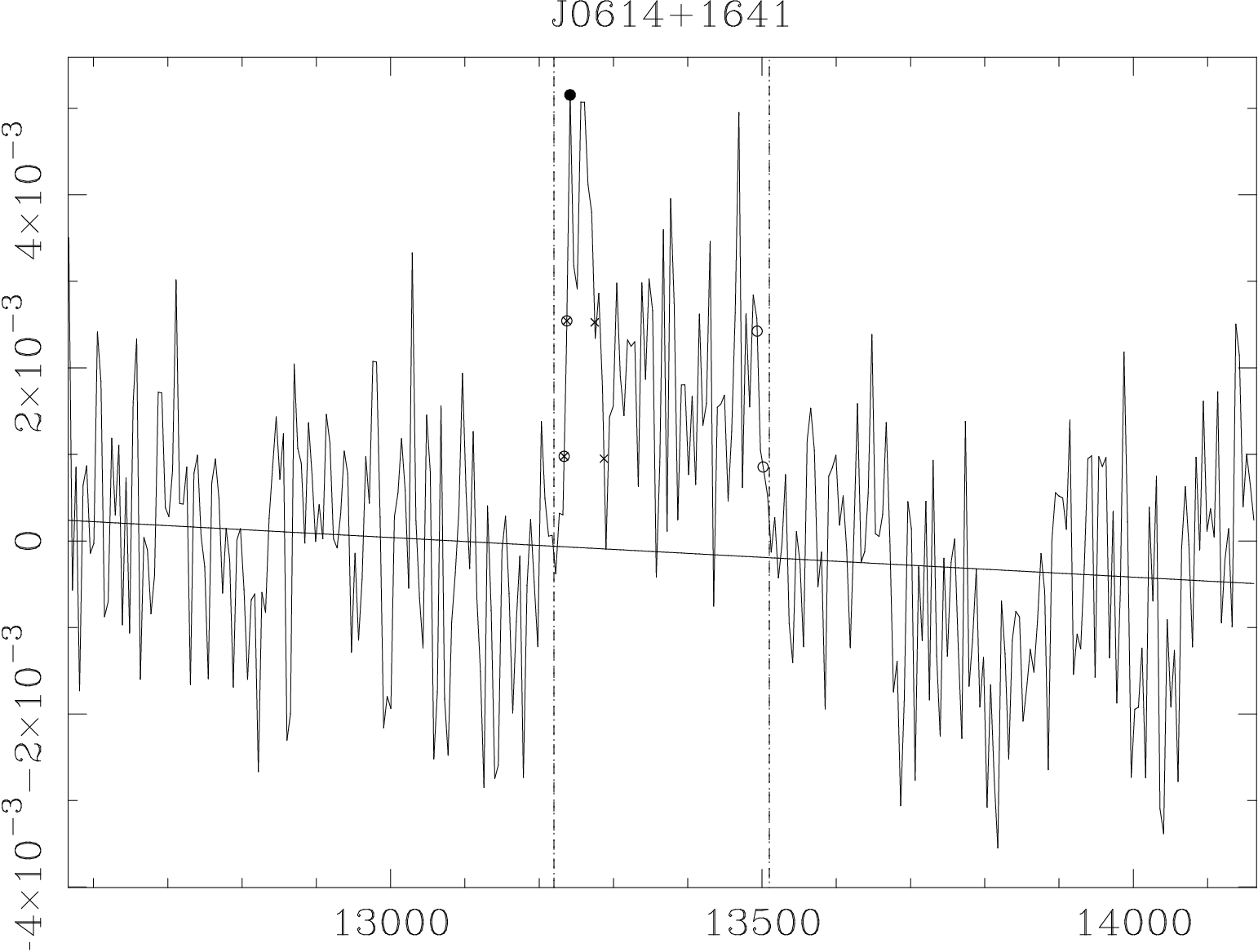}}
\\[1cm]
\leavevmode\smash{\makebox[0pt]{\hspace{0em}
  \rotatebox[origin=l]{90}{\hspace{22em}
    Flux Density (Jy)}%
}}\hspace{0pt plus 1filll}\null
\leavevmode\smash{\makebox[0pt]{\hspace{0em}
  {\hspace{0em}
    Heliocentric Velocity (km s$^{-1}$)}%
}}\hspace{0pt plus 1filll}\null
\caption{H\,{\sc{i}} profiles of ALFA ZOA detections. The description of the profiles is the same as in Figure \ref{J0617+1648.ps}. (Continued)}
\end{figure}
\clearpage

\begin{figure}
\addtocounter{figure}{-1}
\centering{\includegraphics[scale=0.29, angle=270]{fig5_49.ps}
\includegraphics[scale=0.29, angle=270]{fig5_50.ps}
\\[1cm]
\includegraphics[scale=0.29, angle=270]{fig5_51.ps}
\includegraphics[scale=0.29, angle=270]{fig5_52.ps}
\\[1cm]
\includegraphics[scale=0.29, angle=270]{fig5_53.ps}
\includegraphics[scale=0.29, angle=270]{fig5_54.ps}}
\\[1cm]
\leavevmode\smash{\makebox[0pt]{\hspace{0em}
  \rotatebox[origin=l]{90}{\hspace{22em}
    Flux Density (Jy)}%
}}\hspace{0pt plus 1filll}\null
\leavevmode\smash{\makebox[0pt]{\hspace{0em}
  {\hspace{0em}
    Heliocentric Velocity (km s$^{-1}$)}%
}}\hspace{0pt plus 1filll}\null
\caption{H\,{\sc{i}} profiles of ALFA ZOA detections. The description of the profiles is the same as in Figure \ref{J0617+1648.ps}. (Continued)}
\end{figure}
\clearpage

\begin{figure}
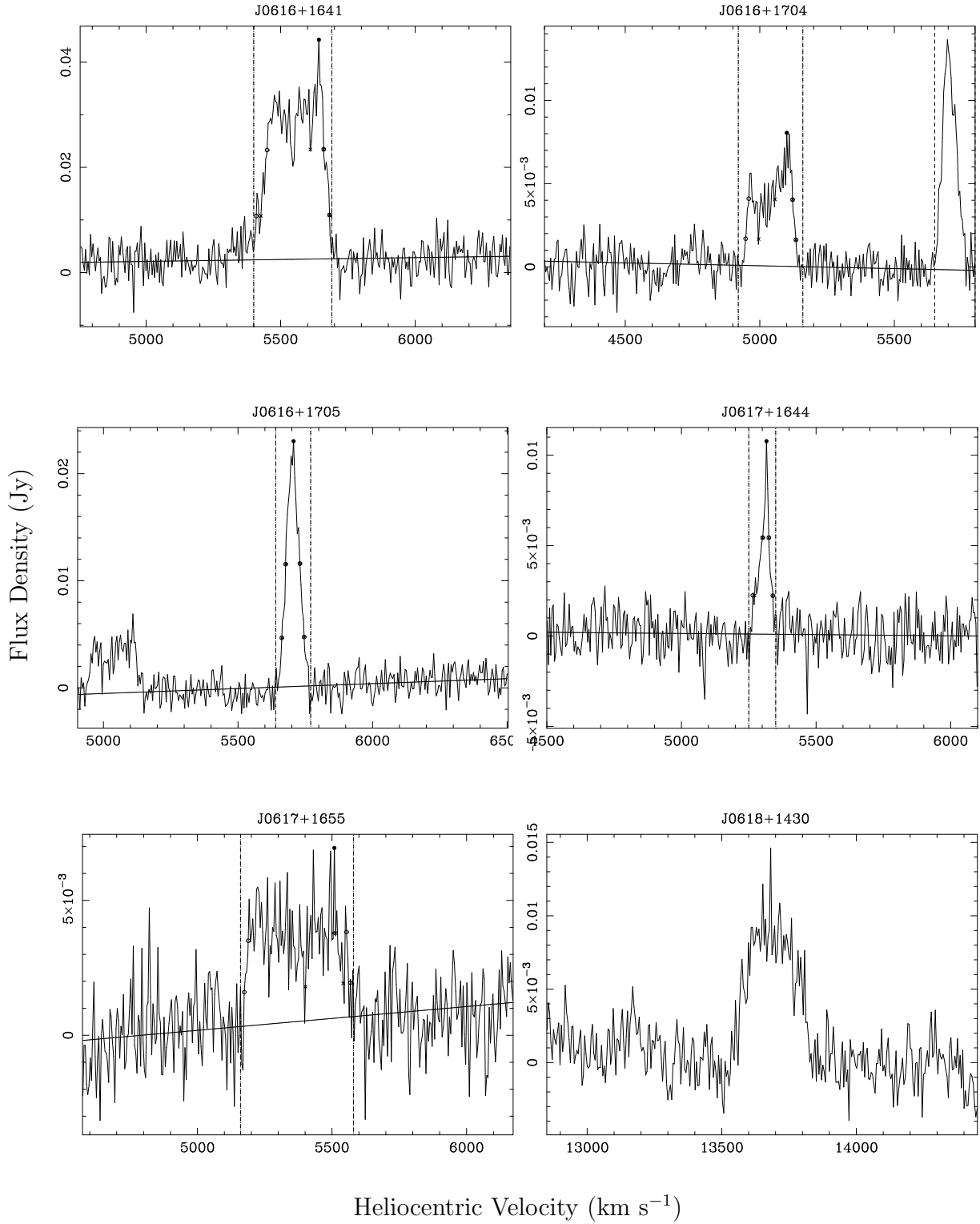

\addtocounter{figure}{-1}
\centering{\includegraphics[scale=0.29, angle=270]{fig5_55.ps}
\includegraphics[scale=0.29, angle=270]{fig5_56.ps}
\\[1cm]
\includegraphics[scale=0.29, angle=270]{fig5_57.ps}
\includegraphics[scale=0.29, angle=270]{fig5_58.ps}
\\[1cm]
\includegraphics[scale=0.29, angle=270]{fig5_59.ps}
\includegraphics[scale=0.29, angle=270]{fig5_60.ps}}
\\[1cm]
\leavevmode\smash{\makebox[0pt]{\hspace{0em}
  \rotatebox[origin=l]{90}{\hspace{22em}
    Flux Density (Jy)}%
}}\hspace{0pt plus 1filll}\null
\leavevmode\smash{\makebox[0pt]{\hspace{0em}
  {\hspace{0em}
    Heliocentric Velocity (km s$^{-1}$)}%
}}\hspace{0pt plus 1filll}\null
\caption{H\,{\sc{i}} profiles of ALFA ZOA detections. The description of the profiles is the same as in Figure \ref{J0617+1648.ps}. (Continued)}
\end{figure}
\clearpage

The uncertainties on $F_{HI}$, $v_{hel}$, $W_{50}$, and $W_{20}$ were calculated in the same way as Koribalski et al. (2004). The error on the flux integral is
\begin{equation}
\sigma(F_{HI}) = 4 (S/N)^{-1} \; (Sp \; F_{HI} \; \delta v)^{1/2},
\end{equation}
where Sp is the peak flux, S/N is the signal-to-noise ratio Sp to $\sigma(Sp)$, $F_{HI}$ is the integrated flux, and $\delta$v is the velocity resolution of the data, 9 km s$^{-1}$. $\sigma(Sp)$ is the error in the peak flux
\begin{equation}
\sigma(Sp)^2 = rms^2 + (0.05 \; Sp)^2.
\end{equation}
The uncertainty in the heliocentric velocity is
\begin{equation}
\sigma(v_{hel}) = 4 \; (S/N)^{-1} \; (P \; \delta v)^{1/2},
\end{equation}
where
\begin{equation}
P = 0.5 \; (W_{20} - W_{50})
\end{equation}
is a measure of the steepness of the profile edges. The uncertainties in the velocity widths are given by
\begin{equation}
\sigma(W_{20}) = 3 \; \sigma(v_{hel}),
\end{equation}
\begin{equation}
\sigma(W_{50}) = 2 \; \sigma(v_{hel}).
\end{equation}
The uncertainties on $D_{LG}$ and $M_{HI}$, whose values rely heavily on cosmological assumptions, are not calculated.

Two sources are located on the edge of the cube, J0555+1531 and J0618+1430, and so their flux could not be completely recovered. Their positions are listed in the catalog followed by a semi-colon in order to indicate that they were not measured through the parameterization process described in the previous section. None of their other parameters are derived and so are indicated with ellipses.

\begin{figure}[h!]
\centering
\includegraphics[scale=.8]{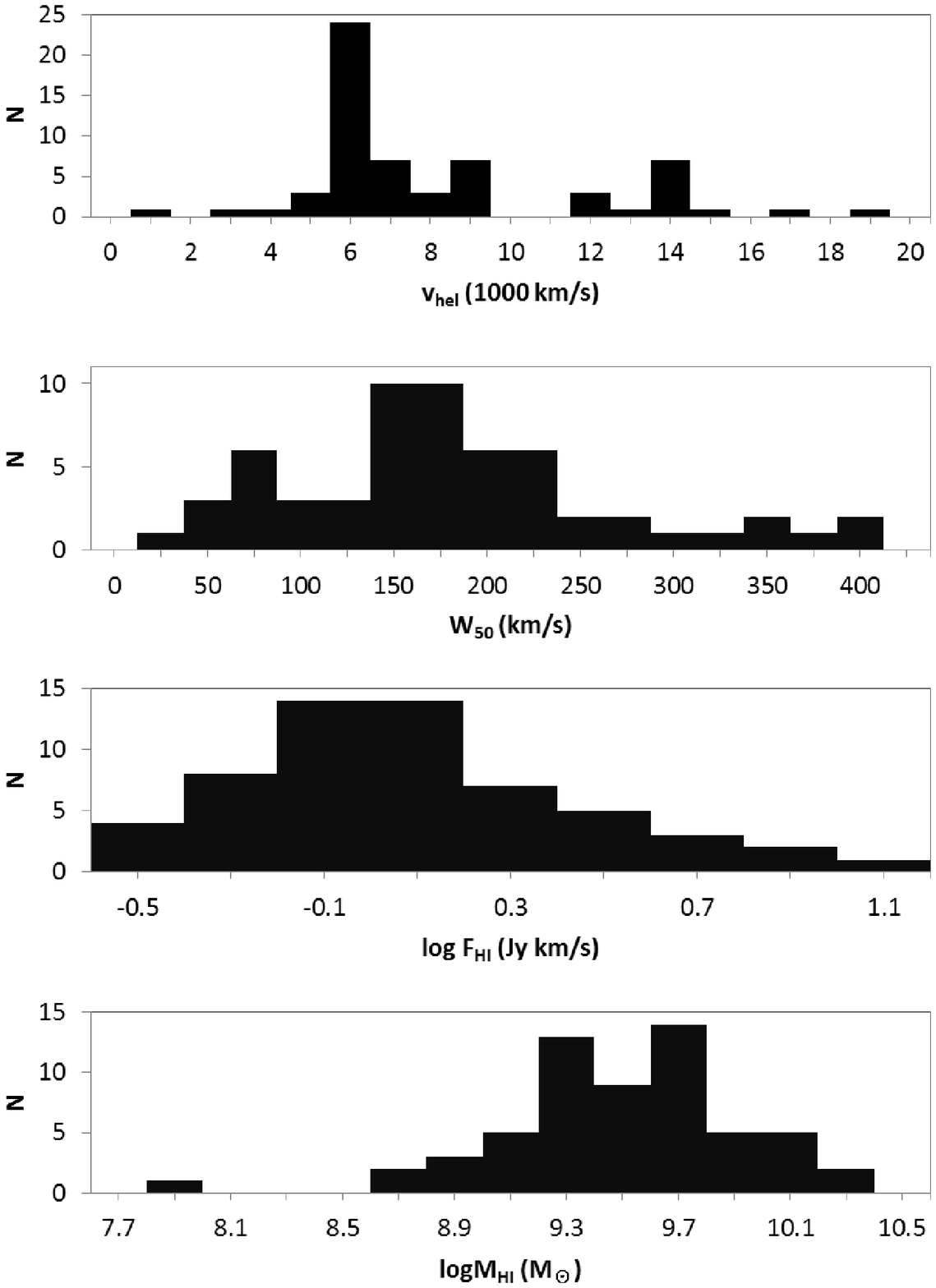}
\caption{\label{deephistograms} \footnotesize Histograms of H\,{\sc{i}} parameters from the ALFA ZOA Deep Survey. From top to bottom: heliocentric velocity, velocity width at 50\% peak flux, integrated flux, logarithm of H\,{\sc{i}} mass.}
\end{figure}

Histograms of parameters from the survey are shown in Figure \ref{deephistograms}. The distribution of heliocentric velocity shows detections out to nearly 19,000 km s$^{-1}$. The velocity widths show detections of dwarf galaxies with $W_{50}$ $\sim$ 30 km s$^{-1}$ as well as large spirals with $W_{50}$ $\sim$ 400 km s$^{-1}$. The distribution of integrated flux ranges from 0.2 Jy km s$^{-1}$ to 150 Jy km s$^{-1}$. The distribution of mass ranges from $M_{HI}$= 10$^{7.8}$ to 10$^{10.4} M_\odot$.

\subsection{\label{Counterparts} Counterparts}
The NASA/IPAC Extragalactic Database (NED) was searched for extragalactic sources within a radius of 2$^{\prime}$ at each galaxy's position in order to find potential counterparts in the literature. NED was also searched for H\,{\sc{i}} extragalactic counterparts within 8$^{\prime}$ and 100 km s$^{-1}$. The counterparts are listed in Table 2, which presents the following information:

	Column (1). ALFA ZOA source name.
	
	Column (2). Galactic longitude, $l$, in degrees.
	
	Column (3). Galactic latitude, $b$, in degrees.
	
	Column (4). Foreground extinction, $A_{B}$, as estimated by Schlafly \& Finkbeiner (2011).
	
	Column (5). Primary name of the counterpart as given by NED.
	
	Column (6). Separation in arcminutes between ALFA ZOA detection and counterpart.
	
	Column (7). Difference in $v_{hel}$ between ALFA ZOA detection and counterpart in km s$^{-1}$.

\begin{figure}[h]
\centering
\includegraphics[scale=0.8]{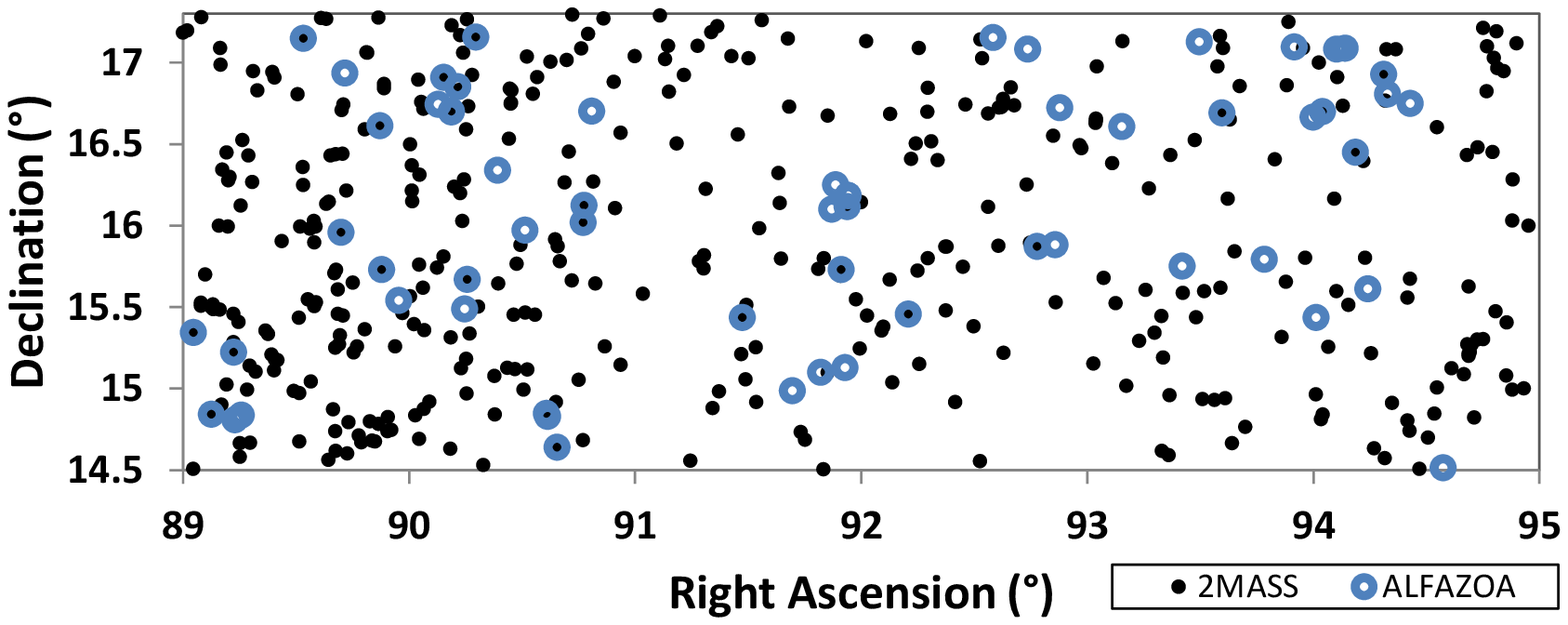}
\caption{\label{deep2mass} \footnotesize Distribution of ALFA ZOA (open blue circles) and 2MASS (black dots) detections within the survey area.}
\end{figure}

\begin{figure}[h]
\centering
\includegraphics[scale=0.8]{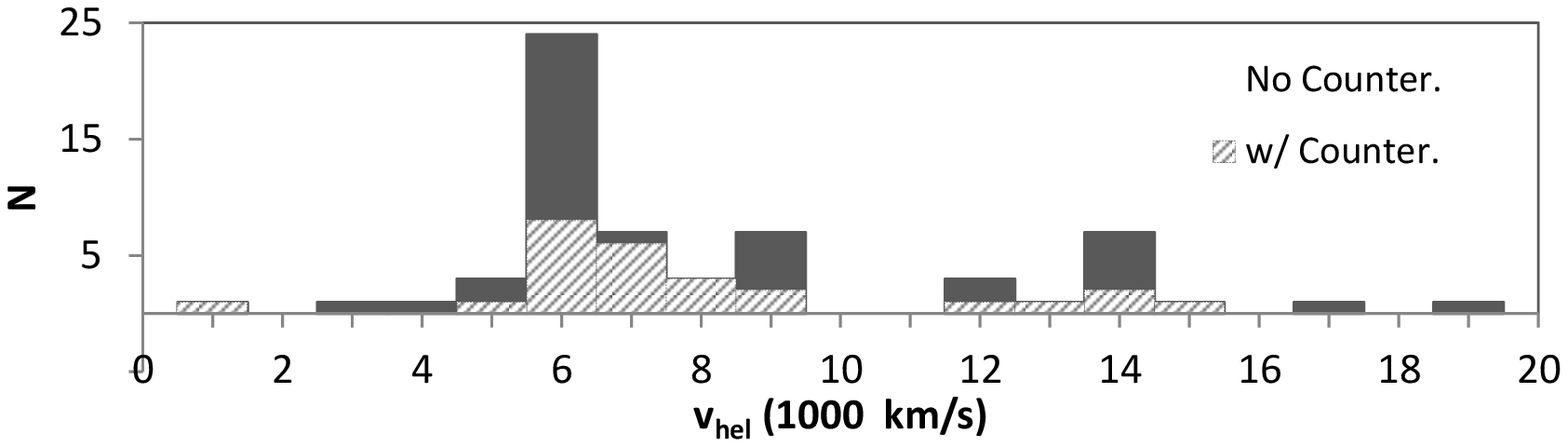}
\caption{\label{deepvelcounters} \footnotesize Histogram of heliocentric velocities for galaxies with no counterpart (solid) and with a counterpart (diagonal).}
\end{figure}

Figure \ref{deep2mass} shows the distribution of ALFA ZOA and 2MASS detections within the data cube. There are 26 detections (43\% of all sources) that have at least one counterpart within 2$^{\prime}$. Two of these have known redshifts. Some positions have more than one counterpart within 2$^{\prime}$, and so there are 30 possible counterparts listed in the table. No attempt is made to distinguish between multiple counterparts for the same position unless the counterpart has a known $v_{hel}$ within 100 km s$^{-1}$ of the ALFA ZOA detection. Every H\,{\sc{i}} source with a counterpart has at least one 2 Micron All Sky Survey (2MASS; Skrutskie et al. 2006) galaxy associated with it. Figure \ref{deepvelcounters} shows the histogram of ALFA ZOA heliocentric velocities. Detections that have a possible counterpart are coded with a diagonal stripe. Counterparts from known overdensities can be seen from the C1 cluster at 2000 km s$^{-1}$, Orion cluster at 6000 km s$^{-1}$, Abel 539 at 8000 km s$^{-1}$, and C21 cluster at 14,000 km s$^{-1}$ (Erdogdu et al. 2006). The percent of detections with a counterpart beyond 10,000 km s$^{-1}$ drops by 20\%.

\subsubsection{Galaxy Classification}

 There is no clear bimodal color distribution for galaxies in NIR like there is in optical (Jarrett 2000) and so 2MASS colors do not reliably allow for the classification of an individual galaxy's morphological type. However, there are parameters other than color indices that indicate morphological type, such as mass and size.

\begin{figure}[h]
\centering
\includegraphics[scale=0.8]{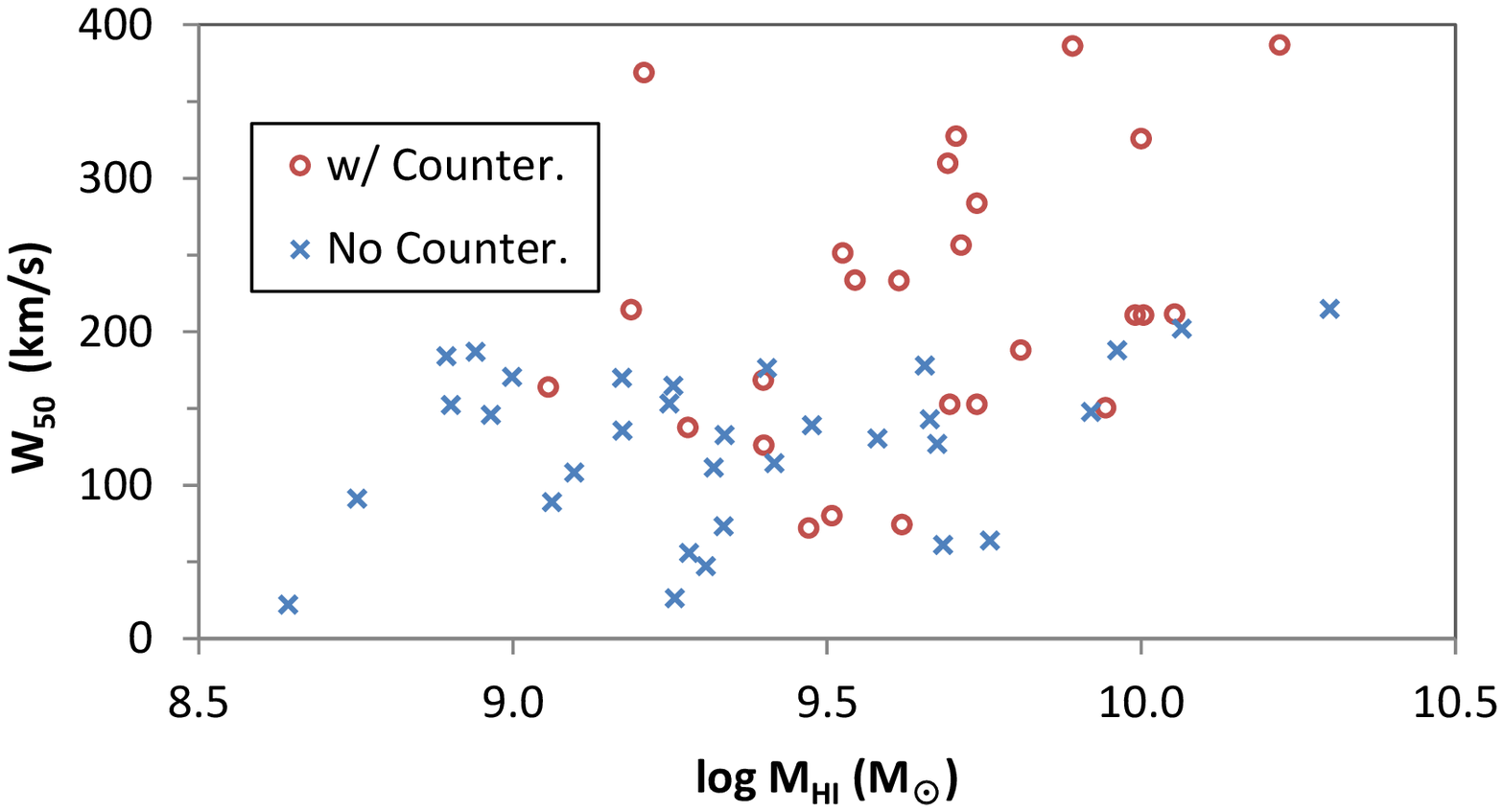}
\caption{\label{deepmorph} \footnotesize $W_{50}$ versus $M_{HI}$ for ALFA ZOA detections, showing both objects with counterparts (red, open circles) and without counterparts (blue, x's).}
\end{figure}

Figure \ref{deepmorph} shows velocity width, $W_{50}$, versus H\,{\sc{i}} mass, $M_{HI}$, for ALFA ZOA detections, marking objects that have a counterpart and those without a counterpart separately. The existence of a counterpart has a strong dependence on velocity width and H\,{\sc{i}} mass.  There are no counterparts with a velocity width below $W_{50}$ = 70 km s$^{-1}$, and every detection with a velocity width above $W_{50}$ = 215 km s$^{-1}$ has a counterpart. The mean velocity width is $W_{50} = 218$ km s$^{-1}$ for detections with a counterpart and 125 km s$^{-1}$ for detections with no counterparts. The mean H\,{\sc{i}} mass is log $<M_{HI}>$ = 9.65 $M_\odot$ for detections with a counterpart versus log $<M_{HI}>$ = 9.31 $M_\odot$ for detections without counterparts, over two times more massive on average. The two subsamples are statistically different in velocity width and H\,{\sc{i}} mass at 4.6$\sigma$- and 3.5$\sigma$-confidence levels, respectively. Nearly every large spiral galaxy (i.e. $W_{50} > 200 $km s$^{-1}$, log $M_{HI} >$ 9.5 $M_\odot$) detected by ALFA ZOA was also detected by 2MASS. The vast majority of smaller spirals and dwarf galaxies (i.e. $W_{50} <$ 200 km s$^{-1}$, log $M_{HI} < 9.5$ $M_\odot$) were not detected by 2MASS.

\section{\label{performance} Survey Performance}

\subsection{Positional Accuracy}

\begin{figure}[h]
\centering
\includegraphics[scale=0.8]{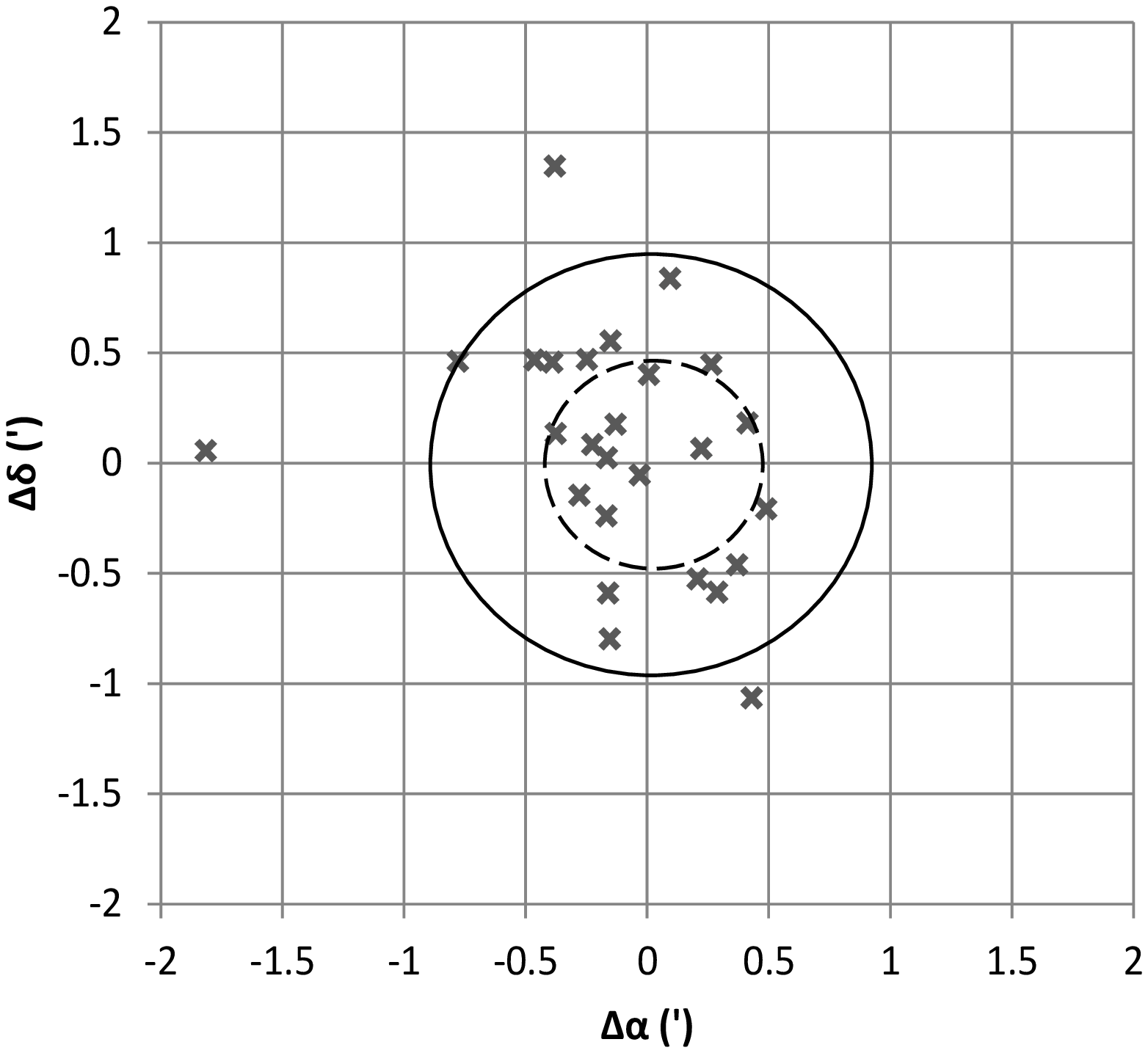}
\caption{\label{deepcountpos} \footnotesize Separations (ALFA ZOA minus literature) in right ascension ($\Delta\alpha$) and declination ($\Delta\delta$) between ALFA ZOA Deep and NED counterparts, in arcminutes. The one (dashed circle) and two (solid circle) standard deviations of the separations are shown.}
\end{figure}

The positional accuracy of ALFA ZOA should be well within the FWHM of the telescope (FWHM = 3.4$^{\prime}$) because the survey is Nyquist sampled. Figure \ref{deepcountpos} shows the positional separation between ALFA ZOA Deep detections and their counterparts. The counterpart with the smallest separation is chosen for this plot when there are multiple possible counterparts for the same source. All of the nearest counterparts are 2MASS galaxies. The positional accuracy of 2MASS is 0.5$^{\prime\prime}$ (Skrutskie et al. 2006), over two orders of magnitude finer than the FWHM of ALFA ZOA, meaning that the distribution of separations should be almost entirely due to the positional uncertainty of the ALFA ZOA Survey. There can be intrinsic offset between a galaxy's neutral hydrogen and stellar structure, though an offset of a kiloparsec only subtends 3$^{\prime\prime}$ at 4000 km s$^{-1}$ (beyond which 25 of 26 counterparts are located).

Assuming that counterparts are real and the separations in Figure \ref{deepcountpos} are due to uncertainty in ALFA ZOA positions, then their standard deviation is an estimate of the positional accuracy of the survey. The standard deviation of separations is $\sigma$ = 0.47$^{\prime}$, and this value is adopted as the survey's positional accuracy. The one (dashed circle) and two (solid circle) $\sigma$ boundaries are overplotted in Figure \ref{deepcountpos}. The distribution of positions shows an average offset in right ascension of $<\Delta\alpha> = 0.12^\prime$ to the east. This offset is not considered to be statistically significant as it is found to be at only a 1.3$\sigma$-confidence level. There is no indication that positions are systematically offset in declination.

\subsection{Sensitivity}

\subsubsection{Noise}

\begin{figure}[h]
\centering
\includegraphics[scale=.4]{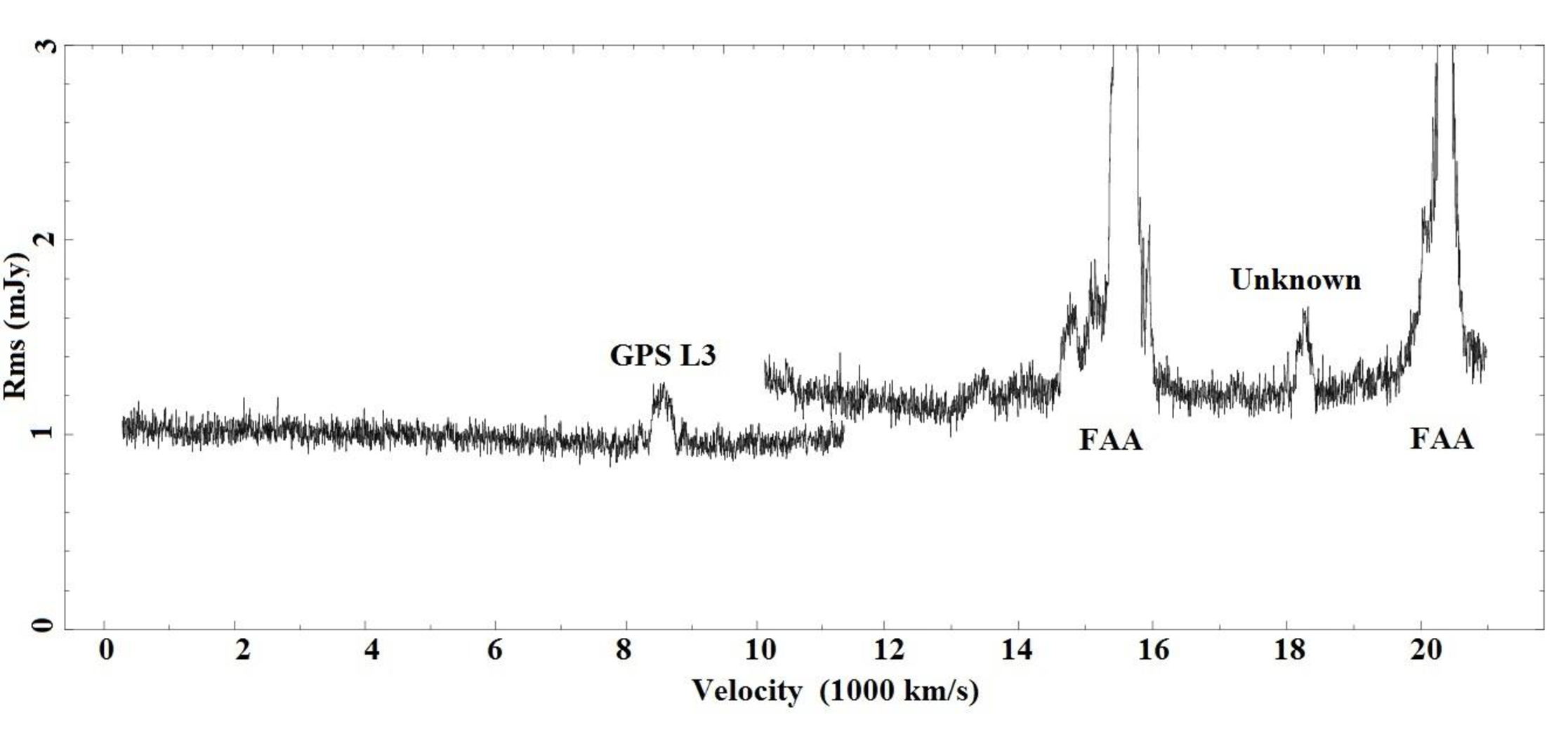}
\caption{\label{cubenoise} \footnotesize Noise in mJy per heliocentric velocity averaged over the inner quarter of the image plane for both the high- and low-frequency sub-bands. Major RFI is labeled.}
\end{figure}

The noise level of the ALFA ZOA Deep data cube reached its expected rms value of 1 mJy (at 9 km s$^{-1}$ velocity resolution). Figure \ref{cubenoise} shows the cube's rms as a function of heliocentric velocity. The rms was averaged over an inner quarter of the image plane, chosen in an area with a relatively low detection rate in order not to pollute the rms map with source flux. The mean noise is rms = 1 mJy for the high frequency sub-band (i.e. $v_{hel} <$ 11,500 km s$^{-1}$), but it increases to rms = 1.2 mJy for the low frequency sub-band (i.e. $v_{hel} >$ 10,000 km s$^{-1}$). The overlap between the two sub-bands can be seen between 10,000 km s$^{-1}$ and 11,500 km s$^{-1}$. It is clear that the radio frequency interference (RFI) prevalent outside of the protected frequencies (i.e. 1400 - 1427 MHz) raises the average system temperature non-negligibly inside the low frequency sub-band.

Known RFI is labeled in Figure \ref{cubenoise} using information provided by Arecibo Observatory. The GPS L3 satellite produces RFI at 1381 MHz that spans velocities from 8400 - 8800 km s$^{-1}$. FAA radar from the airport at Punta Borinquen near Aguadilla and east of San Juan at Pico del Este produces RFI at 1350 MHz and 1330 MHz, covering velocities ranging from 14,600 - 16,000 km s$^{-1}$ and 19,700 - 20,700 km s$^{-1}$, respectively. There is also RFI at 1339 MHz, spanning velocities from 18,100 - 18,400 km s$^{-1}$. There is no RFI source currently known to the Observatory at 1339 MHz.

\subsubsection{\label{detectability} H\,{\sc{i}} Detection Limit}

\begin{figure}[h]
\centering
\includegraphics[scale=0.65]{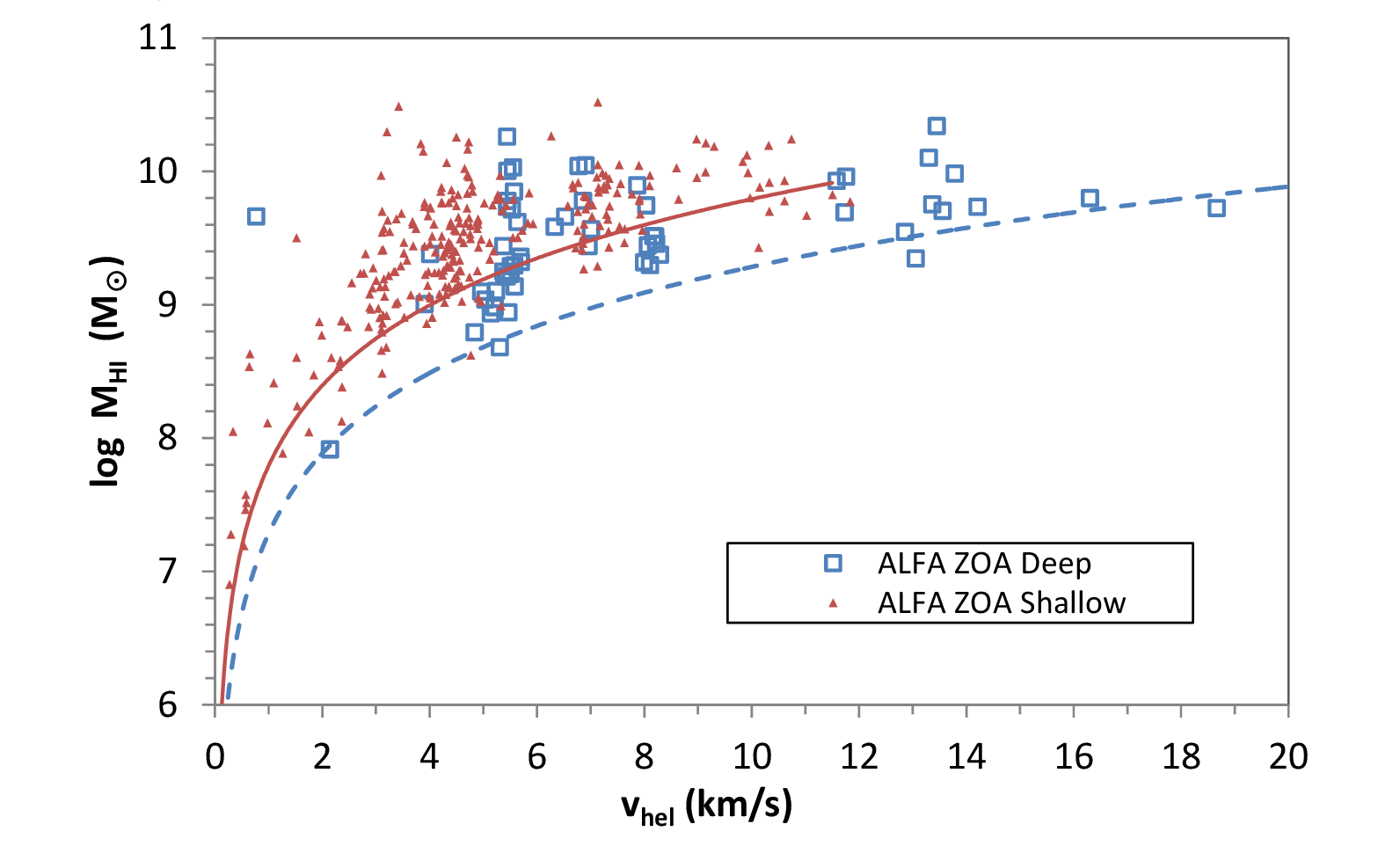}
\caption{\label{deepdetectability} \footnotesize H\,{\sc{i}} mass as a function of heliocentric velocity for both the ALFA ZOA Deep (open blue boxes) and Shallow (closed red triangles) Survey detections. Predicted H\,{\sc{i}} mass detection limit is shown for the Deep (dashed blue line) and Shallow (solid red line) Surveys.}
\end{figure}

Figure \ref{deepdetectability} shows a plot of H\,{\sc{i}} mass as a function of velocity for both the Deep (open boxes) and Shallow (closed triangles; rms = 5.4 mJy; Henning et al. 2010; McIntyre et al. in prep.) Surveys. The sensitivity and depth of the Deep Survey can clearly be seen. Gaps in detections at specific redshifts are due to the RFI discussed above as well as large scale structure, which will be discussed in the next section.

The detection limit of the survey is defined as the farthest distance that a specific H\,{\sc{i}} mass can be detected above a signal-to-noise ratio, S/N = 6.5. A fit of the detection limit of each survey is shown in Figure \ref{deepdetectability} using criteria devised by Giovanelli et al. (2007) for a bivariate signal-to-noise ratio,
\begin{equation}
\label{bivariateSN}
 S/N = \frac{F_{HI}}{rms \; (2\; \delta v \; W_{50})^{1/2}},
\end{equation}
where $F_{HI}$ is the integrated flux in Jy km s$^{-1}$, $W_{50}$ is the half peak velocity width in km s$^{-1}$, rms is the noise in Jy, and $\delta v$ = 9 km s$^{-1}$ is the velocity resolution of the survey. Though sources can be detected below the detection limit as seen in Figure \ref{deepdetectability}, Giovanelli et al. showed empirically that sources above S/N = 6.5 have a 95\% reliability rate, and this estimate is adopted for this paper. Solving for $F_{HI}$ and plugging into equation (\ref{MHI}) gives an expression for the H\,{\sc{i}} mass detection limit of the survey,
\begin{equation}
 M_{HI} = 2.36\times 10^5 \; r^2 \; S/N \; rms \; (2\; \delta v \; W_{50})^{1/2},
\end{equation}
where r is the distance to a detection in Mpc, a value of $W_{50}$ = 200 km s$^{-1}$ is chosen for the plot in Figure \ref{deepdetectability}, and the distance is converted from Mpc to heliocentric velocity using H$_{0}$ = 70 km s$^{-1}$. The typical noise of the survey is rms = 5.4 mJy, 1 mJy for the shallow, deep survey respectively. The superiority of the deep survey at detecting lower mass sources out to heliocentric velocities of 20,000 km s$^{-1}$ can clearly be seen.

\subsection{Completeness}

The completeness limit of the survey is the lowest integrated flux, $F_{HI \; lim}$, above which every galaxy can be detected. One technique for measuring completeness is by fitting a power law with a slope of -3/2 to the histogram of flux. The completeness limit is reached where the histogram begins to deviate from the slope. However, this method only works if galaxies are homogeneous and isotropic, which is not the case for the small area surveyed in our First Results Catalog. An alternative method for determining the survey's completeness is the Test for Completeness, T$_c$, a statistical test derived by Rauzy (2001) for a magnitude-redshift sample, independent of large scale structure.  The method takes a sample of galaxies within a given volume and brighter than a given flux and compares the number of galaxies that are fainter and brighter than every galaxy in the sample. T$_c$ is calculated for H\,{\sc{i}} as in Zwaan et al. (2004).

\begin{figure}[h]
\centering
\includegraphics[scale=.7]{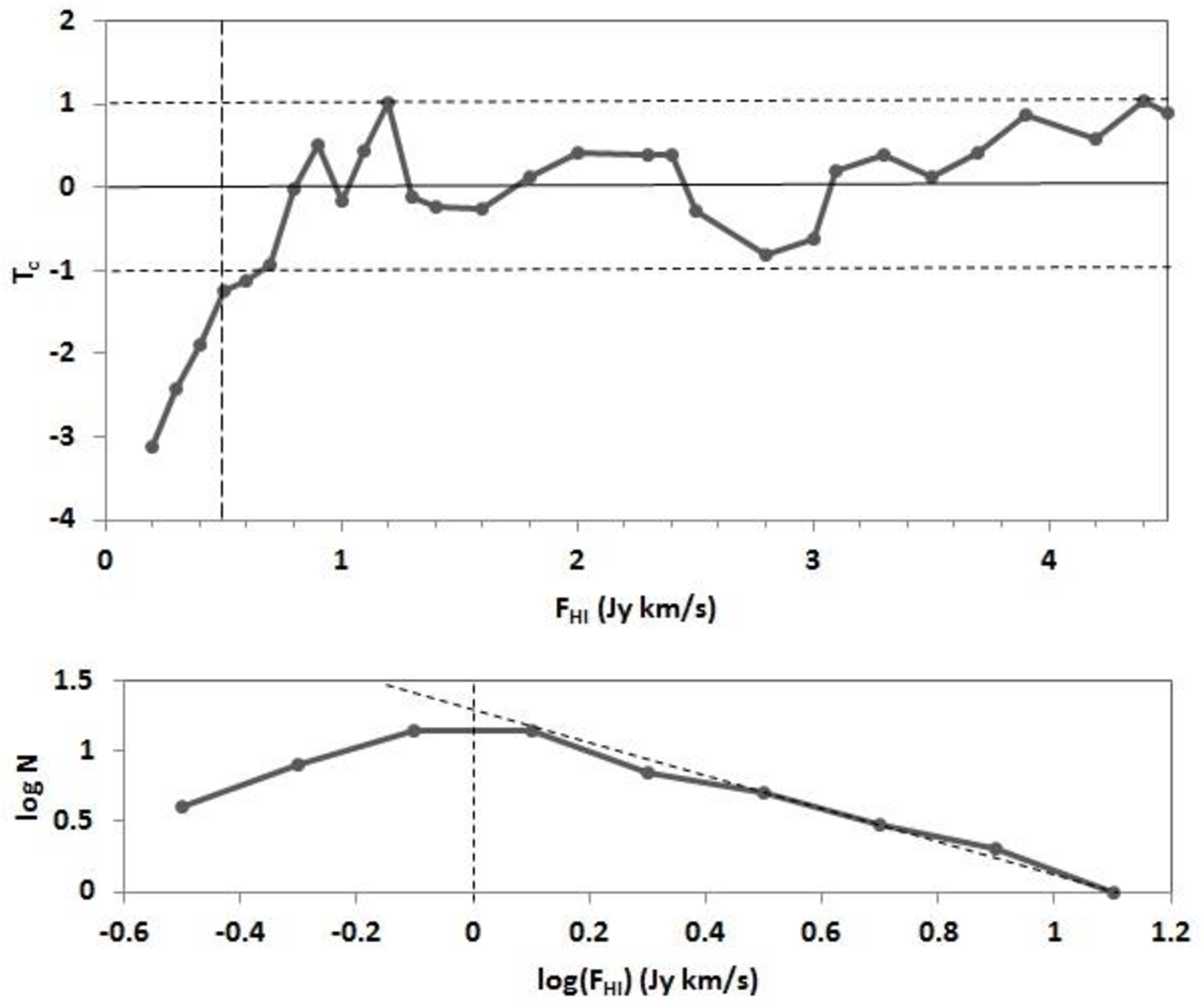}
\caption{\label{deepcomplete} \footnotesize Top Panel. Test of completeness for ALFA ZOA. Horizontal dashed lines indicate unit variance. The vertical dashed line shows the adopted completeness limit. Bottom Panel. Log-log histogram of integrated flux. The -3/2 slope (diagonal dashed line) falls off from the fit somewhere between -0.1 and 0.1, indicating that the completeness is reached somewhere around 1 Jy km s$^{-1}$ (vertical dashed line).}
\end{figure}

T$_c$ should follow a Gaussian distribution with an average value of 0 and unit variance for galaxy samples that are above the completeness limit, and move systematically to negative values for samples that are not complete. A plot of T$_c$ as a function of integrated flux can be seen in Figure \ref{deepcomplete}. T$_c$ begins to fall below -1 at $F_{HI \; lim} <$ 0.7 Jy km s$^{-1}$. It hovers near -1 until $F_{HI \; lim} <$ 0.4 Jy km s$^{-1}$, where it systematically drops below -2. This indicates that the survey is not complete below 0.4 Jy km s$^{-1}$ with about a 98\% confidence level. The completeness limit adopted for the ALFA ZOA Deep Survey is $F_{HI \; lim}$ = 0.5 Jy km s$^{-1}$, where T$_c$ = -1.25 (i.e. the lowest value before the clear systematic drop in T$_c$).

\begin{figure}[h]
\centering
\includegraphics[scale=0.8]{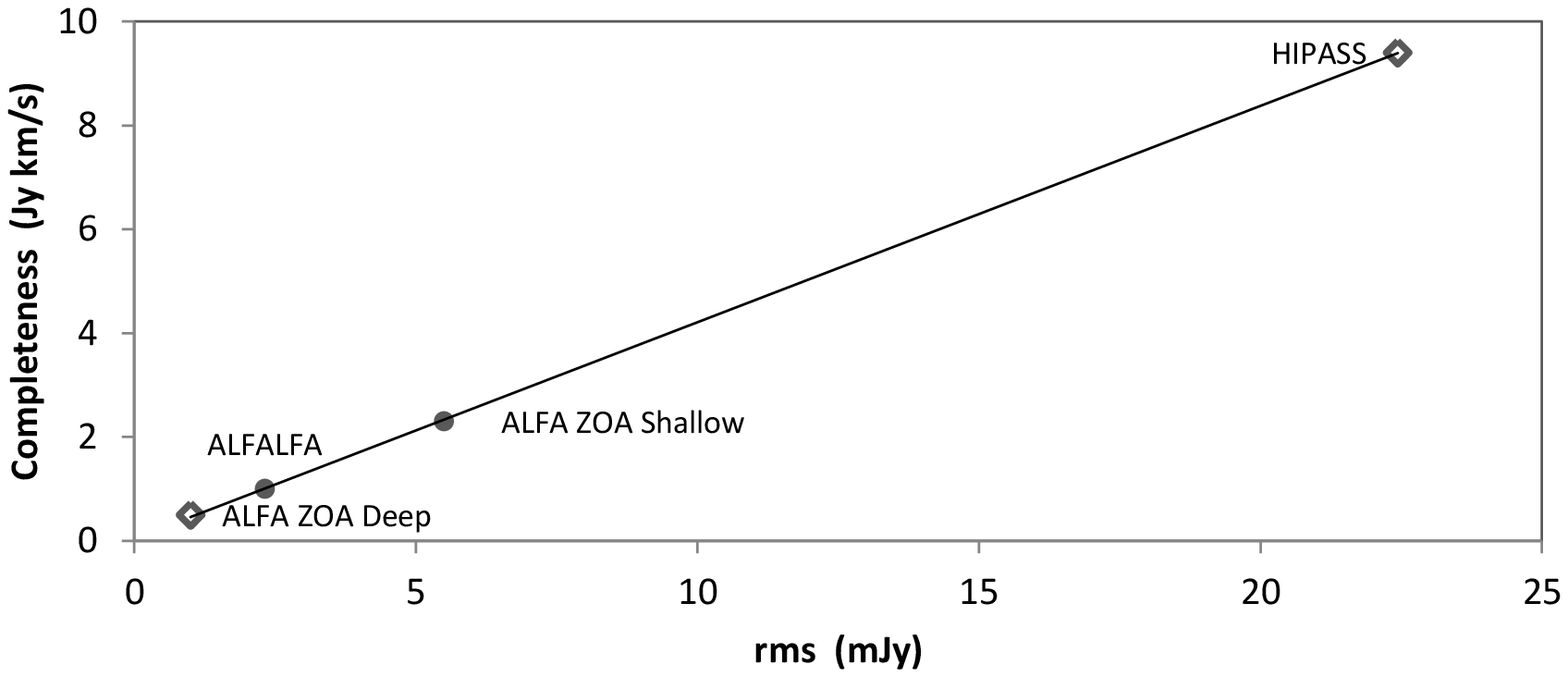}
\caption{\label{completecompare} \footnotesize Integrated flux completeness limit versus rms for major surveys. The open diamonds indicate surveys that used the T$_c$ method and the closed circles are for surveys that fit a -3/2 power law for calculating completeness limit.}
\end{figure}

The bottom panel of Figure \ref{deepcomplete} shows an attempt to measure the completeness from where the flux histogram deviates from a -3/2 power law. The value for completeness using this method is $F_{HI \; lim}$ = 1 Jy km s$^{-1}$, significantly higher than the T$_c$  method. As mentioned above, the power law technique should fail if the distribution of galaxies in the survey area is not homogeneous, which is the case for our small First Results area dominated by large-scale structure.  Further, a completeness limit of $F_{HI \; lim}$ = 0.5 Jy km s$^{-1}$ makes sense compared to recent H\,{\sc{i}} surveys, given differing rms sensitivities, as illustrated in Figure \ref{completecompare}.

For normalized comparison, the noise for ALFALFA (Saintonge 2007) and HIPASS (Zwaan et al. 2004) is scaled to a velocity resolution of 9 km s$^{-1}$. Figure \ref{completecompare} shows a remarkably well fit, linear relationship between H\,{\sc{i}} survey noise and integrated flux completeness limit, indicating that the T$_c$ method is giving an expected value compared to how other surveys are performing. It should be noted that ALFALFA (Haynes et al. 2011) published bivariate completeness as a function of integrated flux, $F_{HI}$, and galaxy velocity width, $W_{50}$, and so the completeness limit for their average $W_{50}$ is used for Figure \ref{completecompare}.

\begin{figure}[h]
\centering
\includegraphics[scale=0.5]{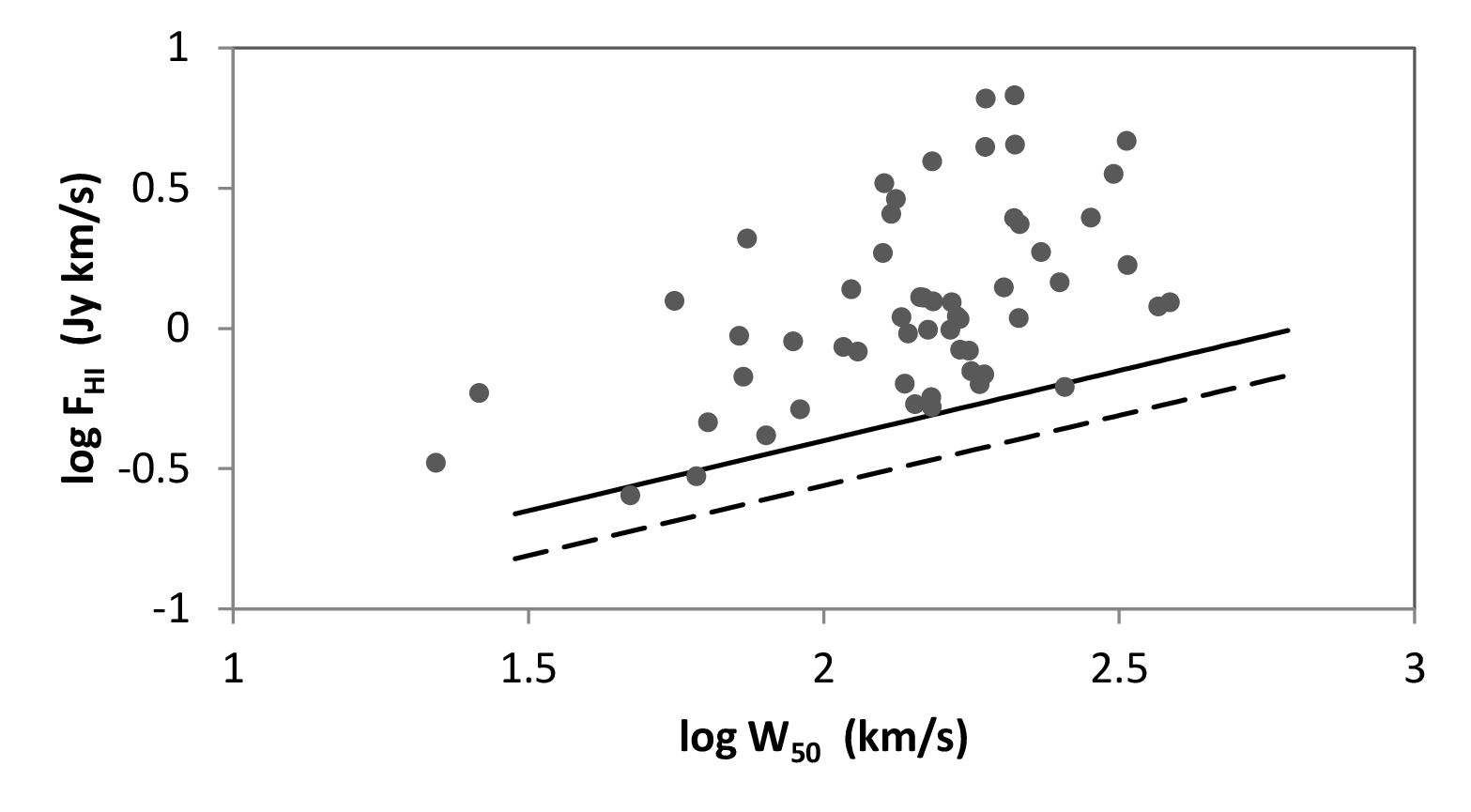}
\caption{\label{deepbivariatecompl} \footnotesize Log-log plot of integrated flux versus velocity width. The detection limit of the survey (dashed line) and the bivariate completeness limit (solid line) are shown.}
\end{figure}

Figure \ref{deepbivariatecompl} shows a log-log plot of integrated flux, $F_{HI}$, versus velocity width, $W_{50}$ for ALFA ZOA. Similar to the bivariate signal-to-noise ratio used to estimate the detection limit of the survey, discussed in Section \ref{detectability}, the relationship between $F_{HI}$ and $W_{50}$ can be used to estimate the bivariate completeness limit, $F_{HI}(W_{50})_{ \; lim}$, above which every galaxy can be detected. $F_{HI}(W_{50})_{ \; lim}$ is estimated by scaling equation (\ref{bivariateSN}) so that $F_{HI \; lim}$ = 0.5 Jy km s$^{-1}$ at the mean $W_{50}$. The mean velocity width is $W_{50}$ = 167 km s$^{-1}$, so the bivariate completeness limit can be estimated as,
\begin{equation}
 log[F_{HI}(W_{50})_{ \; lim}] = 0.5 \; log(W_{50 \; lim}) - 1.4.
\end{equation}
Both the bivariate detection limit (dashed line) and bivariate completeness limit (solid line) are shown in Figure \ref{deepbivariatecompl}. It should be noted that, while three ALFA ZOA Deep detections are located below the detection limit line in Figure \ref{deepdetectability} and none are below the line here, that is because an estimate of $W_{50}$ = 200 km s$^{-1}$ was assumed to make the plot in Figure \ref{deepdetectability}.

\subsection{Zone of Avoidance}

\begin{figure}[h]
\centering
\includegraphics[scale=0.8]{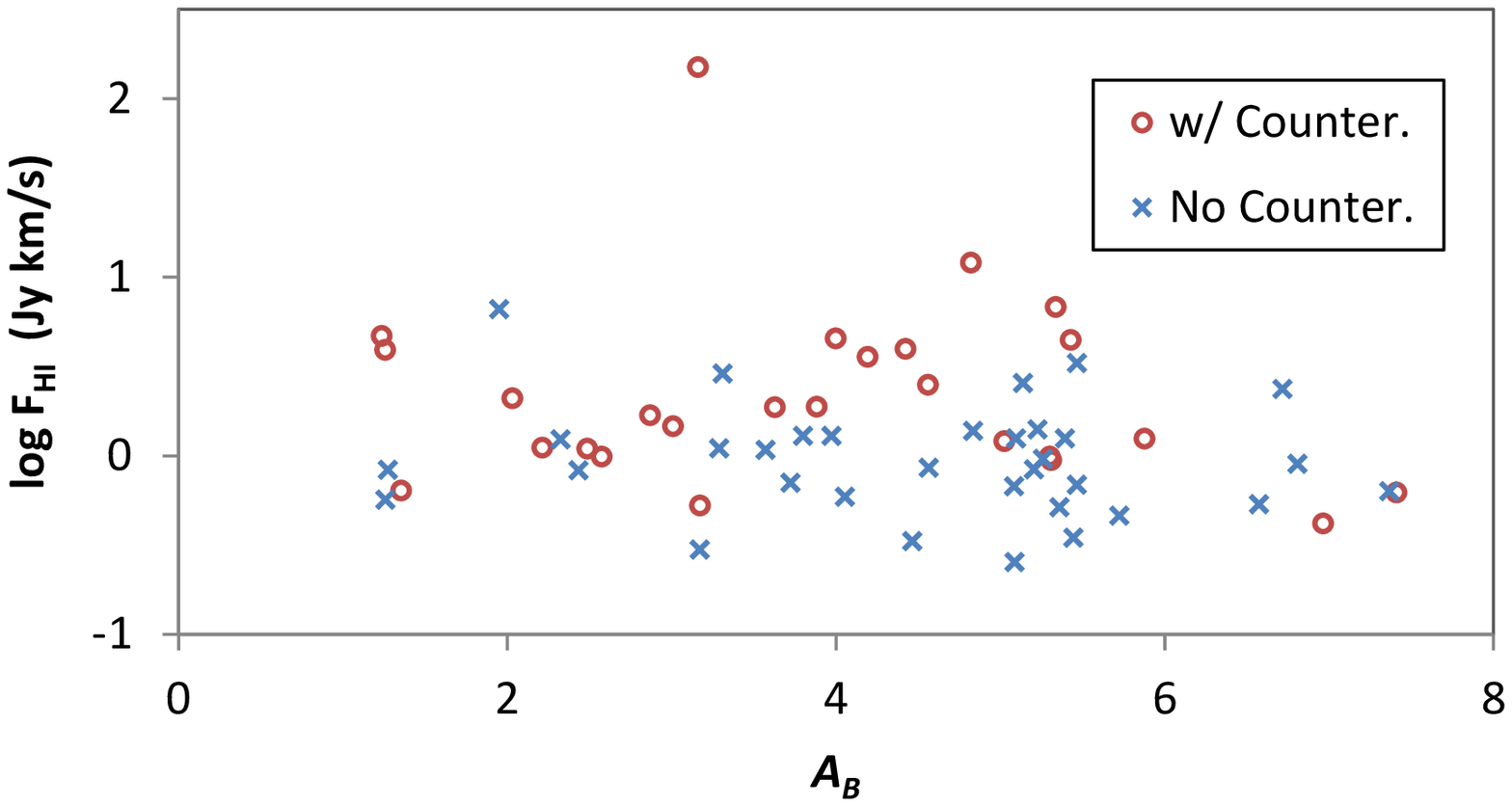}
\caption{\label{deepextinct} \footnotesize Integrated flux as a function of extinction, $A_B$. Detections with a counterpart (open red circles) and with no counterpart (blue x's) are shown.}
\end{figure}

Figure \ref{deepextinct} shows a plot of integrated flux, $F_{HI}$, as a function of foreground extinction, $A_B$. Extinction estimates were taken at the position of each ALFA ZOA detection using NED as described in Section \ref{Counterparts}. It should be noted that estimates of extinction at $|b| < 5^\circ$ are particularly uncertain (Schlegel, Finkbeiner \& Davis 1998). This plot does not show a relationship between $F_{HI}$ and extinction, indicating that our survey does not have a Zone of Avoidance, as expected.

Sources with higher $F_{HI}$ are more likely to have a counterpart, and counterparts are less likely at higher $A_B$. This is expected, as the purpose of ALFA ZOA is to detect galaxies in areas with high extinction, where surveys at higher frequencies cannot. 55\% of detections have a counterpart below $A_B$ = 3.5 magnitudes while only 37\% have counterparts above $A_B$ = 3.5 mag.

\begin{figure}[h]
\centering
\includegraphics[scale=0.7]{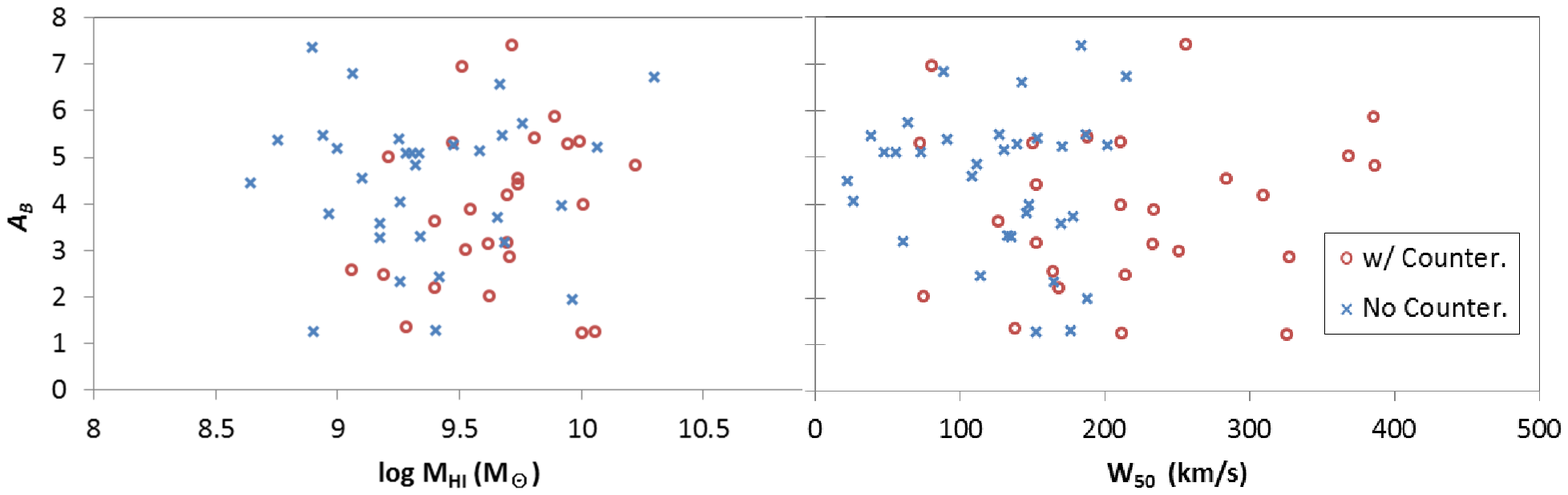}
\caption{\label{deepextinction} \footnotesize Left panel. Extinction versus log$M_{HI}$, color coded for those with counterparts (open red circles) and those without (blue x's). Right panel. Extinction versus velocity width, $W_{50}$, same color scheme as left panel.}
\end{figure}

Because the number of detections is not very high, it is possible that the detections at high extinction happen to come from intrinsically low mass or low velocity width galaxies that are less sensitive to NIR observations. Figure \ref{deepextinction} shows a plot of extinction, $A_B$, as a function of $M_{HI}$ and $W_{50}$, color coded for those with a counterpart (open red circles) and those without (blue x's). There is a near uniform distribution of $M_{HI}$ and $W_{50}$ at all extinctions, indicating there is nothing intrinsic about the galaxies being detected that causes a lack of counterparts at high extinction. The even distribution of $W_{50}$ and $M_{HI}$ across extinction is another indicator of the efficacy of ALFA ZOA at penetrating the Zone of Avoidance. For instance, ALFA ZOA J0610+1709 has the highest $M_{HI}$, and 15th highest $F_{HI}$ in the catalog, with a large velocity width $W_{50}$ = 215 km s$^{-1}$, but it has no counterpart in the literature, even though its measured parameters indicate a massive spiral that should be easily detected by a deep NIR survey. The most likely reason why it has gone undiscovered is because it is located in an area of high extinction, $A_B$ = 6.7 ($A_J$ = 1.3) or high stellar crowding, both of which affect low-latitude optical and NIR surveys but are of no concern to 21-cm surveys.

\section{\label{LSS} Large Scale Structure}

\begin{figure}[h]
\centering
\includegraphics[scale=.6]{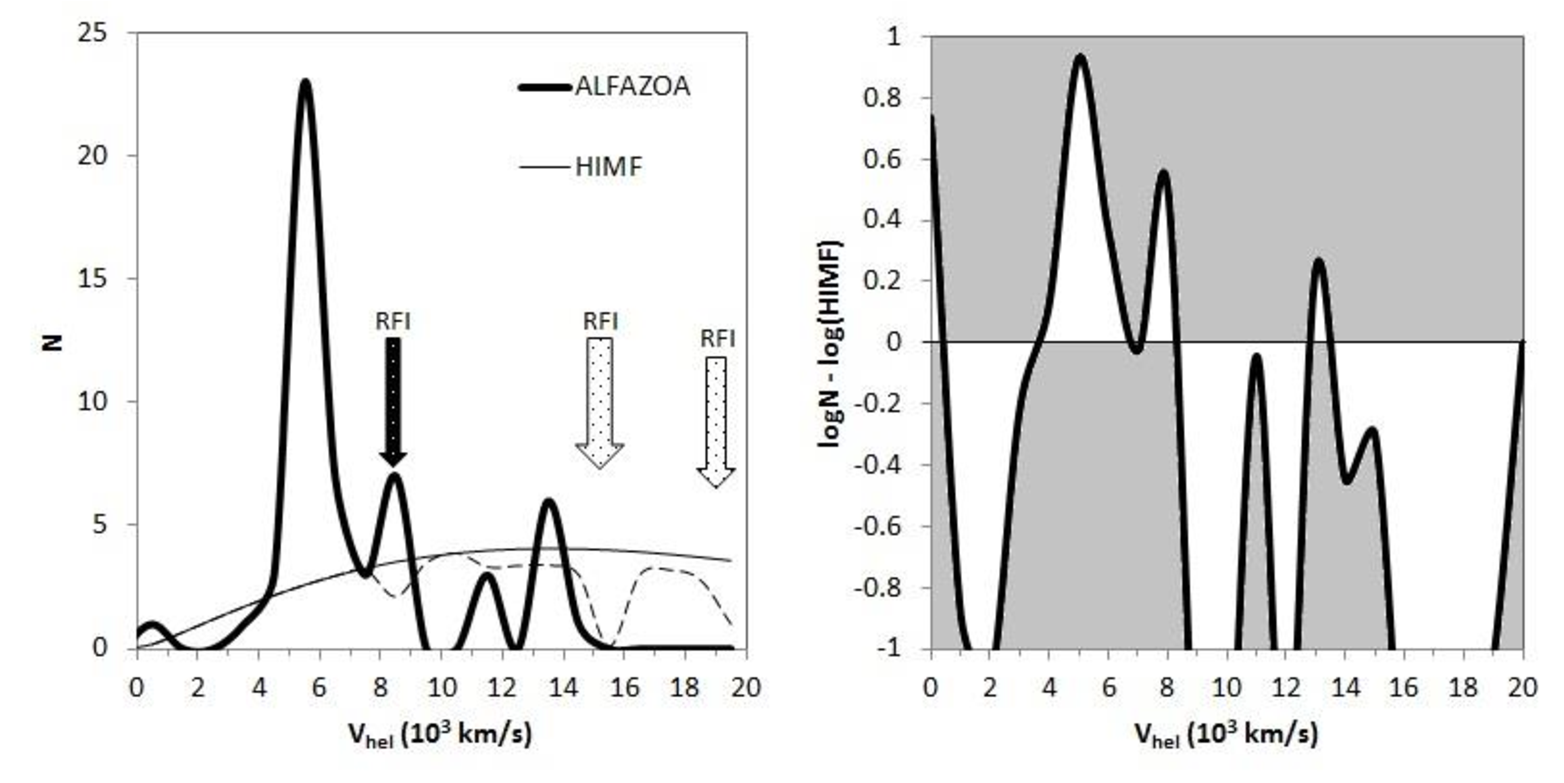}
\caption{\label{deepLSScompl} \footnotesize Left panel. Distribution of ALFA ZOA detections (thick solid line) that are above the completeness limit, as a function of redshift. The expected distribution (thin solid line) from integrating the HIMF (Martin et al. 2010) is shown along with a more robust expectation (dashed line) from taking into account RFI and rms = 1.2 mJy for the low sub-band. Velocities with significant RFI are labeled; the arrow with white polkadots indicates the redshift and bandwidth of the GPS L3 Satellite, black polkadots are for FAA radar. Right panel. The log difference between ALFA ZOA and the robust expected distribution. Observed large scale structure is shaded white. All lines are smoothed for visualization purposes using a cubic Bezier curve.}
\end{figure}

While it is not quite a ``pencil beam'' survey the first results area is narrow enough that it does not fairly probe large scale structure (LSS) in angular dimensions. Therefore, examining a histogram of redshifts is a useful way to explore large scale structure in redshift space. Figure \ref{deepLSScompl} shows detections per velocity above the completeness limit (thick line) along with the expected number of detections per velocity (thin line). The expected number of detections is calculated by integrating the HIMF at the completeness limit using values from Martin et al. (2010). As we are expecting to detect every galaxy above the completeness limit, the difference between the two lines is a measure of the over- and underdensity of the large scale structure cut across by the survey. Redshifts where significant RFI occurs are labeled. RFI can limit the sensitivity of the survey over the volume affected by its bandwidth. A more robust estimate of the expected distribution of detections for ALFA ZOA is shown (dashed line) by taking into account the effect of RFI on survey volume as well as the increase in noise in the low sub-band (i.e rms = 1.2 mJy for $v_{hel} >$ 11,500 km s$^{-1}$). The lines in Figure \ref{deepLSScompl} are smoothed using a cubic Bezier curve for the sake of visualization. The smoothing is not used for analysis.

\subsection{Comparison to Predicted Large Scale Structure}

\begin{figure}[h!]
\centering
\includegraphics[scale=.9]{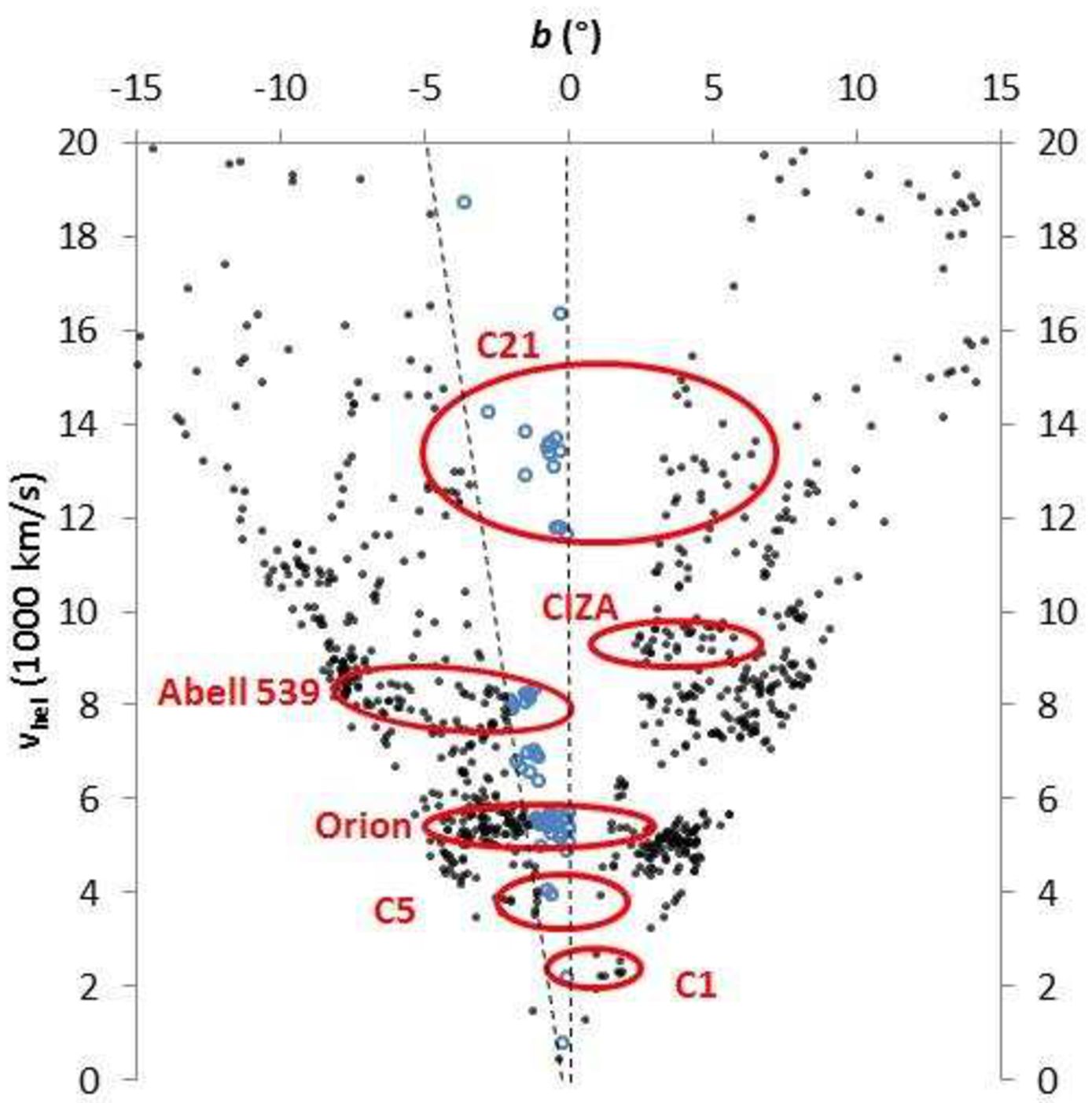}
\caption{\label{deepLSSwedge} \footnotesize 2MRS galaxies (black dots) and ALFA ZOA detections (blue open circles). Overdensities predicted from 2MRS (Erdogdu et al. 2006) are circled in red and the Erdogdu et al. naming convention is preserved. The field of view of the first results survey is shown (dotted line). In the plot, CIZA is shorthand for CIZA J0603.8+2939.}
\end{figure}

Erdogdu et al. (2006) created density reconstruction maps out to 16,000 km s$^{-1}$ from the Two Micron Redshift Survey (2MRS; Huchra et al. 2012). The Zone of Avoidance for 2MRS is $|b| < 5^\circ$ in the outer Galaxy, and so Erdogdu et al. were forced to predict the extent of large scale structure across the ZOA. Comparing ALFA ZOA detections with the Erdogdu density maps is a good check for the effectiveness of predicting LSS in the ZOA. Figure \ref{deepLSSwedge} shows a wedge plot of ALFA ZOA detections (open circles) alongside 2MRS detections (black dots) and indicates major overdensities from the density reconstructions of Erdogdu et al. inside the red ovals. ALFA ZOA confirms the continuation across the ZOA of much of the structure predicted by Erdogdu et al., and contradicts some of the predictions as well.

\begin{figure}[h]
\centering
\rotate
\includegraphics[scale=0.45]{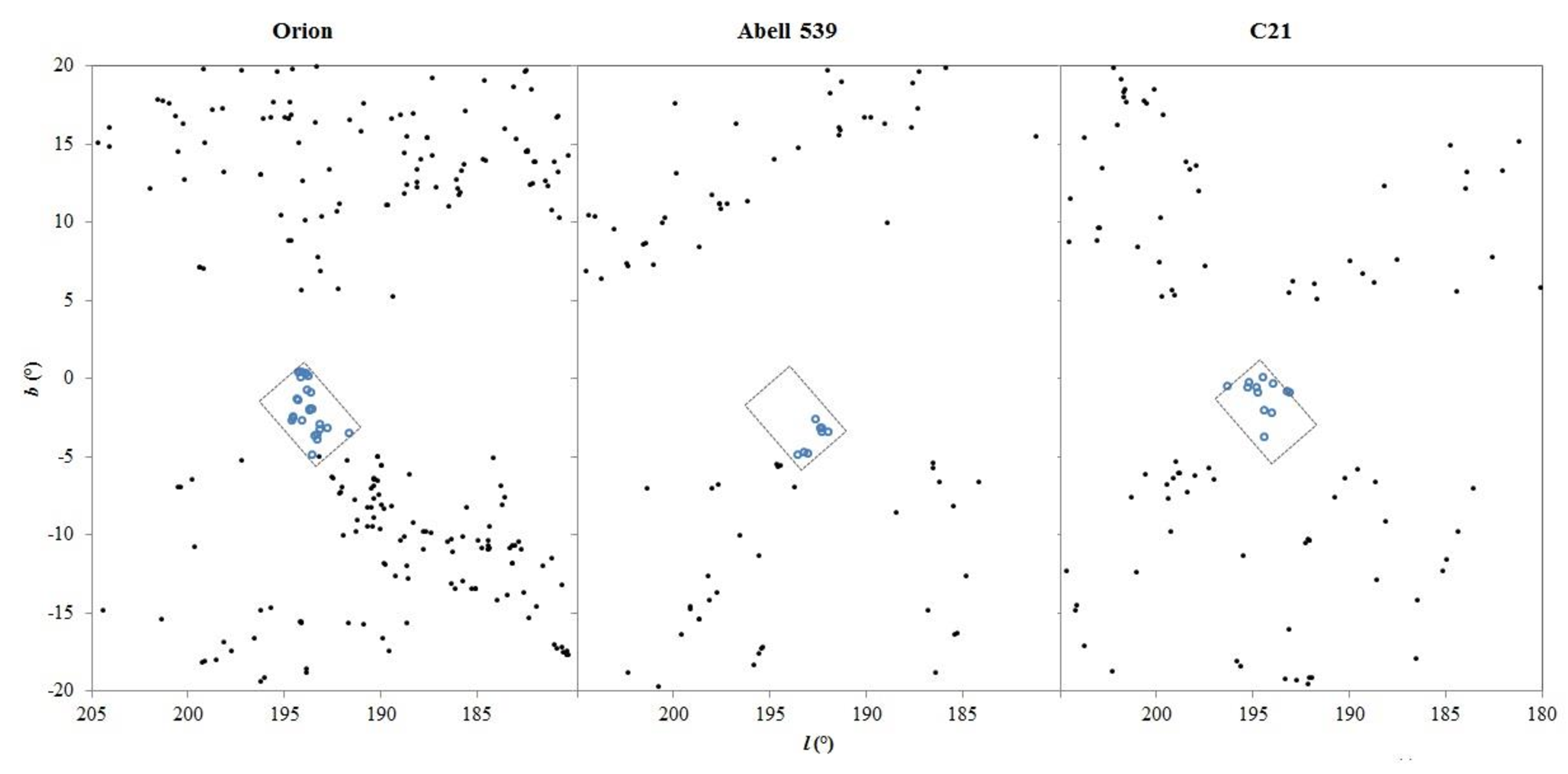}
\caption{\label{deepLSSslice} \footnotesize Sky distribution plots showing 2MRS (black dots) and ALFA ZOA (open blue circles) detections for the three major overdensities detected by ALFA ZOA. The overdensities are labeled above each plot and the field of view of the first results survey is shown (dotted line). Orion covers the velocity range 5000 to 6000 km s$^{-1}$. Abell 549 ranges from 7500 to 8500 km s$^{-1}$. C21 ranges from 12,000 to 15,000 km s$^{-1}$.}
\end{figure}

Referring to Figures \ref{deepLSScompl} and \ref{deepLSSwedge}, the prediction of the overdensity C1 and its extension through C5 from 2000 km s$^{-1}$ to 4000 km s$^{-1}$ is neither confirmed nor unconfirmed, though galaxies at the location of C1 and C5 are detected. The Orion overdensity at 6000 km s$^{-1}$ is clearly confirmed to extend into the ZOA, as is Abell 539 at 8000 km s$^{-1}$. Abell 539 is most likely an even greater overdensity than detected here because of RFI from the GPS L3 satellite affecting velocities 8400 km s$^{-1}$ - 8800 km s$^{-1}$. Galaxies between Orion and Abell 539 can be seen around $v_{hel}$ $\sim$ 7000 km s$^{-1}$, though whether there is a connection between Orion and Abell 539 in redshift space cannot be confirmed. CIZA J0603.8+2939 is confirmed not to extend below $b = 0^\circ$. A significant underdensity is confirmed between 9000 and 12,000 km s$^{-1}$, though this underdensity appears to be greater and extend farther in redshift than predicted. Large scale structure from C21 is confirmed as an overdensity starting at 13,500 km s$^{-1}$ but its extent through 16,000 km s$^{-1}$ cannot be confirmed because of FAA radar interference. The very strong RFI presence due to FAA radar from 14,600 to 16,000 km s$^{-1}$ most likely contributes to a lack of detections of C21 galaxies in that velocity range. ALFA ZOA detects an underdensity from 16,000 to 20,000 km s$^{-1}$, at velocities beyond the range of 2MRS and the Erdogdu density reconstruction maps. There are no detections above the completeness limit in that velocity range though an average universe would contain 10 galaxies above the completeness limit in that region. This is a statistically significant underdensity at about a 99\% confidence level. There is narrow RFI around 18,500 km s$^{-1}$ that affects sensitivity at that velocity and the FAA radar starting at 20,000 km s$^{-1}$ reduces sensitivity to zero at the edge of the cube's velocity cutoff. Figure \ref{deepLSSslice} shows sky distribution plots of ALFA ZOA and 2MRS detections for the three volumes where ALFA ZOA detects an overdensity. While near infrared maintains a clear Zone of Avoidance, ALFA ZOA easily traces large scale structure through the lowest Galactic latitudes.

\section{\label{conclusion} Conclusion}

This first data release of the ALFA ZOA Deep Survey demonstrates that the survey achieves its expected noise level of 1 mJy (at 9 km s$^{-1}$ resolution) and it is complete above $F_{HI}$ = 0.5 Jy km s$^{-1}$. First results display the ability to detect galaxies out to nearly $v_{hel}$ = 19,000 km s$^{-1}$ and at extinctions that surveys at other wavelengths struggle to penetrate. ALFA ZOA Deep has completed 15 square degrees in the outer Galaxy and is continuing to take data, intending to cover about 300 square degrees in both the inner (30$^\circ \le l \le 75^\circ; b \le |2^\circ|$) and outer (175$^\circ \le l \le 207^\circ; -2^\circ \le b \le +1^\circ$) Galaxy, over the next several years.

As results from the full survey come in, asymmetry in large scale structure beyond 100 h$^{-1}$ Mpc, a possible explanation for the discrepancy in the mass density dipole compared to the CMB dipole, will be probed in regions of the inner and outer Galaxy. The ability to do so has been demonstrated by the survey's first results. Also shown is a detection rate that indicates the full survey's ability to detect over a thousand galaxies, which can be used to measure of the low mass end of the H\,{\sc{i}} Mass Function out to farther distances than have been measured previously by large, blind H\,{\sc{i}} surveys.

\section{Acknowledgments}

The authors would like to thank former UNM graduate student G. Vaive-Barron for contributing to source detection efforts. The Arecibo Observatory is operated by SRI International under a cooperative agreement with the National Science Foundation (AST-1100968), and in alliance with Ana G. Mendez-Universidad Metropolitana, and the Universities Space Research Association. This work was supported in part by the Cornell NAIC Predoctoral Fellowship and the NASA New Mexico Space Grant program. The National Radio Astronomy Observatory is a facility of the National Science Foundation operated under cooperative agreement by Associated Universities, Inc. This research has made use of the NASA/IPAC Extragalactic Database (NED) which is operated by the Jet Propulsion Laboratory, California Institute of Technology, under contract with the National Aeronautics and Space Administration.

\end{document}